\documentclass[iop]{emulateapj}
\usepackage{natbib}
\usepackage{subfigure}

\newcommand{\myemail}{kafrank@purdue.edu}

\newcommand{\emkTalpha}{2.69\pm0.24}
\newcommand{\emkTbeta}{65.45\pm0.15}
\newcommand{\lowemkTalpha}{3.96\pm0.98}
\newcommand{\lowemkTbeta}{65.03\pm0.35}
\newcommand{\nolargeemkTalpha}{2.18\pm0.29}
\newcommand{\nolargeemkTbeta}{65.82\pm0.19}

\newcommand{\kTsigmaalpha}{0.20}
\newcommand{\kTsigmabeta}{1.08}
\newcommand{\kTsigmalinearchi}{242}
\newcommand{\kTsigmaconstant}{1.38}
\newcommand{\kTsigmaconstantchi}{333}
\newcommand{\kTsigmadeltachi}{91}

\newcommand{\intscatter}{$\sim0.55$ keV}

\newcommand{\nclusters}{62 } 

\shorttitle{ICM Temperature Distributions}
\shortauthors{Frank et al.}

\slugcomment{{\sc Accepted to Apj:} December 10, 2012}

\begin{document}

\title{Characterization of ICM Temperature Distributions of 62 Galaxy Clusters with XMM-Newton}

\author{K. A. Frank, J. R. Peterson}
\affil{Department of Physics, Purdue University, 525 Northwestern Ave, West Lafayette, IN 47907}
\email{\myemail}
\author{K. Andersson}
\affil{Department of Physics, Ludwig-Maximilians-Universit\"at, Scheinerstr.~1, 81679 M\"unchen, Germany} 
\and
\author{A. C. Fabian, J. S. Sanders} 
\affil{Institute of Astronomy, Madingley Road, Cambridge CB3 OHA, UK}

\begin{abstract}

We measure the intracluster medium temperature distributions for 62 galaxy clusters in the HIFLUGCS, an X-ray flux-limited sample, with available X-ray data from XMM-Newton.  We search for correlations between the width of the temperature distributions and other cluster properties, including median cluster temperature, luminosity, size, presence of a cool core, AGN activity, and dynamical state.  We use a Markov Chain Monte Carlo analysis which models the ICM as a collection of X-ray emitting smoothed particles of plasma. Each smoothed particle is given its own set of parameters, including temperature, spatial position, redshift, size, and emission measure. This allows us to measure the width of the temperature distribution, median temperature, and total emission measure of each cluster. We find that 
none of the clusters have a temperature width consistent with isothermality.  Counterintuitively, we also find that the temperature distribution widths of disturbed, non-cool-core, and AGN-free clusters tend to be wider than in other clusters.  A linear fit to $\sigma_{kT}-kT_{med}$ finds $\sigma_{kT}\sim\kTsigmaalpha kT_{med}+\kTsigmabeta$, with an estimated intrinsic scatter of \intscatter, demonstrating a large range in ICM thermal histories.

\end{abstract}

\keywords{Galaxies: clusters: general --- Galaxies: clusters: intracluster medium --- X-rays: galaxies: clusters}

\section{Introduction}
\label{section:intro}

Galaxy clusters have proven to be invaluable tools for cosmological studies.  The largest gravitationally bound objects in the Universe, clusters trace their origins to fluctuations in the primordial density field.  As a result, the evolution of galaxy clusters is intimately linked with the growth of large scale structure. Specifically, the cluster mass function, $n(M,z)$, is sensitive to $\Omega_M$ and $\Omega_\Lambda$, the matter and dark energy densities of the universe, and $\sigma_8$, the mass fluctuation amplitude on scales of $\sim8h^{-1}$ Mpc.  Thus measuring the mass of clusters allows us to probe large scale structure in the universe. Mass measurements for a range of redshifts can then help to trace the formation and evolution of this structure.  The viability of using clusters as cosmological probes relies on obtaining accurate cluster masses.  Typically, this requires using an observable mass proxy, such as X-ray temperature or luminosity, which scales predictably with total cluster mass.  However, significant scatter in such scaling relations is a major source of error in cluster mass measurements.  It has been found that cluster X-ray morphology and temperature substructure, particularly the presence of dense, cool cores, significantly affects the scatter in X-ray scaling relations \citep{Markevitch1998,Arnaud1999,Allen1998a,Zhang2007,Ventimiglia2008}.  A better grasp of cluster physics is clearly essential to understanding the scaling relations and thus obtaining more precise cosmological measurements.  

The complex physics of galaxy clusters results in a wealth of temperature substructure.  Cluster-cluster interactions, including mergers, radiative cooling, and AGN feedback can induce shocks and cold fronts \citep{Markevitch2007}, dense cool cores, X-ray cavities, and soft X-ray filaments \citep[e.g.,][]{Crawford2005,Fabian2011}.  Additionally, the cold gas expected from traditional cooling flow models is largely missing from cluster cores, observations instead revealing a cutoff in the temperature distributions at $T\sim T_{max}/3$ \citep{Peterson2003a}, despite some evidence of minor cooling flow activity in the form of star-formation \citep{Bildfell2008}, H$\alpha$ filaments \citep{McDonald2010}, and a small amount of cold X-ray gas \citep[e.g., ][]{Sanders2008,Sanders2009,Sanders2009a,Sanders2010,Peterson2003a}.  Thus it appears that while cooling flows are not completely quenched, some amount of heating is necessary to explain the missing low-temperature gas in cluster cores.  There are many proposed heating mechanisms, some of the most common being turbulence, thermal conduction, or some form of AGN feedback.  The most popular is AGN feedback via inflation of cavities \citep{McNamara2007}.  It has been shown that there is enough energy available in X-ray cavities to adequately balance radiative cooling \citep{Rafferty2006,Birzan2004}, but the exact nature of how this energy is distributed throughout the ICM is unknown. 

It is clear that cluster temperature structure is strongly tied to the dynamical history of the cluster.  Thus, a more thorough understanding of cluster temperature distributions may lead to reduced scatter in the X-ray scaling relations used for cosmology measurements and shed light on the cooling flow problem, cluster heating, and cluster-cluster interactions.  Most measurements of cluster temperature distributions treat the ICM as a multiphase plasma, with anywhere from one (isothermal) to four temperature components.  Typically, this also involves assumptions on the spatial distribution of the multi-temperature plasma.  Using the Smoothed Particle Inference (SPI) method introduced by \citet{Peterson2007} it is possible to avoid such spectral and spatial assumptions by treating the ICM as a collection of plasma parcels, each emitting X-rays according to an independent set of parameters, including emission measure, temperature, spatial position, and size.  These photons are then propagated through the XMM-Newton instrument response and the model parameters adjusted via Markov Chain Monte Carlo.  The resulting distribution of smoothed particle parameters provides a good description of the cluster.  In this way, we are able to measure the full temperature distribution, across all temperatures, while at the same time allowing for both temperature and luminosity substructure and asymmetry.  We aim to characterize cluster temperature distributions in several ways.  First, we investigate how well we are able to recover the shape of the temperature distribution.  Second, we measure the width of the distribution, $\sigma_{kT}$, as well as the cluster median temperature, total emission measure, and $r_{2500}$.  Third, we search for correlations of $\sigma_{kT}$ with other cluster properties, such as median temperature.  Information on each cluster in our sample is gathered from the literature in order to determine if cluster dynamical state, central AGN activity, or cool-core status has any effect on the temperature distributions.

Assumed cosmology throughout the paper is $H_0 = 70$ km s$^{-1}$Mpc$^{-1}$, $\Omega_M = 0.3$, and $\Omega_{\Lambda} = 0.7$.

\section{Data}
\label{section:data}

\subsection{Data Sample}
\label{section:datasample}
Our cluster sample is the HIghest X-ray FLUx Galaxy Cluster Sample (HIFLUGCS) compiled by \citet{Reiprich2002a}.  HIFLUGCS is a statistically complete flux-limited sample, containing the 63 X-ray brightest clusters in the sky (excluding the Galactic plane).  The HIFLUGCS also contains a variety of cluster morphologies, including merging clusters, cool core and non-cool core clusters,  and both galaxy clusters and galaxy groups.  The clusters have redshifts in the range $0.003\lesssim z\lesssim0.2$ and span a wide range in temperature, $0.8\mbox{ keV}\lesssim T_X\lesssim 13\mbox{ keV}$.  We selected this sample because we wanted to both span the range of cluster types and have enough photons in the data to allow detailed measurements.  

Of the 63 clusters in the sample, 62 have been observed with XMM-Newton (all except Abell 2244).  For clusters with multiple available observations, we choose the longest observation (after flare screening) that is nearest the cluster center.  We use the cluster optical redshifts as given in the NASA/IPAC Extragalactic Database (NED).  Galactic hydrogen column densities are inferred from the 21cm  radio measurements of \citet{Dickey1990}, as given in \citet{Reiprich2002a}.  The observation details are given in Table \ref{table:observations}.

\subsection{Data Reduction}
\label{section:datareduction}
Our analysis requires only a photon event list and exposure map for each XMM-Newton detector (MOS1, MOS2, pn, RGS1, and RGS2).  These were created using the SAS 10.0 pipeline tasks \textit{emchain}, \textit{epchain}, and \textit{rgsproc}.  Light curves were created from the resulting event files to identify periods affected by soft proton flares, which were then removed from the data for the remainder of the analysis.  Net exposure times, after flare screening, are listed in Table \ref{table:observations} for each observation.  In addition, EPIC event files were filtered to include only photon event patterns $0-12$ (to exclude non-X-ray events) and photon energies in the range $0.3-10.0$ keV (MOS) and $1.1 - 10.0$ keV (pn) \citep{Andersson2004}.

\section{Analysis}
\label{section:analysis}
We employ the smoothed particle inference method presented in Peterson, Marshall, and Andersson (2007) as an alternative to traditional X-ray analysis procedures for galaxy clusters. The smoothed particle method models the intracluster medium as a collection of X-ray emitting parcels of plasma. We refer to these parcels as smoothed particles, and each represents a large volume of plasma in the ICM.  Each smoothed particle is characterized by a set of parameters, including temperature, emission measure, gaussian spatial width, redshift, and abundance. The smoothed particles emit photons according to the chosen spectral and spatial model ($\S$\ref{section:model}), which are then propagated through the XMM-Newton instrument response. The resulting simulated data are then iteratively compared with the XMM-Newton data via a Markov Chain Monte Carlo (MCMC) process ($\S$\ref{section:spi}). The EPIC and RGS data are processed separately for each observation and the results combined prior to measuring cluster properties ($\S$\ref{section:rgsepiccombine}).  We use the combined MCMC results to measure the cluster radius, emission measure, median temperature, and temperature distribution width, $\sigma_{kT}$ ($\S$\ref{section:measurements}).

\subsection{Smoothed Particle Model}
\label{section:model}
The model for each smoothed particle consists of multiple components: the ICM model, a Galactic and extragalactic X-ray background, and the instrumental background.  Though some parameters are global (the same for all smoothed particles), the four-component model applies independently to each smoothed particle.  The fractions of photons going to each model component, as well as the relative normalizations, are free parameters.  The model parameters are summarized in Table \ref{table:modelparams}.  

\subsubsection{ICM Model}
\label{section:icmmodel}
For our analysis, the smoothed particles are spatially modeled as gaussians.  The gaussian width as well as the two-dimensional position of each smoothed particle is allowed to vary independently.  The ICM spectrum of each smoothed particle is produced by a WABS \citep{Morrison1983} absorbed, isothermal MEKAL (\citealt{Mewe1985}, \citeyear{Mewe1986}; \citealt{Kaastra1992}, \citealt{Liedahl1995}) model.  Individual smoothed particles thus have a single temperature which may be different from every other smoothed particle. The prior for the distribution of smoothed particle temperatures is logarithmically uniform between $0.1\mbox{ keV} \le kT \le 10.0\mbox{ keV} $, except for EPIC observations  of clusters with $kT_{avg} \ge 4$ keV.  Since these clusters may have some gas at temperatures $>10$ keV, which EPIC can detect, the prior for these clusters is set to $0.1\mbox{ keV}  \le kT \le 19.5\mbox{ keV} $.  The absorbing column density, $n_H$, and the smoothed particle metallicity, $Z$, are also allowed to vary; however they are global parameters.  Galactic column densities and redshifts are those listed in Table \ref{table:observations}. The smoothed particle metallicities are with respect to the solar abundances of \citet{Anders1989} and use a uniform prior from $0.0Z_\odot - 2.0Z_\odot$.  

\subsubsection{X-ray Background}
\label{section:xraybackground}
The XMM-Newton data include not only photons emitted by the ICM, but also photons from the X-ray background. The X-ray background has three components: a soft Galactic X-ray background, a hard extragalactic X-ray background, and the instrumental background. To account for these background photons, additional components are added to the model described in $\S$\ref{section:icmmodel}.  The normalization of each component is a free parameter.

The Galactic X-ray background consists primarily of soft X-ray photons of energies $\lesssim1$ keV with a spectrum consisting of several thermal components from $0.05-0.5$ keV \citep{Lumb2002,Kuntz2000a,Snowden2008}.  Following the model of \citet{Andersson2009}, based on analysis of blank sky data files, we use a spatially uniform, unabsorbed, isothermal MEKAL model to describe the Galactic X-ray emission. For all smoothed particles, the soft X-ray background temperature is fixed at $0.15$ keV, the redshift is fixed at $z=0$, and the metallicity is fixed at $Z=1.0Z_\odot$.

Unresolved extragalactic sources, such as AGN, are responsible for hard (energy $\gtrsim1$ keV) X-ray background emission. The spectrum is well described by a power-law with an index of $\Gamma = 1.47$ and is spatially uniform. A model component with a spatially uniform, WABS absorbed, power law spectrum (with $\Gamma = 1.47$) is used to represent the extragalactic background. The absorbing column density is fixed to the value of $n_{H}$ for each cluster (Table \ref{table:observations}).

The RGS and EPIC instrumental backgrounds (described in \citealt{denHerder2001} and \citealt{Lumb2002}, respectively), consist of several components.  A soft proton background is modeled as a power law with $\Gamma=-0.205~(-0.35)$ for EPIC (RGS).  Electronic noise is modeled as an exponential, $F\propto e^{-E/E_i}$, with $E_i=158~(500)$ eV for EPIC (RGS).  We also model instrumental line emission.  In addition, RGS has four calibration sources, which are modeled as 2D spatial gaussians.  For further discussion on our implementation of the RGS and EPIC instrumental backgrounds, see \citet{Peterson2003a} and \citet{Andersson2007}.  The parameters for the instrumental background component are fixed for all smoothed particles, and are the same for all observations. They only differ between the EPIC and RGS instruments.

\subsection{Smoothed Particle Inference}
\label{section:spi}
The four spectral and spatial model components described in $\S$\ref{section:model} are combined for the MCMC analysis. For each smoothed particle, photons are emitted according to this model, which are then propagated through the instrument response (either EPIC or RGS) and compared with the input XMM-Newton data.  

Some photons in the data are removed based on spatial position prior to beginning the MCMC analysis. The input images are $30\times30$ square arcminutes for EPIC and $5$ arcminutes by $30$ arcminutes for RGS.  EPIC photons outside a $20'\times20'$ square centered on the cluster are removed, or outside a $30'\times30'$ region for clusters in which the core radius, as given in \citet{Reiprich2002a}, extends farther than 10' from the observation center (only applies to 4 clusters).  RGS photons are removed if more than $3'$ ($2.5'$) from the center in the dispersion (cross-dispersion) direction.  Photons are also removed according to energy. EPIC MOS1 and MOS2 photons are only kept if they have energies $0.3\mbox{ keV} < E < 10.0$ keV. Similarly, pn photons in the energy range $1.1\mbox{ keV} < E <10.0$ keV are kept \citep[cf.][]{Andersson2004}.  The number of EPIC photons is increased by less than 2\% if the pn energy range is extended down to 0.3 keV, and thus the difference in MOS and pn energy cuts does not have a significant affect on analysis.  Wavelength cuts for RGS are $5\mbox{\AA}<\lambda<65\mbox{\AA}$ ($0.2\mbox{ keV} - 2.5$ keV).  There are several observations in which the cluster is outside of the RGS field of view and therefore only the EPIC data are used for these observations.

The number of smoothed particles ($N_{sp}$) and the oversim ratio ($r$, the ratio of simulated photons to data photons) are chosen such that 
\begin{equation}
\label{eqn:ratio}
\frac{N_p r}{4N_{sp}} = 10^4,
\end{equation}
where $N_p$ is the number of photons in the data, after making the cuts discussed in the above paragraph. The reasoning for this criteria is described in $\S$\ref{section:isf}.  $r$ is typically chosen to be $10$.  However, to mitigate the already large computing time, if $N_p > 1.5\times10^6$, we use $r=1$ and 10 times fewer smoothed particles, in accordance with Eqn. \ref{eqn:ratio}.  We also set a lower limit to the number of smoothed particles at $N_{sp} \ge 10$.  This limit comes into effect for observations with very few photons (typically RGS observations).  In these cases, $r=100$ and $N_{sp}$ is increased accordingly.  Values of $N_{sp}$ range from $10$ to several hundred.

The MCMC typically converges by iteration 500; however to be conservative, we automatically disregard all iterations prior to 1000.  Convergence is verified by binning the iterations (300 iterations per bin) and measuring the average $\chi^2$ and primary cluster parameters (as described in $\S$\ref{section:measurements}), except cluster size, for each bin.  The chain is considered to be converged when these values agree within $1\sigma$ for $10$ neighboring bins (the average $\chi^2$ must be within 2\%).  The MCMC is allowed to run until a minimum number of iterations (after convergence), $N$, is met.  This minimum depends on $N_{sp}$ for the observation, such that the total number of smoothed particles to be used in our analysis, $N_{sp}N$, is $3\times10^5$ for EPIC and $1\times10^5$ for RGS.  However, a minimum of $2000$ iterations are required for all observations, regardless of $N_{sp}$.  For high quality observations with many smoothed particles, this results in using iterations 1000 to 3000 for our analysis.  Computation times vary from several days to a few months.  We run the MCMC on a pool of parallel machines.  

\subsection{Measurement of Cluster Properties}
\label{section:measurements}

The desired cluster properties can be calculated directly from the MCMC results by analyzing the distribution of individual smoothed particle parameters. The primary properties of interest are median cluster temperature ($kT_{med}$), the width of the temperature distribution ($\sigma_{kT}$), a characteristic cluster radius ($r_{2500}$), and the total emission measure ($EM$) or luminosity.  Uncertainties for most of these properties are also relatively simple to calculate using the standard deviations of the MCMC results ($\S$\ref{section:staterrors}).  

\subsubsection{Combining EPIC and RGS}
\label{section:rgsepiccombine}
\begin{figure*}[p]
\begin{center}
\includegraphics[width=0.6\columnwidth]{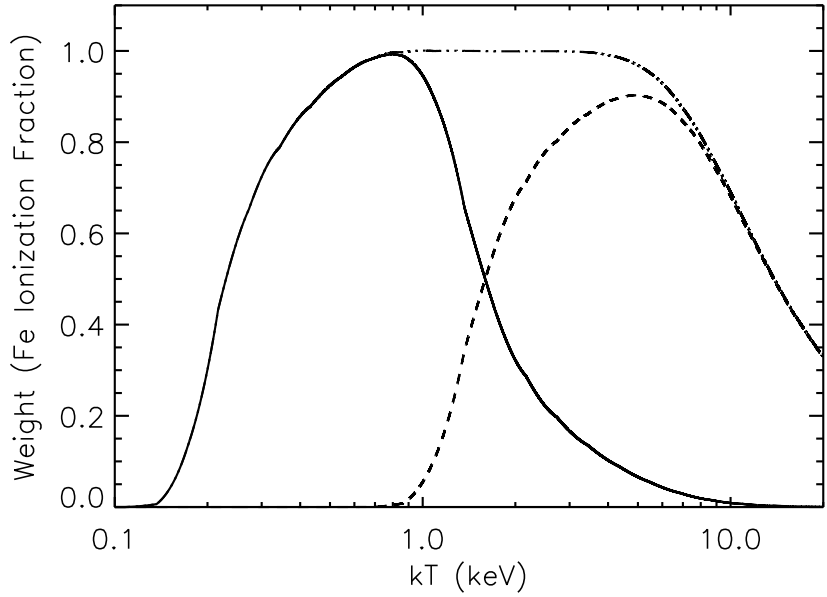}
\caption{\footnotesize Ionization fractions of Fe K and Fe L as a function of temperature.  These ionization fractions are used to weight the RGS and EPIC MCMC results as a function of temperature before combining.  RGS smoothed particles are weighted by Fe L (solid, $W_{\mbox{\tiny{RGS}}}$).   $W_{\mbox{\tiny{EPIC}}}$ is a piecewise function, such that EPIC smoothed particles which are within the RGS field of view are weighted by Fe K (dashed) and EPIC smoothed particles which do not overlap with RGS are weighted with the Fe K + Fe L ionization fractions (dot-dash).}
\label{figure:weights}
\end{center}
\end{figure*}

To take full advantage of the different energy sensitivities of RGS and EPIC, the MCMC results from EPIC and RGS observations are combined for the purpose of measuring cluster properties.  In fact, to measure the entire X-ray temperature distribution, the use of both is essential.  However, the EPIC field of view is larger than that of RGS (the RGS field of view is a 5' strip across the center of the EPIC field of view), and so care must be taken when combining the smoothed particle parameter distributions.  Within the RGS field of view, where there are smoothed particles from both the RGS and EPIC MCMC results, the aim is to give more weight to the RGS observation at lower temperatures and to the EPIC observation at higher temperatures.  The RGS instrument determines temperature primarily via K- and L-shell emission lines of C, N, and O, as well as Fe L- and M-shell emission lines.  EPIC is primarily sensitive to Fe K-shell emission lines.  The temperature sensitivity of RGS and EPIC can therefore be roughly characterized by the ionization fraction of Fe L and Fe K, respectively, as a function of temperature.  Note that the shape of the bremsstrahlung continuum is important when using EPIC to measure a single cluster temperature.  However, it is difficult to disentangle multiple temperature components based on the continuum; consequently, it is more appropriate to use the Fe K emission lines to characterize the ability of EPIC to measure the temperature distribution.  In the overlapping region, we therefore weight the RGS results (the individual smoothed particle parameters) by the Fe L ionization fractions and the EPIC results by the Fe K ionization fractions (see Figure \ref{figure:weights}).  This ensures that where RGS data is available, it is used preferentially for determining the temperature distribution where it is more sensitive than EPIC, namely at temperatures $\lesssim 1.76$ keV, and similarly, that EPIC data is always preferred at higher temperatures.  Outside of the RGS field of view, the EPIC results are weighted by the Fe K + Fe L ionization fractions.  In this way we are able to take advantage of the greater sensitivity of RGS at low temperatures, while at the same time utilizing the higher temperature sensitivity and larger field of view of EPIC.  Note that RGS data contains very little spatial information, other than that based on its field of view, and therefore this method should not be applied if investigating the spatial dependence of the temperature distribution, e.g. for making temperature maps.  For the purpose of this work, it is only relevant whether or not RGS detects gas at a given temperature, not where this gas is located.  

\subsubsection{$r_{2500}$}
\label{section:R}
The characteristic radius of each cluster is chosen to be $r_{2500}$, the radius within which the mean mass density is $2500$ times the critical density of the universe at that redshift. 
To determine $r_{2500}$, we employ the $M - T_X$ scaling relation of \citet{Arnaud2005a},
\begin{equation}
\label{eqn:ArnaudMT}
h(z)M_{2500}=1.69\left(\frac{kT}{5\mbox{ keV}}\right)^{1.70},
\end{equation}
and calculate the radius within which this is true for the measured cluster temperature,
\begin{equation}
\label{eqn:rM}
r_{2500} = \left( \frac{3*M_{2500}}{4\pi\delta\rho_c}\right)^{1/3}
\end{equation}
The temperature used in \citet{Arnaud2005a} is the core-excised spectroscopic temperature from $0.1-0.5r_{200}$.  However, for the purposes of this work, it is not crucial to have a precise estimate of $r_{2500}$, and we use here the emission-weighted median temperature of all smoothed particles (all of which are within the spatial cuts \S\ref{section:spi}).  Excising the core prior to measuring the temperature, as is done in \citet{Arnaud2005a}, only changes our $r_{2500}$ measurements by a few percent.  $r_{2500}$ is the first cluster property measured, and all cluster properties are measured within this radius by only considering smoothed particle emission within $r_{2500}$.


\subsubsection{Median Temperature}
\label{section:kTmed}
There are many different ways to define a single temperature for a cluster.  We choose to use a temperature $kT_{med}$ defined by the median $log~kT$ of the smoothed particles.  Smoothed particles from both RGS and EPIC, within $r_{2500}$, are used.  For each smoothed particle, $log kT$ is weighted by both its emission measure and $W_{\mbox{\tiny{RGS}}}(kT)$ or $W_{\mbox{\tiny{EPIC}}}(kT)$ (where $kT$ is the smoothed particle's temperature and $W_{\mbox{\tiny{RGS}}}(kT)$ or $W_{\mbox{\tiny{EPIC}}}(kT)$ is the ionization fraction weight for a smoothed particle of temperature $kT$).  Recall that for EPIC smoothed particles which are outside the RGS field of view, $W_{\mbox{\tiny{EPIC}}} $ is equal to the sum of both the Fe K and Fe L ionization fractions (Figure \ref{figure:weights}), regardless of the smoothed particle temperature.  The median of this weighted distribution then defines the cluster's median temperature, $kT_{med}$.

\subsubsection{Width of Temperature Distribution}
\label{section:sigmakT}
All of the other primary cluster properties - physical size, total emission measure, and temperature - can be found using traditional analysis methods.  The main advantage of our method is that by constructing a continuous temperature distribution we can also measure the width of this distribution, $\sigma_{kT}$.  

Both RGS and EPIC smoothed particles (again within $r_{2500}$) are used in the determination of $\sigma_{kT}$.  As with the $kT_{med}$ measurement, each smoothed particle's temperature is weighted by its emission measure and $W_{\mbox{\tiny{RGS}}}(kT)$ or $W_{\mbox{\tiny{EPIC}}}(kT)$.  The standard deviation of the weighted temperature distribution is then taken to be $\sigma_{kT}$, as given in Eqns. \ref{eqn:sigmakT} and  \ref{eqn:avgkT}.  Here, $kT_j$ and $EM_j$ are the temperature and emission measure of the $j^{th}$ smoothed particle.  $W_j$ is either $W_{\mbox{\tiny{RGS}}}(kT_j)$ or $W_{\mbox{\tiny{EPIC}}}(kT_j)$.  The sum is over all RGS and EPIC smoothed particles.  This value is then corrected for the isothermal spread function, as described in $\S$\ref{section:isf}.

\begin{equation}
\label{eqn:sigmakT}
\sigma_{kT}^2=\frac{\displaystyle\sum_j^{N_{RGS}+N_{EPIC}} {[kT_j - \overline{kT}]^2  EM_jW_j}}{\displaystyle\sum_j^N EM_jW_j}
\end{equation} 

\begin{equation}
\label{eqn:avgkT}
\overline{kT}=\frac{\displaystyle\sum_j^{N_{RGS}+N_{EPIC}} kT_j   EM_jW_j}{\displaystyle\sum_j^N EM_jW_j}
\end{equation}  

\subsubsection{Emission Measure}
\label{section:EM}
Rather than explicitly calculating the luminosity, we choose to use a closely related quantity, emission measure, given by Eqn. \ref{eqn:emdef}.  
\begin{equation}
\label{eqn:emdef}
EM=\int n_e n_i dV
\end{equation}
It is directly related to the X-ray luminosity through the cooling function, $\Lambda(T)$.  For most clusters, $\Lambda\sim T^{1/2}$, i.e. only bremsstrahlung emission is significant, but for lower temperature clusters the effect of line emission begins to alter the cooling function.  Therefore when comparing clusters over a wide range of temperatures, where the cooling function is not constant, it seems more appropriate to use the emission measure rather than the luminosity.

It is relatively straightforward to calculate the emission measure for each smoothed particle (and thus also for the entire cluster) from the MCMC results.  One of the smoothed particle parameters is the MEKAL normalization ($\S$\ref{section:icmmodel}).  This is related directly to the emission measure (Eqn. \ref{eqn:emissionmeasure}).
\begin{equation}
\label{eqn:emissionmeasure}
norm_{MEKAL}=\frac{10^{-14}}{4\pi D_A^2(1+z)^2} EM
\end{equation} 
The total emission measure for the cluster is simply the weighted sum of the smoothed particle emission measures, averaged over the number of iterations used, as in Eqn. \ref{eqn:totalem}, 
\begin{equation}
\label{eqn:totalem}
EM_{total}= \frac{1}{N_{\mbox{\tiny{RGS}}}}{\displaystyle\sum_j ^{N_{\mbox{\tiny{RGS}}}}  W_jEM_j} + \frac{1}{N_{\mbox{\tiny{EPIC}}}}{\displaystyle\sum_k ^{N_{\mbox{\tiny{EPIC}}}}  W_kEM_k}
\end{equation} 
where the sums are over all the RGS ($j$) and EPIC ($k$) smoothed particles, and $N_{\mbox{\tiny{RGS}}}$ and $N_{\mbox{\tiny{EPIC}}}$ are the number of RGS and EPIC iterations.  The sum over RGS smoothed particles yields the contribution to the total emission measure of cool gas within the RGS field of view.  The sum over EPIC smoothed particles comprises the contributions of the higher temperature gas within the RGS field of view and gas of all temperatures outside the RGS field of view.

\subsubsection{Statistical Uncertainties}
\label{section:staterrors}
One of the advantages of using  a Markov Chain Monte Carlo is that it comes with a built-in estimate of statistical uncertainties.  Once the chain has converged, the parameters from any individual iteration comprise a statistically acceptable representation of the cluster.  The (weighted) contribution of RGS to $EM$, $kT_{med}$, and $\sigma_{kT}$ is calculated separately for each RGS iteration and the standard deviation of these values then comprises the RGS contribution to the error.  The same is done for EPIC, and error on the combined quantity is then calculated using standard propagation of error techniques.  In the case of $r_{2500}$, the error is calculated using propagation of error on Equations \ref{eqn:ArnaudMT} and \ref{eqn:rM}, including the errors on $kT_{med}$ and the two $M-T_X$ fit parameters from \citet{Arnaud2005a}.


\subsection{Isothermal Spread Function}
\label{section:isf}
There is an intrinsic limit to how well the temperature distribution can be resolved.  This limit is affected by two major factors.  The first is the statistical ability of the data to determine the temperature distribution, given the imperfect instrument response and finite number of photons.  Secondly, even with a perfect instrument and infinite number of photons, the spectroscopic information from atomic transitions has an incomplete set of information about the temperature distribution.  The effect of this limit is to introduce extra broadening to the temperature distribution, similar to the effect of a point-spread function, so that the measured temperature width of even an isothermal cluster will be nonzero.  We must correct for this 'Isothermal Spread Function' (ISF) in order to obtain a measurement of the cluster's intrinsic temperature width.  We quantify the ISF by simulating a set of isothermal clusters, whose true $\sigma_{kT}=0$.  The simulated clusters are analyzed as described in $\S$\ref{section:measurements} and their $\sigma_{kT}$ values are taken to be measurements of the ISF, $\sigma_{ISF}$.  

Clusters were simulated according to an isothermal MEKAL spectral model.  The spatial distribution of each cluster is described by a $\beta$-model, with $r_c=60"$ and $\beta=0.8$.  $\rho_0$ is chosen to get the desired number of photons, $N_p$.  As a measure of the ability to measure ICM temperatures, the ISF is a spectral phenomenon, and is therefore not expected to be sensitive to spatial characteristics of the X-ray emission (e.g. $r_c$ or $\beta$).  Energy and spatial cuts are the same as  in $\S$\ref{section:spi}.    We determined the effect of four parameters on the ISF: the cluster temperature ($kT$), the number of smoothed particles ($N_{sp}$), the number of photons in the data ($N_p$), and the ratio of simulated photons to photons in the data (the 'over-simulate factor' $ r$).  Combining these effects, we find the ISF can be approximated as $\sigma_{ISF}= \sigma_{1}(N_p r/N_{sp}) \sigma_{2}(kT)$.  

\subsubsection{$N_p$, $N_{sp}$, $ r$}
\label{section:isfratio}
The first group of simulated clusters determined the effect of $N_p$, $N_{sp}$, and $ r$ on the ISF.  The temperature of all simulated clusters in this group was set to $1$ keV.  The values of  $N_p$, $N_{sp}$, and $ r$ for the model were chosen from the following: $N_p=\{10^3,10^4,10^5\}$. $N_{sp}=\{10,100,1000\}$, and $ r=\{1,10,100\}$.  All possible combinations of these parameters were simulated for both RGS and EPIC (with the exception of the RGS $ r=100$, $N_p=10^5$, $N_{sp}=1000$ and EPIC $ r=100$, $N_p=10^5$, $N_{sp}=10$ simulations, due to the long computation time required),  resulting in 26 measurements of the ISF for EPIC and 26 for RGS. 

\begin{figure*}[p]
\begin{center}
\includegraphics[width=0.6\columnwidth]{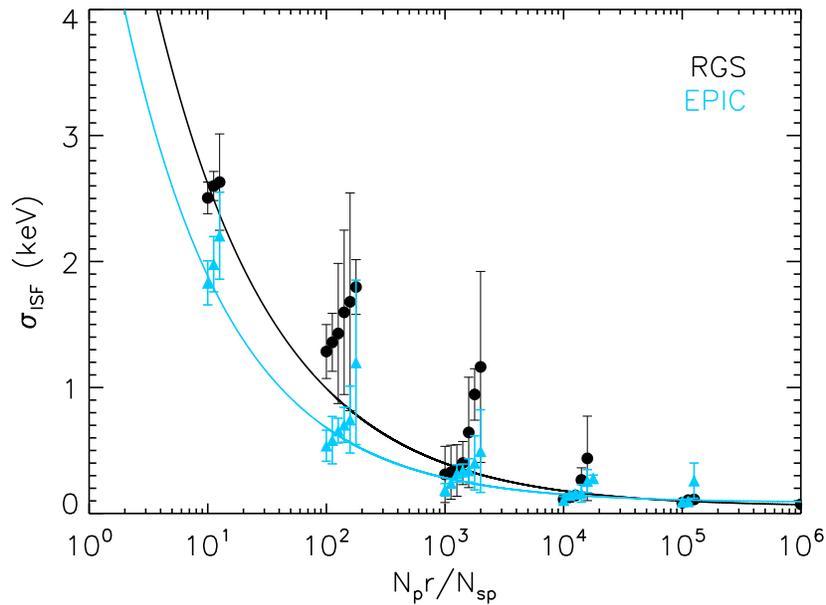}
\caption{\footnotesize{ISF measurements for EPIC (blue) and RGS (black) as a function of the ratio $N_p r/N_{sp}$ and the resulting best fits, as given in Eqns. \ref{eqn:ratiorgs} and \ref{eqn:ratioepic}.  Note that the ratio values have been incrementally shifted to the right to differentiate the error bars; all ratios are $10^1,10^2,10^3,10^4,10^5$, or $10^6$. }}
\label{figure:isfratio}
\end{center}
\end{figure*}

We found that the ISF depends on the ratio $N_p r/N_{sp}$, which is the average number of simulated photons per smoothed particle (Figure \ref{figure:isfratio}).  This ratio ranged from $1$ to $10^5$ in our simulations.  The best fit as a function of $N_p r/N_{sp}$ for RGS is

\begin{equation}
\label{eqn:ratiorgs}
\sigma_{1}(\frac{N_p R}{N_{sp}})\sigma_{2}(1\mbox{ keV})=6.88\left(\frac{N_p r}{N_{sp}}\right)^{-0.43}+0.05,
\end{equation}
and for EPIC is 
\begin{equation}
\label{eqn:ratioepic}
\sigma_{1}(\frac{N_p r}{N_{sp}})\sigma_{2}(1\mbox{ keV})=5.50\left(\frac{N_p r}{N_{sp}}\right)^{-0.49}+0.09.
\end{equation}

Unsurprisingly, the more simulated photons per smoothed particle, the better the temperature distribution can be constrained.  There proved to be too few simulated photons per smoothed particle in the $ r=1$, $N_p=10^3$, and $N_{sp}=1000$ simulations to constrain the temperature distribution (these were excluded from the $\sigma_{ISF}$ fits for this reason).  Thus, it appears we must have $N_p r/N_{sp}>1$, and given the swift increase in the size of the ISF, $10^3\le N_p r/N_{sp}\le10^4$ is preferable.  Higher values may decrease the ISF somewhat; however, they also drastically increase the already large computation time. We therefore choose $N_{sp}$ and $r$ for each dataset such that $N_p r/N_{sp}\approx10^4$.  Thus we are able to minimize that contribution to the ISF of the number of photons in the data by a choosing an appropriate number of smoothed particles and over-simulate factor.

\subsubsection{$kT$}
\label{section:isfkT}

\begin{figure*}[p]
\begin{center}
\includegraphics[width=0.6\columnwidth]{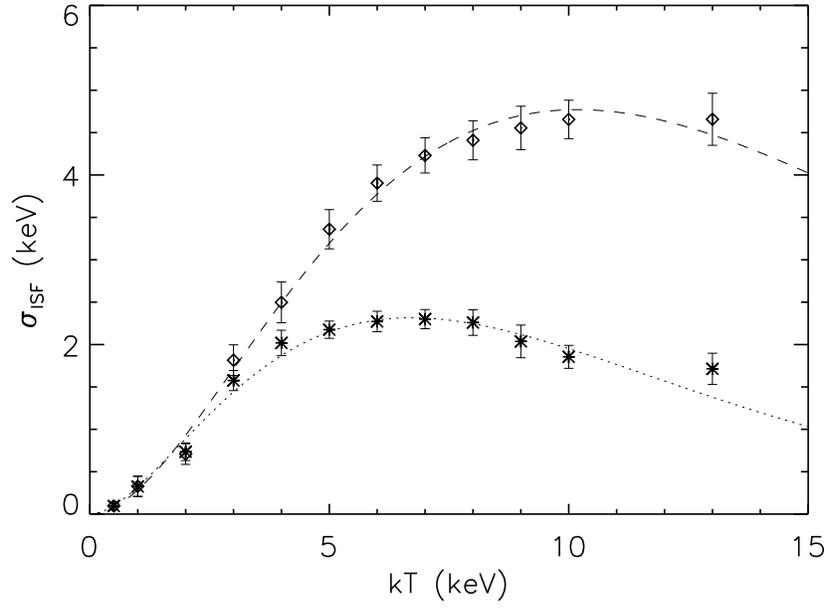}
\caption{\footnotesize{ISF measurements for temperature priors of $0.1\le kT\le 10.0$ keV (asterisks) and $0.1\le kT\le 19.5$ keV  (diamonds) as a function of the cluster temperature and the resulting best fits (dotted and dashed lines correspond to the $0.1\le kT\le 10.0$ keV and $0.1\le kT\le 19.5$ keV priors, respectively), as given in Eqns. \ref{eqn:tempcool} and  \ref{eqn:temphot}.  }}
\label{figure:isftemp}
\end{center}
\end{figure*}

We  minimize the contribution to the ISF due to the number of photons in the as described above. The cluster temperature is thus primarily responsible for determining the ISF.  We therefore simulated clusters with temperatures (in keV) of $kT=\{0.5,1,2,3,4,5,6,7,8,9,10,13\}$.  For each temperature, we simulated the same cluster (with a given $\rho_0$, $r_c$, and $\beta$) three times, with the following instruments and temperature priors: 1) RGS, $0.1\mbox{ keV}\le kT\le10.0$ keV, 2) EPIC, $0.1\mbox{ keV}\le kT\le10.0$ keV, and 3) EPIC $0.1\mbox{ keV}\le kT\le19.5$ keV.  $\rho_0$ was chosen at each temperature such that $N_p=10^5$ for EPIC.  $N_{sp}$ and $r$ were then chosen individually for each simulation according to $\S$\ref{section:spi}, such that $N_p r/N_{sp}=10^4$.  Each EPIC simulation was then combined with the RGS simulation of the same $kT$ and $\sigma_{ISF}$ measured according to $\S$\ref{section:sigmakT}.  The resulting $\sigma_{ISF}$ as a function of cluster temperature, measured in keV, for the EPIC temperature prior of $0.1\mbox{ keV}\le kT\le 10.0\mbox{ keV}$ is 
\begin{equation}
\label{eqn:tempcool}
\sigma_{ISF}(kT)=0.41kT^{2.13}e^{-0.35kT},
\end{equation} 
and for the $0.1\mbox{ keV}\le kT\le 19.5\mbox{ keV}$ temperature prior is
\begin{equation}
\label{eqn:temphot}
\sigma_{ISF}(kT)=0.43kT^{2.00}e^{-0.22kT}.
\end{equation} 
Both functions are shown in Figure \ref{figure:isftemp}.

\subsubsection{ISF Correction}
\label{section:correction}

For the MCMC analysis, the $N_p r/N_{sp}$ contribution to the ISF if minimized by setting $r$ and $N_{sp}$ such that $N_p r/N_{sp}\approx10^4$, where $N_p$ is estimated as one quarter of the total photons in the data (since there are four model components, and only one component represents the ICM).  The ISF is then calculated using Eqn. \ref{eqn:tempcool} or \ref{eqn:temphot}.  For the temperature, we use the combined median temperature as described in $\S$\ref{section:kTmed}.  The ISF is subtracted from the raw $\sigma_{kT}$ ($\S$\ref{section:sigmakT}) according to Eqn. \ref{eqn:sigmacorrection}.  All reported $\sigma_{kT}$ are corrected for the ISF in this manner.  An additional systematic error term is incorporated in the reported $\sigma_{kT}$ errors to account for uncertainties in the ISF measurements and any residual error due to the number of photons. 

\begin{equation}
\label{eqn:sigmacorrection}
\sigma_{kT} = \sqrt{\sigma_{kT,raw}^2 - \sigma_{ISF}^2}
\end{equation}

\section{Results}
\label{section:results}

\subsection{Cluster Morphology and Dynamical State}
\label{section:clustermorph}
In order to investigate possible connections between our results and cluster morphology or dynamical state, we have taken advantage of the well-studied nature of the HIFLUGCS sample and found information on each cluster's X-ray and radio morphology in the literature (see references in Table \ref{table:clusterparameters}).  For each cluster, its radio features are classified according to the taxonomy of \citet{Kempner2004}, the major categories being radio features that are associated with AGN and features that are associated with the ICM, which are usually indicative of merging activity (e.g. giant radio halos).  X-ray results for each cluster are used to determine if the cluster is known to have X-ray cavities (with or without coincident radio emission) and whether it is known to have a disturbed morphology (such as filamentary structures, secondary X-ray brightness peaks, or significant isophotal centroid shifts, e.g. as in \citet{Vikhlinin2009a}) or a regular, symmetric X-ray morphology.  It is also noted if the cluster is known to be undergoing a merger based on detailed X-ray observations.  

Based on this information, we divide the cluster sample two different ways.  First, we have separated clusters which are known to be relaxed from those which are disturbed in any way.  A cluster is designated as disturbed if it has any of the following features: evidence of merging activity (from detailed X-ray observations or large-scale radio emission), or a disturbed X-ray brightness morphology.  A cluster is designated as relaxed only if it has none of the latter and has a regular, symmetric X-ray morphology (which may include cavities).  A cluster typically satisfies several of the criteria for its designated category.  For all but six cases, this classification matches the disturbed/undisturbed designations of the HIFLUGCS clusters used in \citet{Vikhlinin2009a} and \citet{Zhang2011}, which is based solely on the X-ray brightness morphology.  These are all cases in which \citet{Vikhlinin2009a} classify a cluster as undisturbed, but more detailed X-ray (and in some cases radio) observations show strong evidence of merging activity.  Second, the clusters are classified according to whether or not they have a central AGN, according to \citet{Mittal2009}.  
Additionally, we use the results of \citet{Hudson2010}, which divide the entire HIFLUGCS sample into strong, weak, and non-cool core clusters (SCC, WCC, and NCC) based on central cooling time, providing a third perspective from which to view the results.  See Table \ref{table:clusterparameters} for each cluster's classification and the associated references.  
  
\subsection{Temperature Distributions}
\label{section:tempdists}
Continuous temperature distributions for all clusters in our sample are shown in Figures \ref{figure:tempdists0} - \ref{figure:tempdists3}.  The measured temperature distribution is a combination of the intrinsic ICM temperature distribution and the ISF.  To determine the relative contribution of the ISF to the shape of the temperature distributions, we directly compared the temperature distributions of the simulated clusters to clusters from the HIFLUGCS sample (Figure \ref{figure:tempdistscomp}).  Least squares fitting to both the simulated cluster and HIFLUGCS distributions show they are both approximately log-normal.  Attempts to fit the distributions with a single common function yielded similar $\chi^2$ values.  Low temperature peaks are seen in many of the temperature distributions from the HIFLUGCS clusters, but we also see such secondary peaks in a few of the simulations, indicating at least some of these peaks may be artifacts introduced in the MCMC, with the possible exception of the binary clusters A399 and A401.  However, the temperature distributions of the HIFLUGCS clusters tend to be slightly wider than that of similar simulated distributions.  Thus, while the shape of the temperature distributions are clearly dominated by the ISF, we are capable of measuring extra broadening beyond the ISF.  It is also worth noting that there are some variations in the temperature distributions from one cluster to the next; each cluster is unique at some level.  

\subsection{Cluster Parameters}
\label{section:clusterparameters}

We measured the median temperature, total emission measure, $r_{2500}$, and $\sigma_{kT}$ for each cluster in our sample (Table \ref{table:clusterparameters}).  The median temperatures are also illustrated on the cluster temperature distributions (Figures \ref{figure:tempdists0} - \ref{figure:tempdists3}).

\begin{figure*}[p]
\begin{center}
\subfigure{\includegraphics[width=0.3\columnwidth]{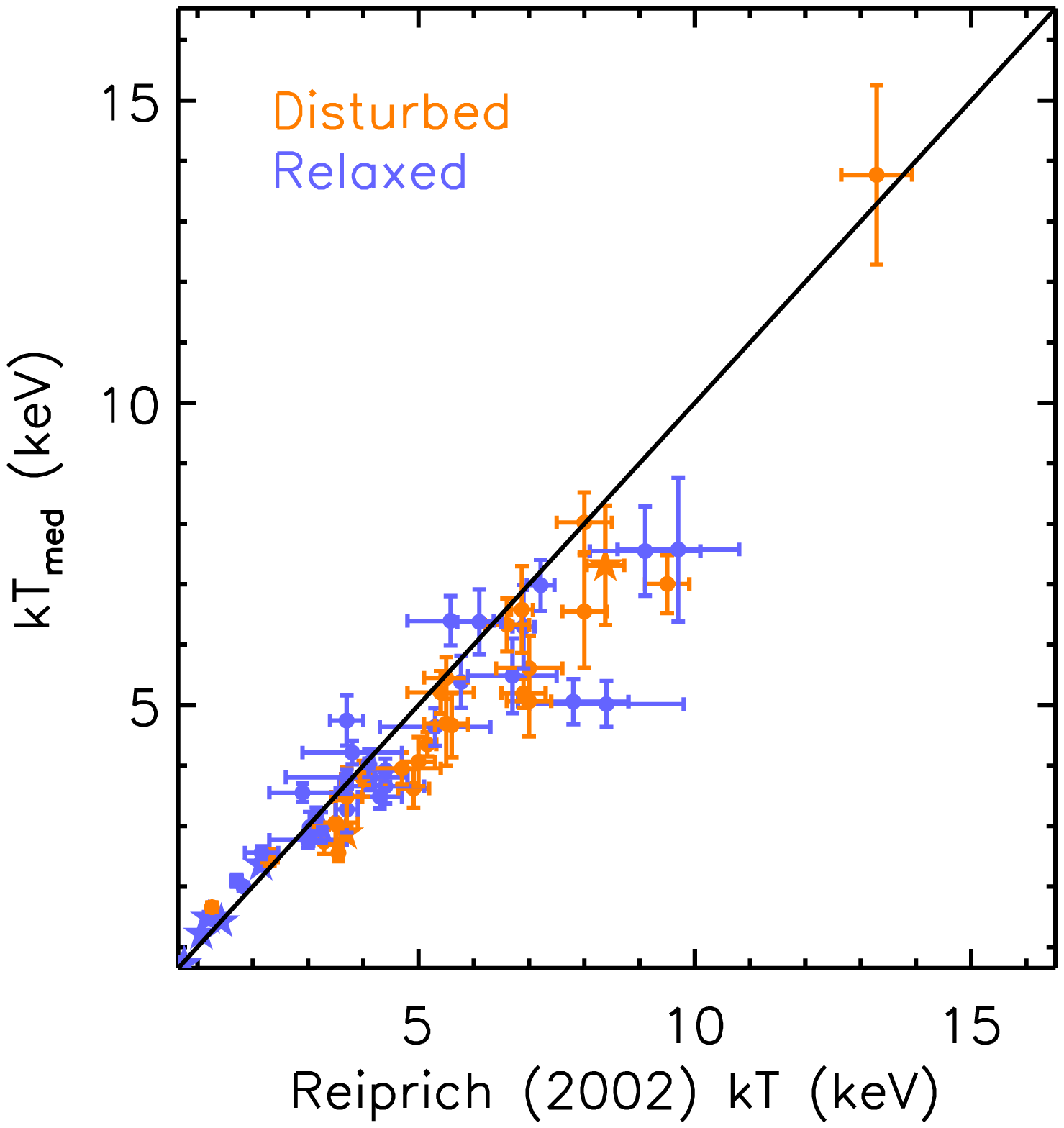}}
\subfigure{\includegraphics[width=0.3\columnwidth]{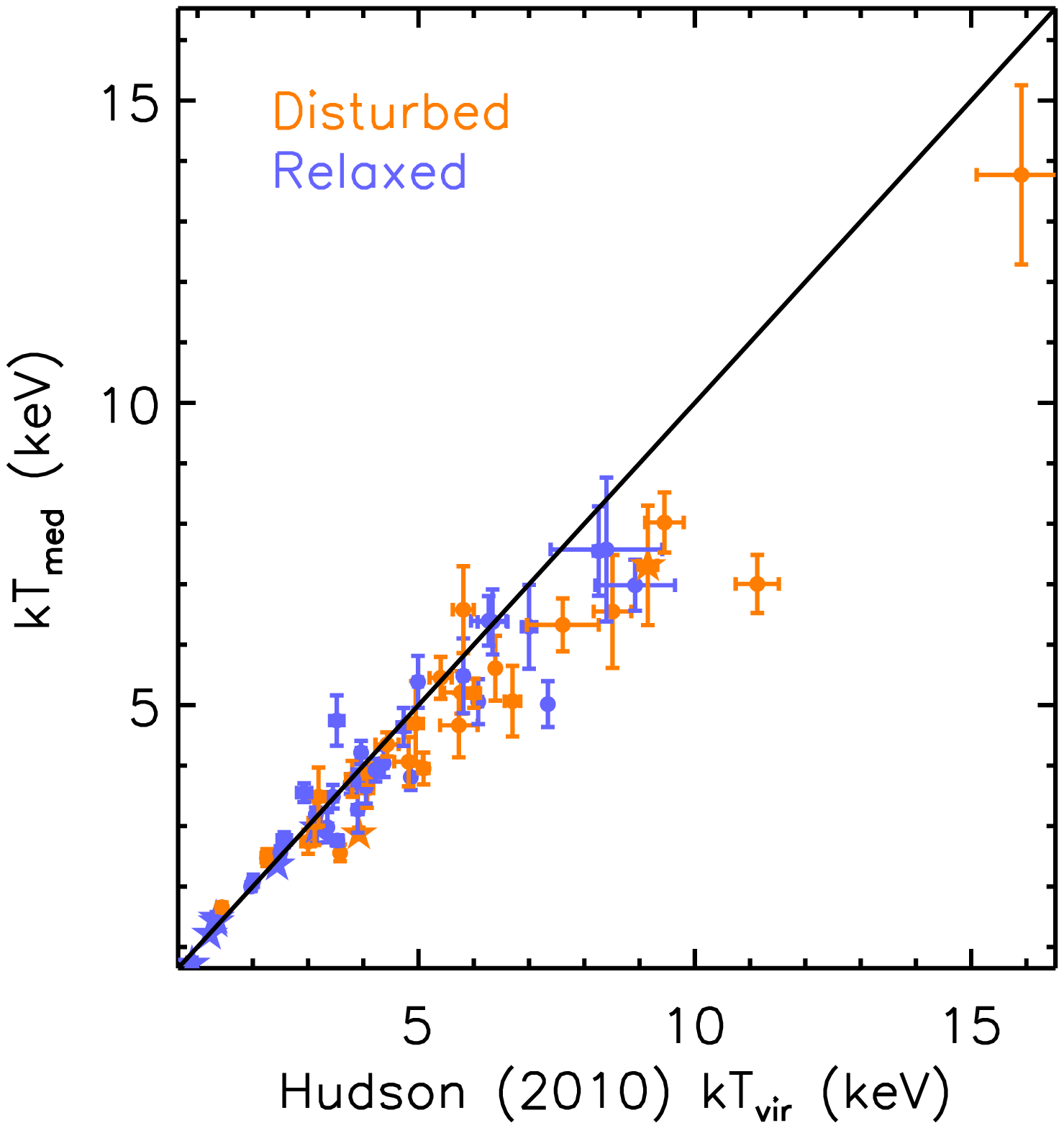}}
\subfigure{\includegraphics[width=0.3\columnwidth]{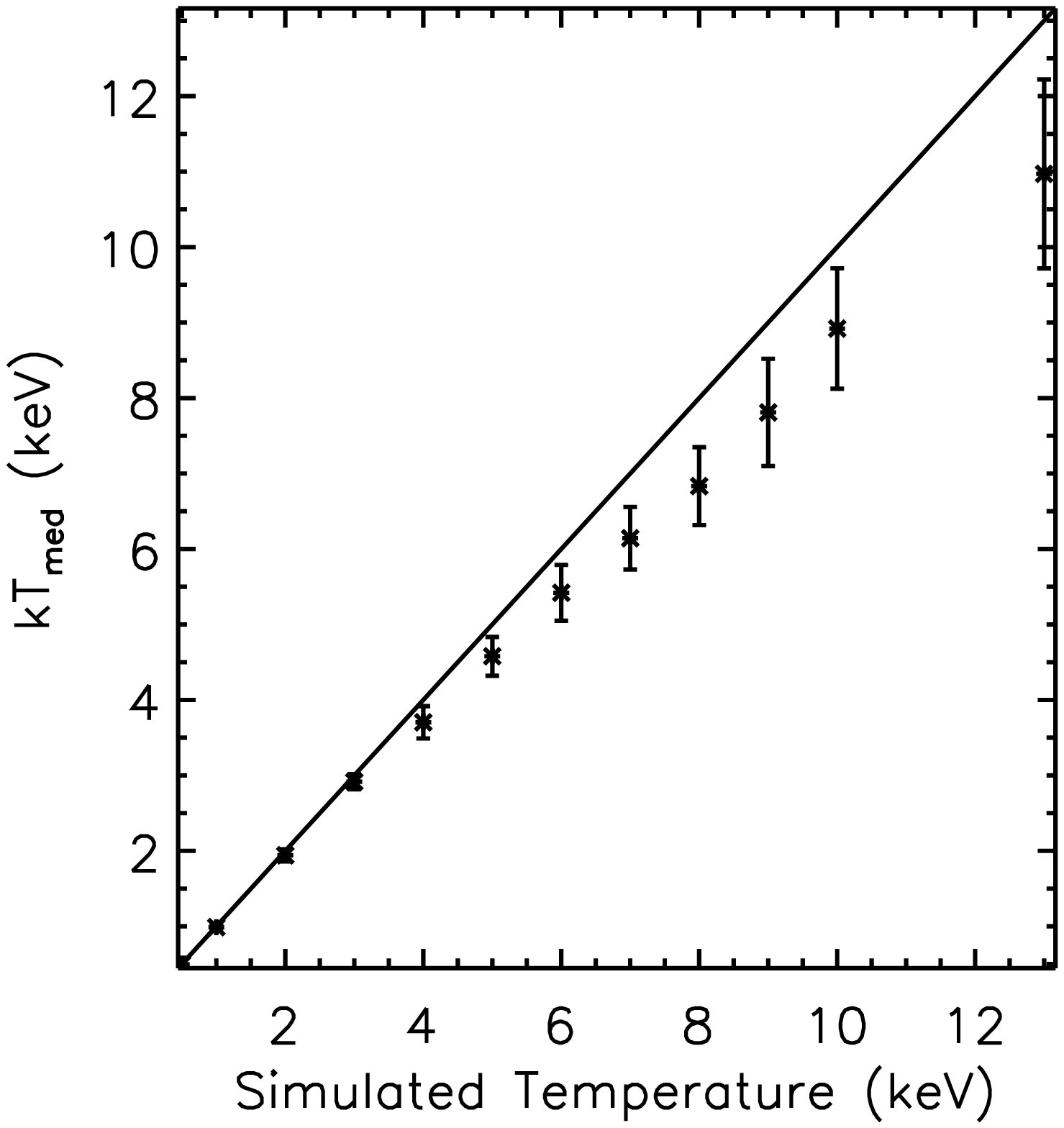}}
\caption{\footnotesize Comparisons of $kT_{med}$ from our sample with the average cluster temperatures reported in \citet{Reiprich2002a} (left) and \citet{Hudson2010} (middle), and for the simulated clusters, $kT_{med}$ with the simulated cluster temperature (right).  The lines represent complete agreement between the two quantities (slope=1).}
\label{figure:tempcomparisons}
\end{center}
\end{figure*}

To evaluate the measurements of $kT_{med}$, we measure the median temperatures of the simulated isothermal clusters to test how well the known cluster temperature can be recovered (Figure \ref{figure:tempcomparisons}).  For clusters with $kT\gtrsim5$ keV, $kT_{med}$ is biased slightly low.  This is likely due in part to the high temperature side of these clusters' log-normal temperature distributions (see $\S$\ref{section:tempdists}) extending beyond the allowed temperature range for the smoothed particles, and the resulting asymmetry in the measured temperature distribution artificially lowers $kT_{med}$.  In addition, EPIC is not as sensitive to gas at such high temperatures, making it more difficult to constrain the temperature distributions.  For real clusters, these two effects combine to lower the measured value of $kT_{med}$.  Comparing $kT_{med}$ with the more traditional temperatures reported in \citet{Reiprich2002a} and \citet{Hudson2010} (Figure \ref{figure:tempcomparisons}), we find reasonable agreement, although again for higher temperature clusters $kT_{med}$ tends to be slightly lower than the \citet{Reiprich2002a} temperatures.  In addition to the effect described above, the temperatures of \citet{Reiprich2002a} and \citet{Hudson2010} are obtained using completely different methods than this work, and may not represent the same physical temperature, and are also measured within larger apertures than $r_{2500}$, which is probably the main source of the discrepancy in the measured temperatures.  There are several clusters (the seven nearest clusters and the Coma Cluster) for which $r_{2500}$ is completely beyond the spatial cuts described in \S\ref{section:spi}; $kT_{med}$, $\sigma_{kT}$, and $EM$ for these clusters are therefore measured within a region smaller than $r_{2500}$, using emission from all available smoothed particles (within the spatial cuts).  Excluding these clusters has little effect on the overall results (with the exception of the slope of the $EM-kT_{med}$ relation, see below), so they are included in the remainder of the analysis.  Similarly, there are an additional 18 clusters for which $r_{2500}$ is only partially enclosed within the square region defined by the spatial cuts, either because $r_{2500}$ is slightly larger than 10' or 15', or because the cluster observation was off center.  However, in these cases some emission out to $r_{2500}$ is included, and the amount of missing emission area is at most $\sim5-10\%$.  We therefore treat these clusters normally.

\begin{figure*}[p]
\begin{center}
\subfigure{\includegraphics[width=0.3\textwidth]{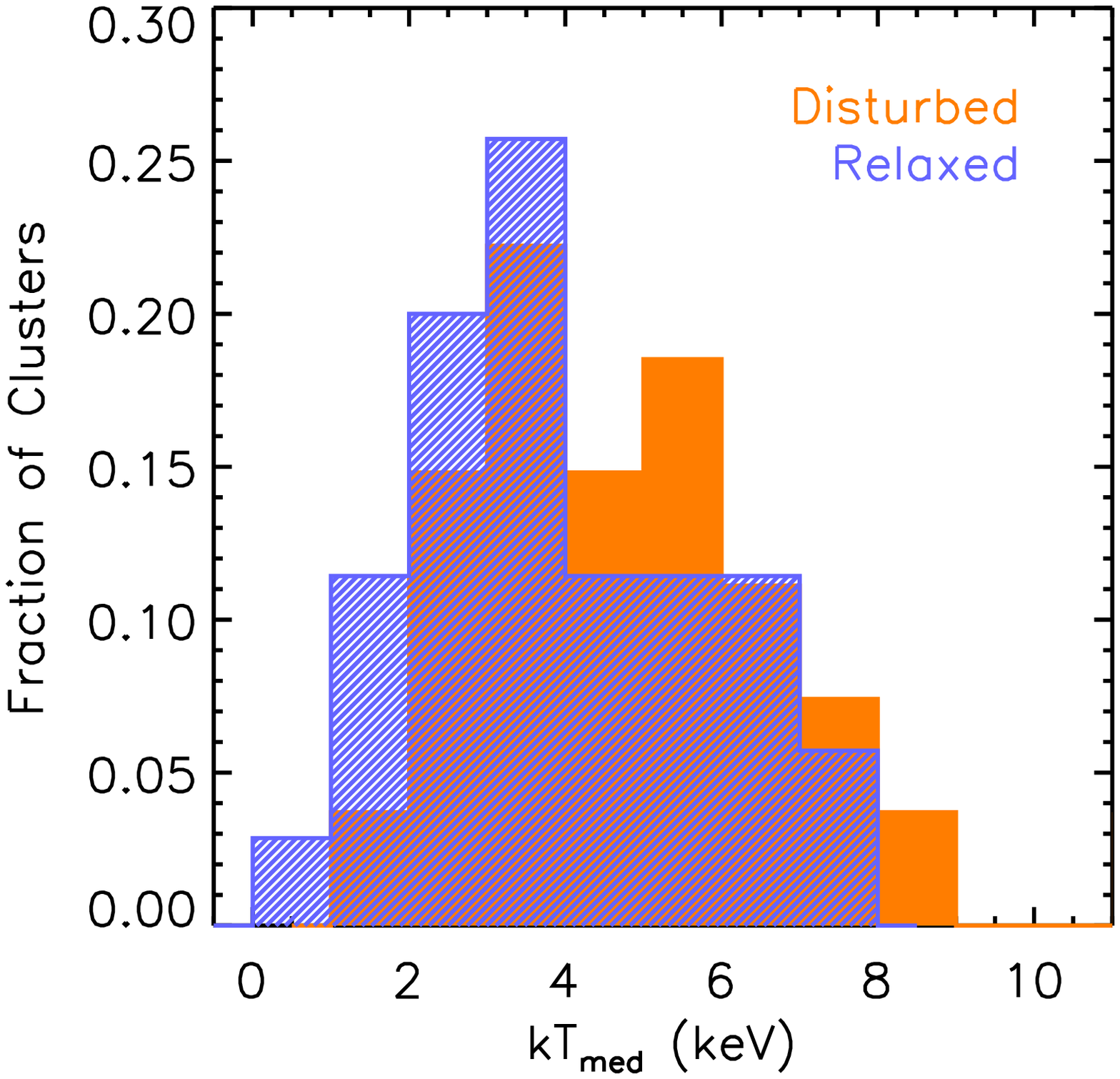}}
\subfigure{\includegraphics[width=0.3\textwidth]{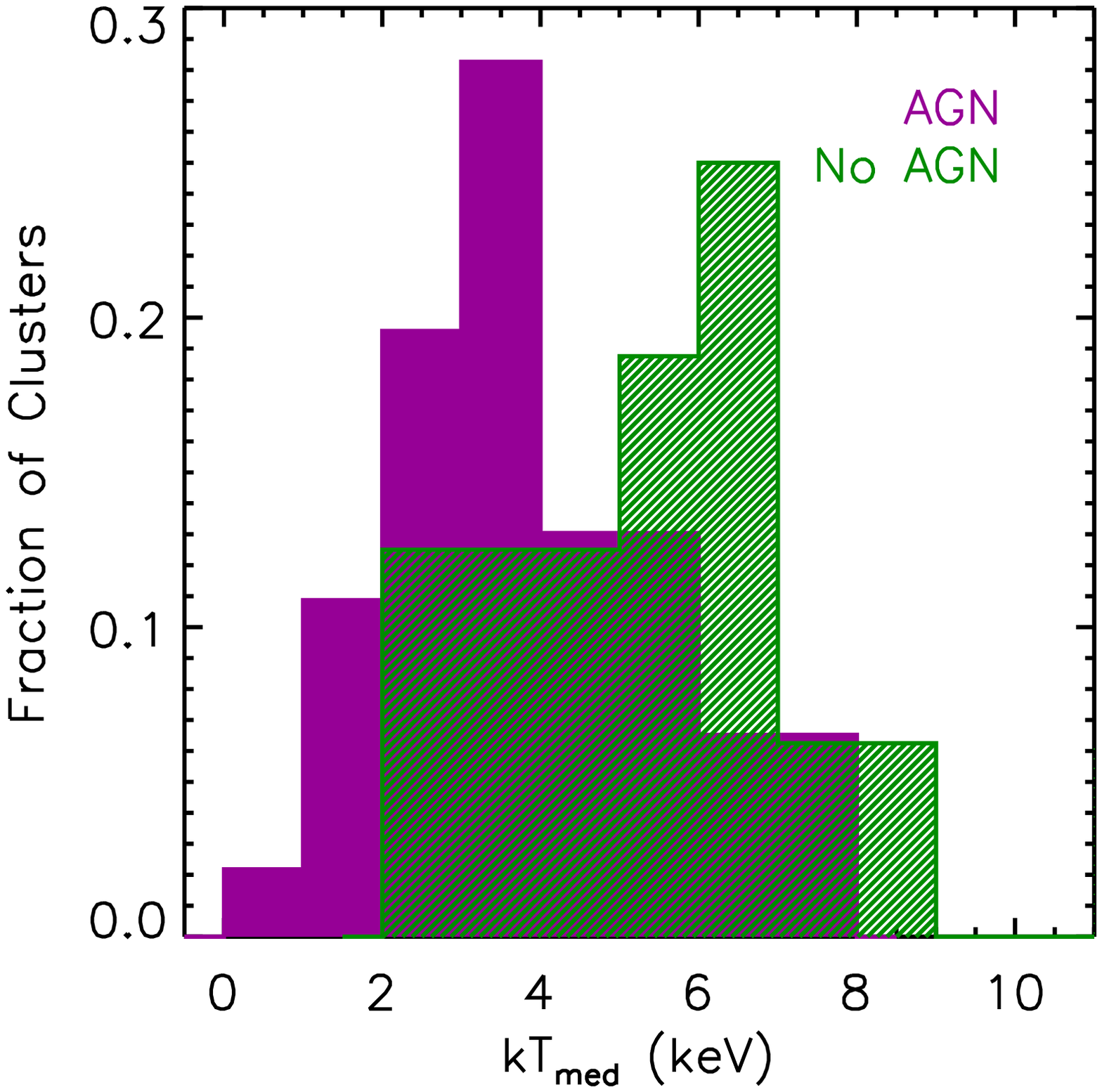}}
\subfigure{\includegraphics[width=0.3\textwidth]{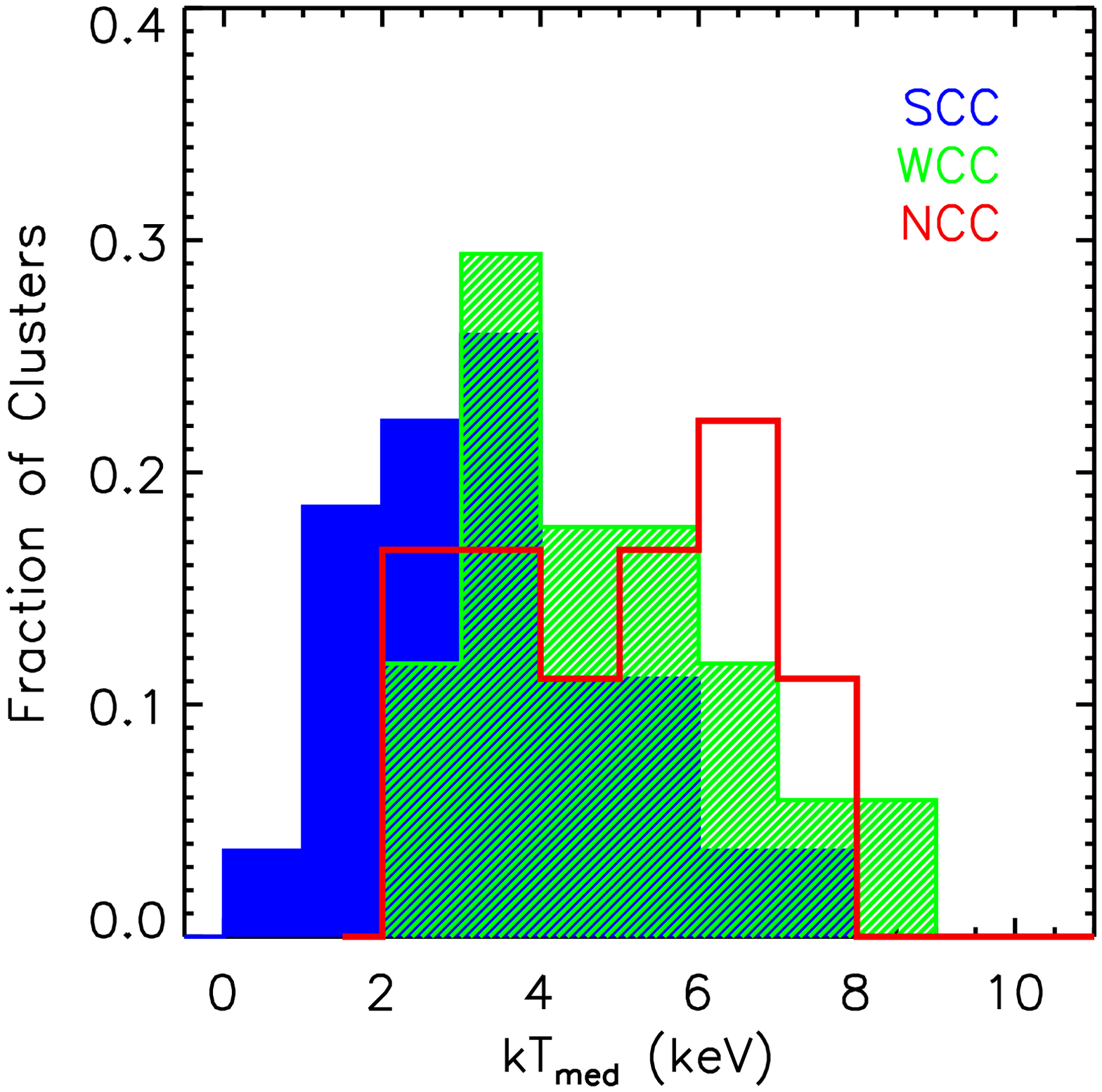}}
\caption{\footnotesize Histograms of $kT_{med}$ for each cluster subsample.  Left: relaxed clusters (blue) and disturbed clusters (orange).  Center: clusters with (purple) and without (green) central AGN activity.  Right: SCC (blue), WCC (green), and NCC (red) clusters.}
\label{figure:kThists}
\end{center}
\end{figure*}

\begin{figure*}[p]
\begin{center}
\subfigure{\includegraphics[width=0.3\textwidth]{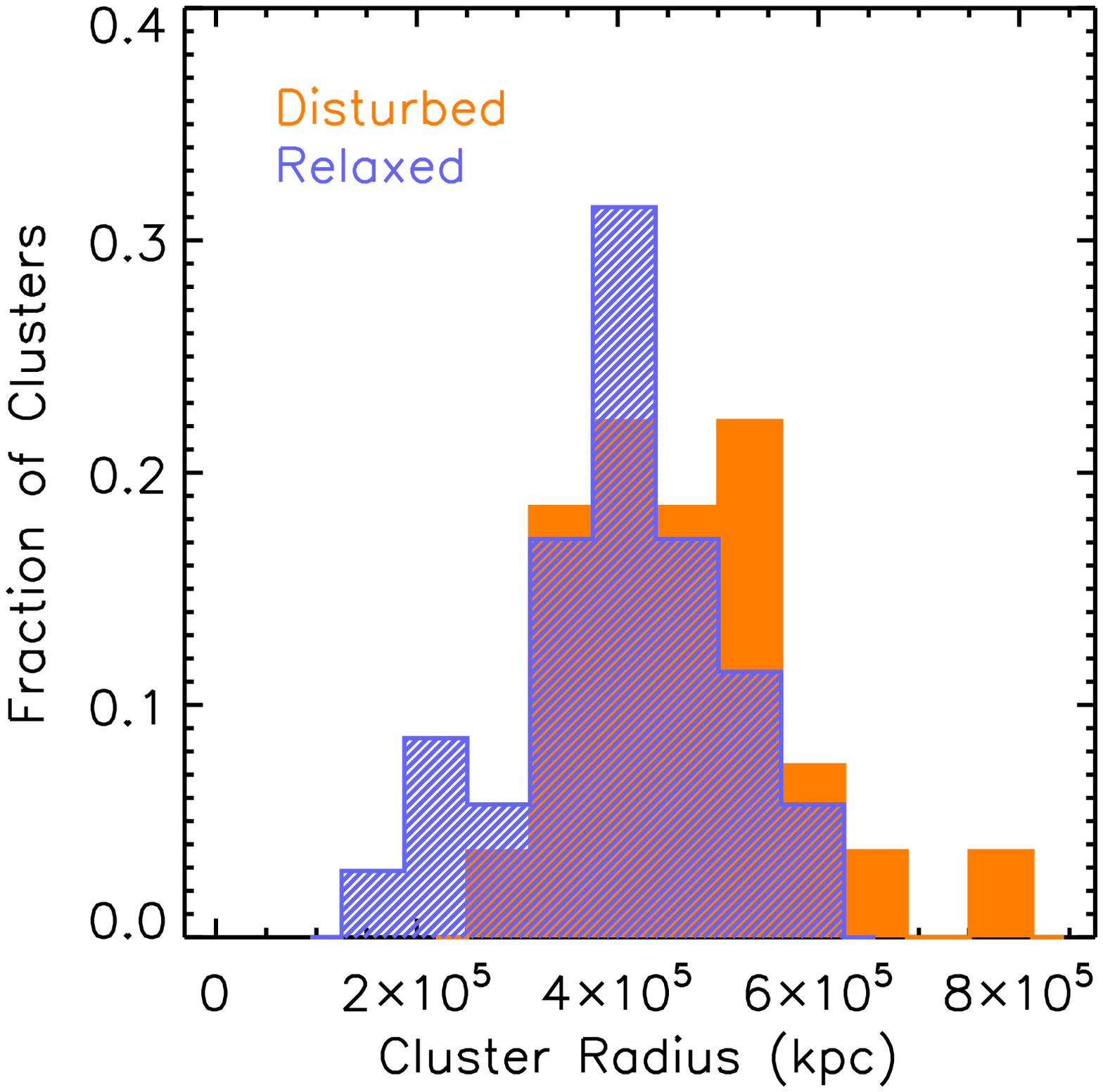}}
\subfigure{\includegraphics[width=0.3\textwidth]{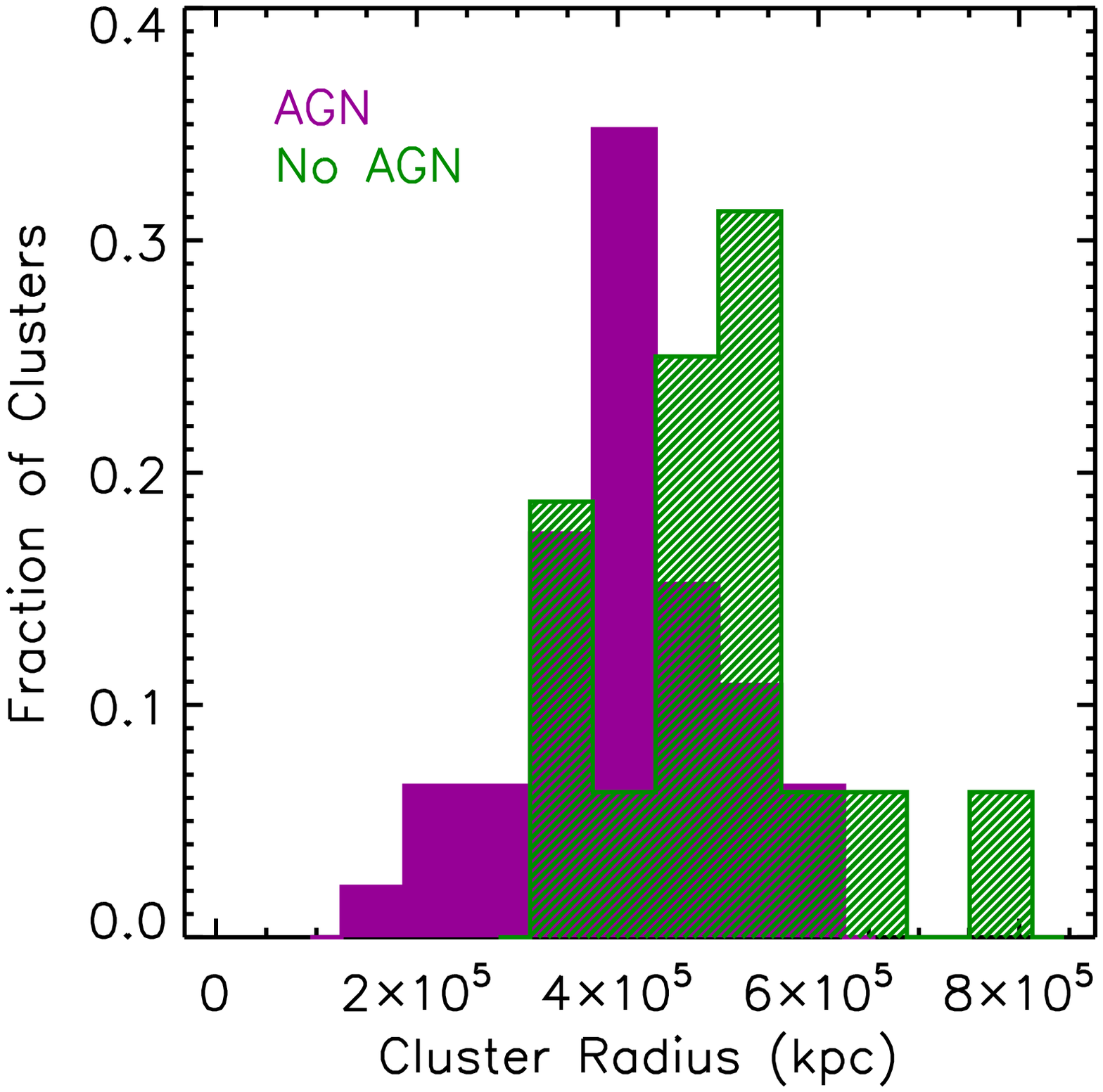}}
\subfigure{\includegraphics[width=0.3\textwidth]{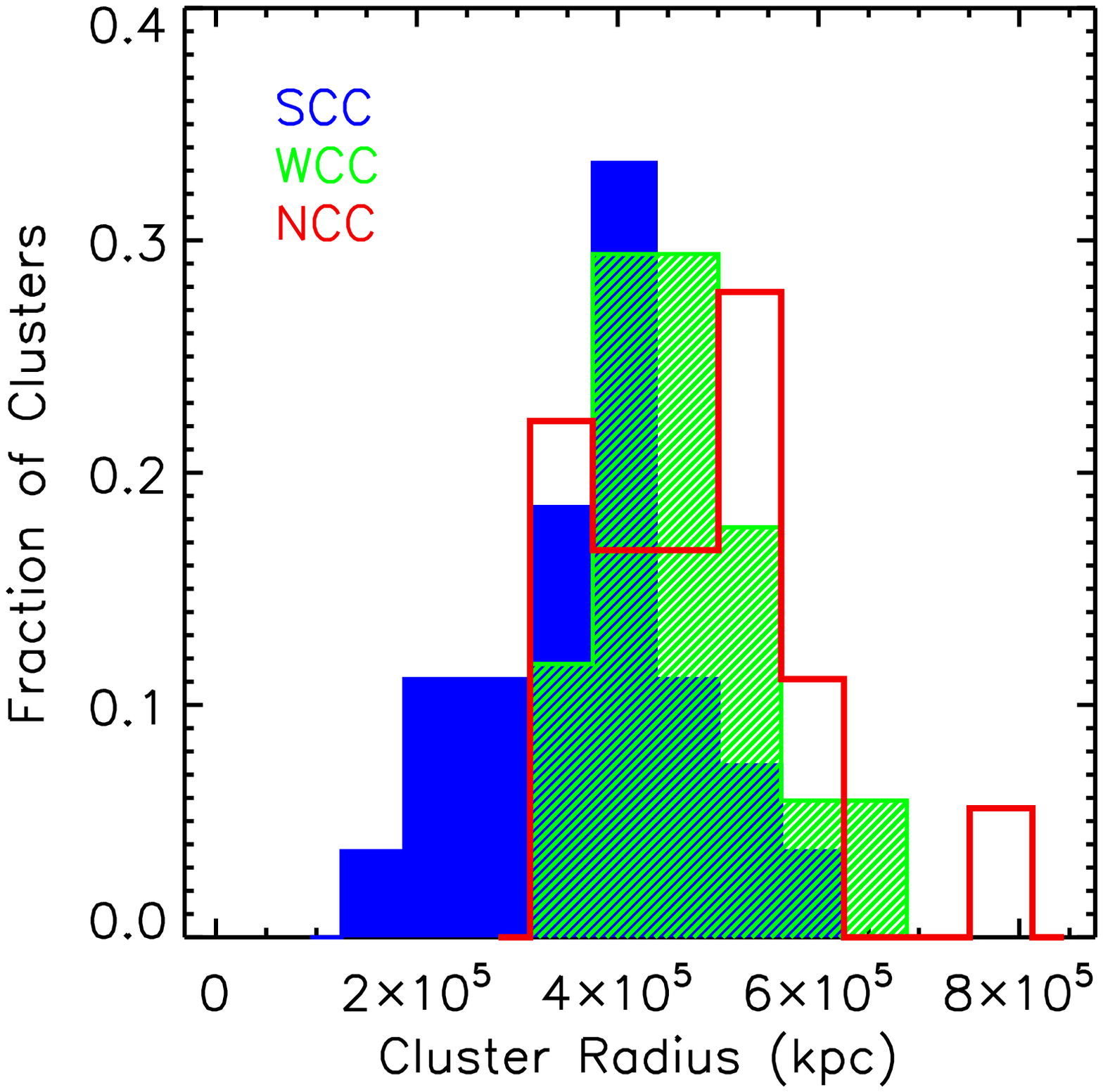}}
\caption{\footnotesize Histograms of $r_{2500}$ for each cluster subsample.  Colors are the same as Figure \ref{figure:kThists}.}
\label{figure:Rhists}
\end{center}
\end{figure*}

From the perspective of our classification schemes, there are some notable differences in $kT_{med}$ between types of clusters.  As found in other works, including \citet{Chen2007,Burns2008a,Mittal2009}, we find that relaxed, SCC, and AGN clusters (groups which largely overlap), all tend to have lower median temperatures, as well as smaller radii, than disturbed, NCC, and non-AGN clusters, as can be seen in Figures \ref{figure:kThists} and \ref{figure:Rhists}.  

\begin{figure*}[p]
\begin{center}
\subfigure{\includegraphics[width=0.3\textwidth]{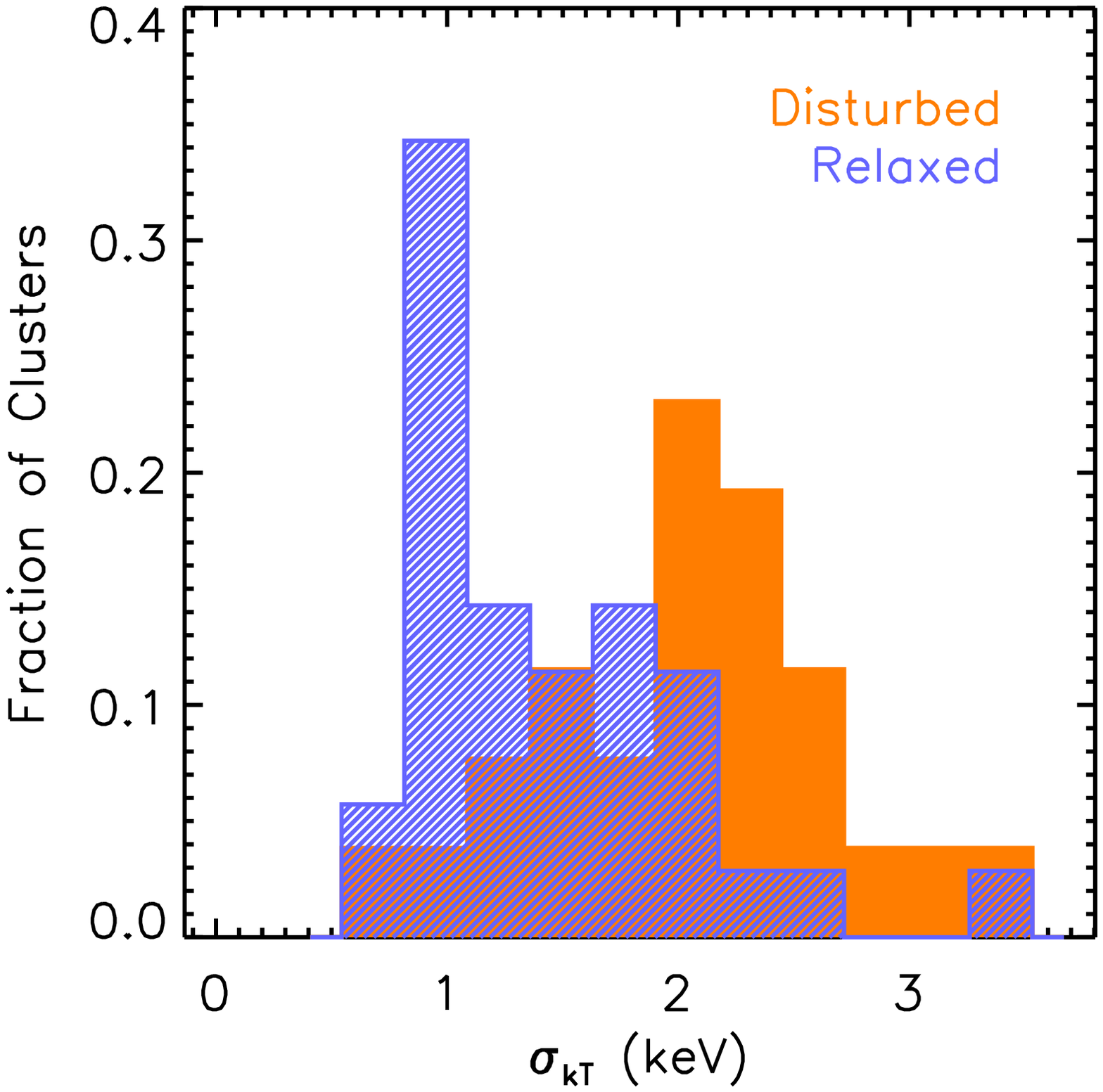}}
\subfigure{\includegraphics[width=0.3\textwidth]{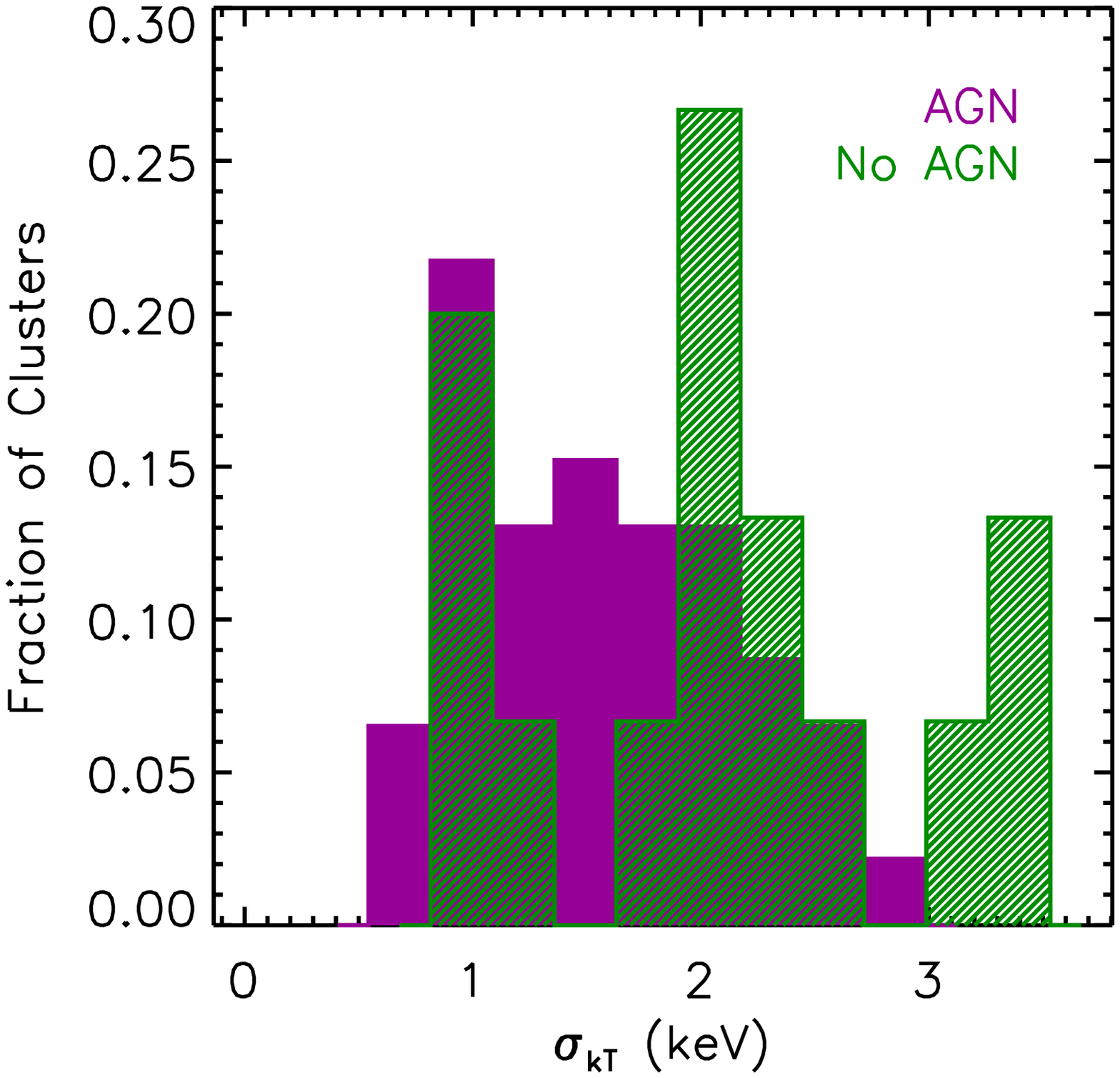}}
\subfigure{\includegraphics[width=0.3\textwidth]{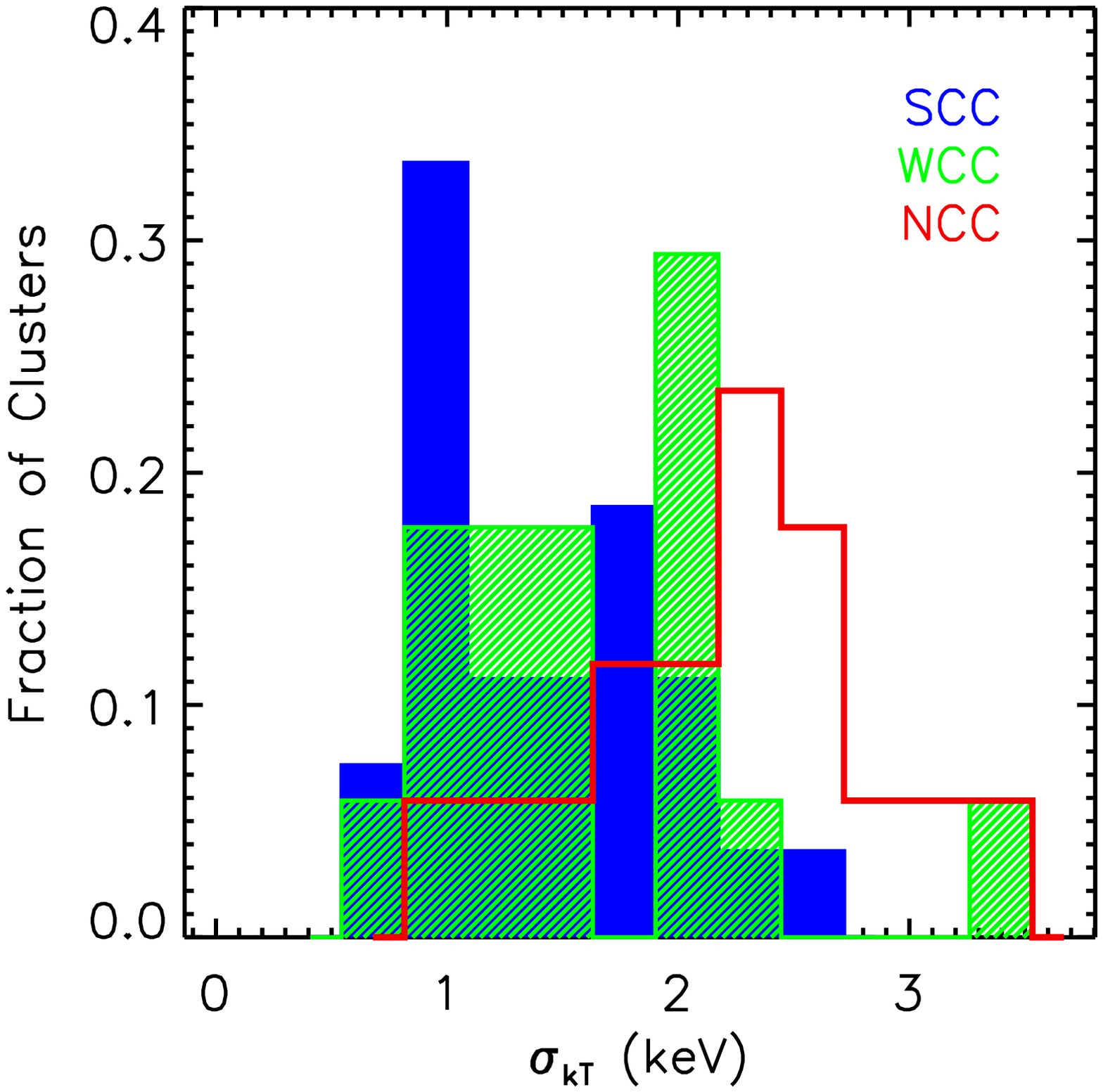}}
\caption{\footnotesize Histograms of $\sigma_{kT}$ for each cluster subsample.  Colors are the same as Figure \ref{figure:kThists}.}
\label{figure:sigmakThists}
\end{center}
\end{figure*}

For a given cluster, $\sigma_{kT}$ quantifies any departure of the cluster ICM from isothermality.  For all \nclusters clusters $\sigma_{kT}$ rules out isothermality by at least 1$\sigma$, and in many cases much higher significance.  Using the classification schemes introduced in \S\ref{section:clustermorph}, it can be seen from Figure \ref{figure:sigmakThists} that the $\sigma_{kT}$ appears to behave differently for different types of clusters.  Relaxed, SCC, WCC, and AGN clusters are more likely to have narrower temperature distributions (smaller $\sigma_{kT}$), whereas disturbed, NCC, and non-AGN clusters are more likely to have wider temperature distributions.  This effect is strongest for the disturbed/relaxed cluster classification.

\subsection{Scaling Relations}
\label{section:scalingrelations}

\begin{figure*}[p]
\begin{center}
\subfigure{\includegraphics[width=0.3\textwidth]{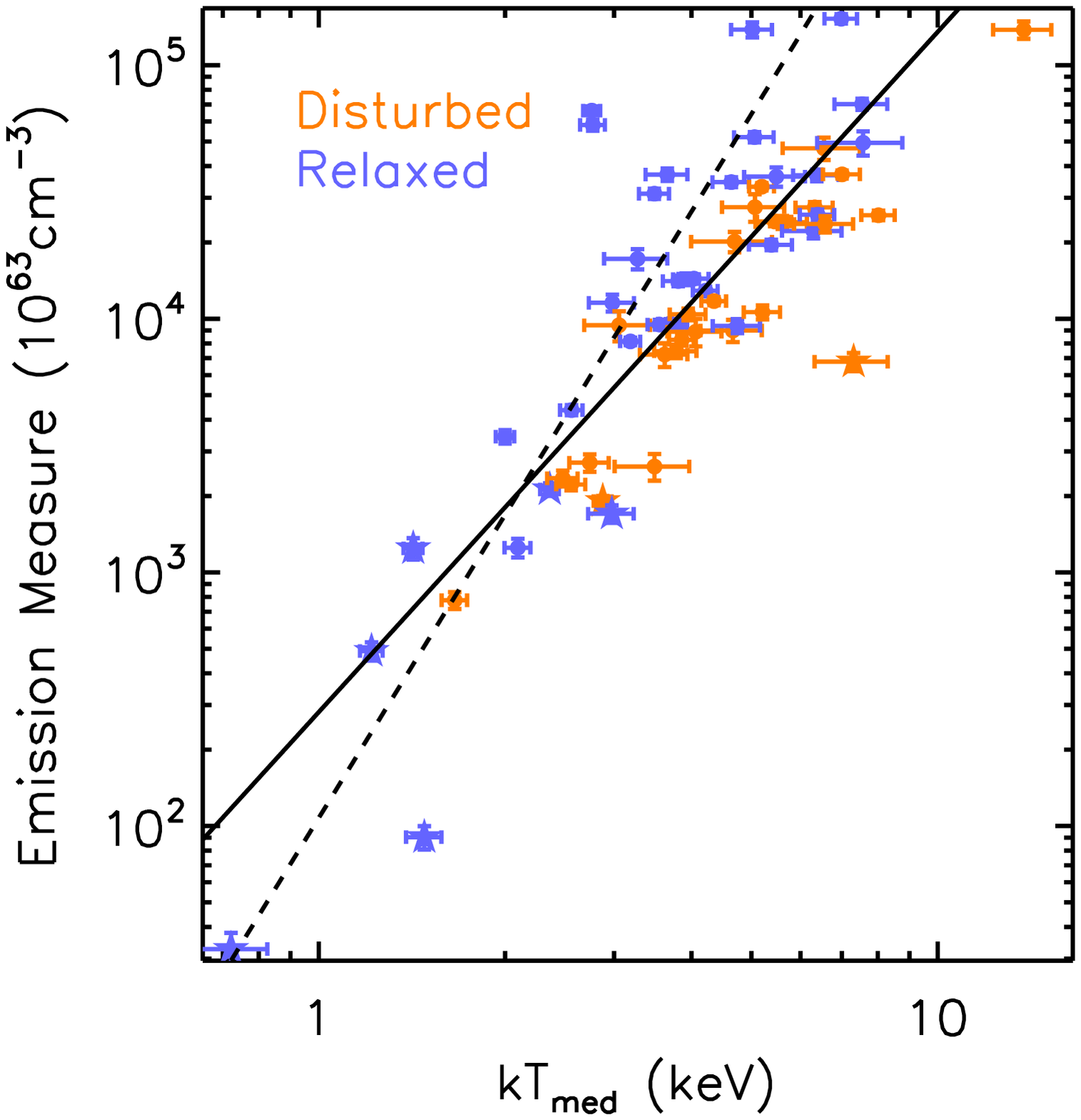}}
\subfigure{\includegraphics[width=0.3\textwidth]{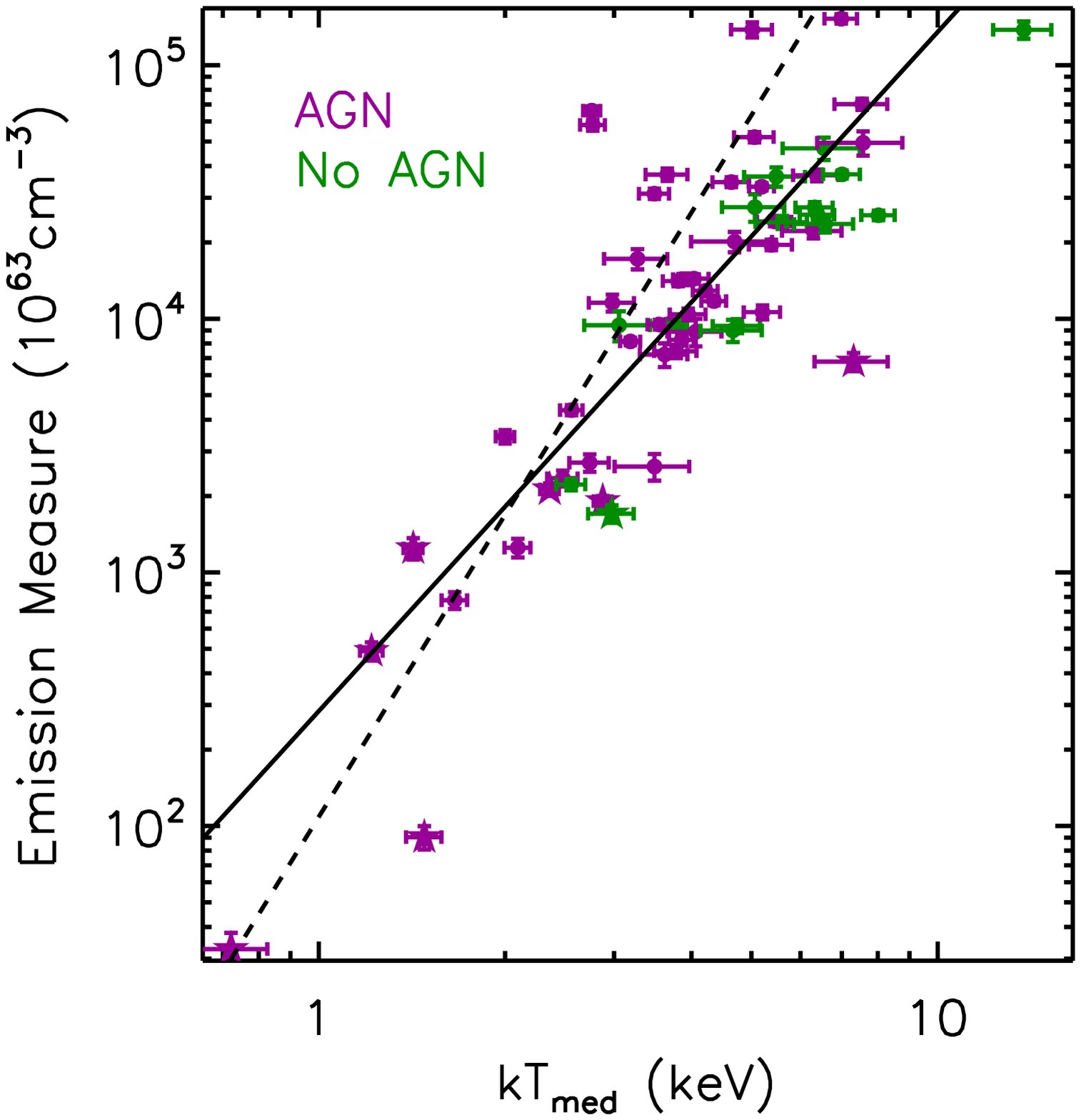}}
\subfigure{\includegraphics[width=0.3\textwidth]{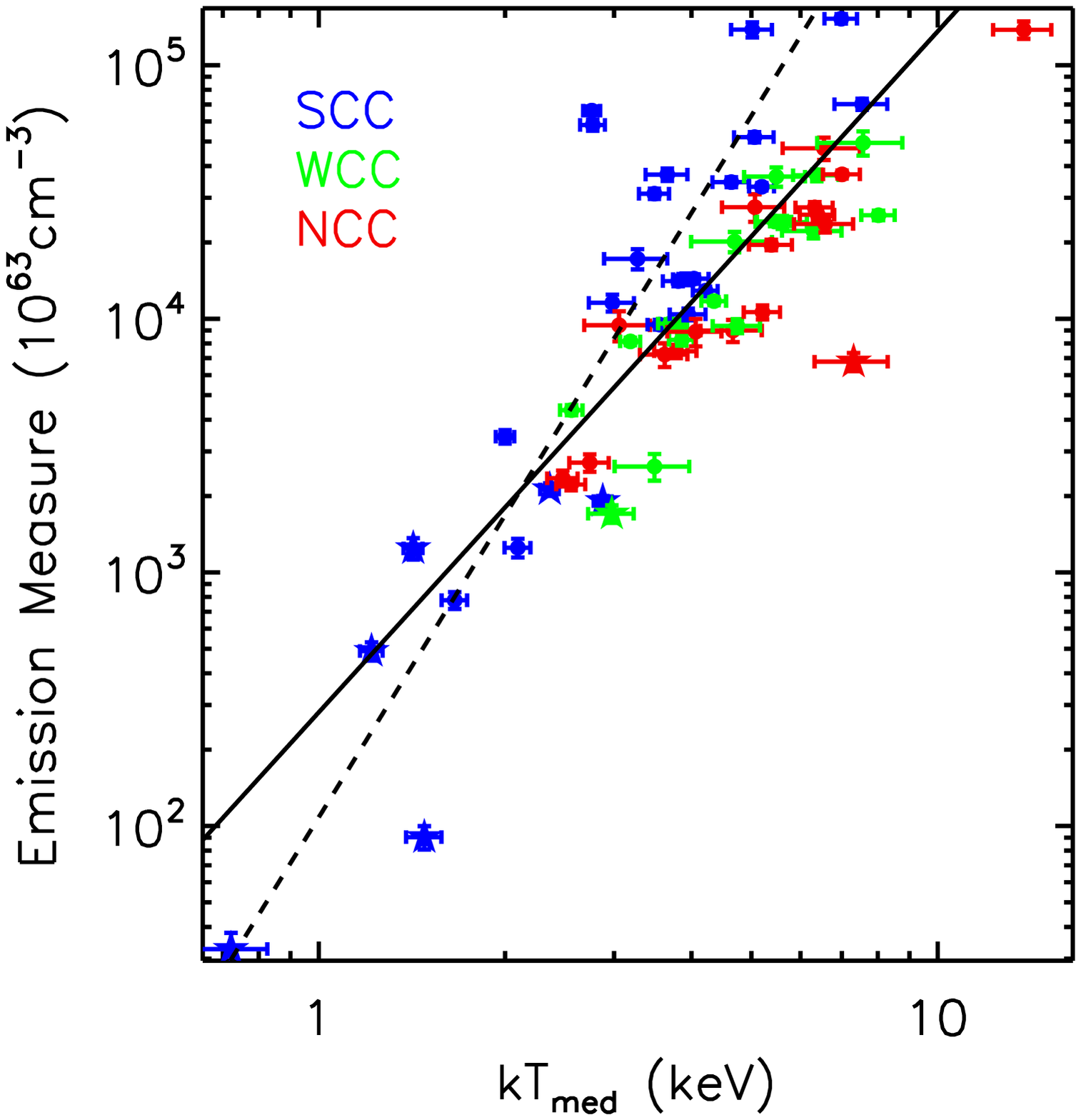}}
\caption{\footnotesize $EM-kT$ scaling relation for all clusters in the sample, for each classification scheme.  Stars represent the eight largest clusters for which $r_{2500}$ is beyond the field of view.  The powerlaw best fit to all clusters, $logEM=\alpha logkT_{med}+\beta$, is $\alpha=\emkTalpha$, $\beta=\emkTbeta$ (solid black).  Including only clusters with $kT_{med}<3 keV$, the best fit slope and intercept are $\alpha=\lowemkTalpha$ and $\beta=\lowemkTbeta$, respectively (dashed).  Excluding the eight largest clusters, the best fit is shallower, $\alpha=\nolargeemkTalpha$, $\beta=\nolargeemkTbeta$.  Colors are the same as Figure \ref{figure:kThists}.}
\label{figure:emkT}
\end{center}
\end{figure*}

We plot $EM - kT_{med}$, the analog of the $L_X - T$ relation, in Figure \ref{figure:emkT}.  Assuming a powerlaw relation $logEM=\alpha logkT + \beta$, a fit to all clusters using the IDL MCMC fitting procedure \textit{linmix\_err} \citep{Kelly2007} yields $\alpha=\emkTalpha$, $\beta=\emkTbeta$, in agreement with other works.  The slope is significantly steeper if only clusters with $kT_{med}<3$ keV are included; $\alpha=\lowemkTalpha$.  As expected, relaxed, SCC, and AGN clusters tend to lie above and to the left of the best fit.  If the eight largest clusters (for which $r_{2500}$ is not visible) are excluded, the effect is to essentially remove most of the coolest clusters, which results in a shallower $EM - kT_{med}$ slope of $\alpha=\nolargeemkTalpha$. 

\begin{figure*}[p]
\begin{center}
\includegraphics[width=0.3\textwidth]{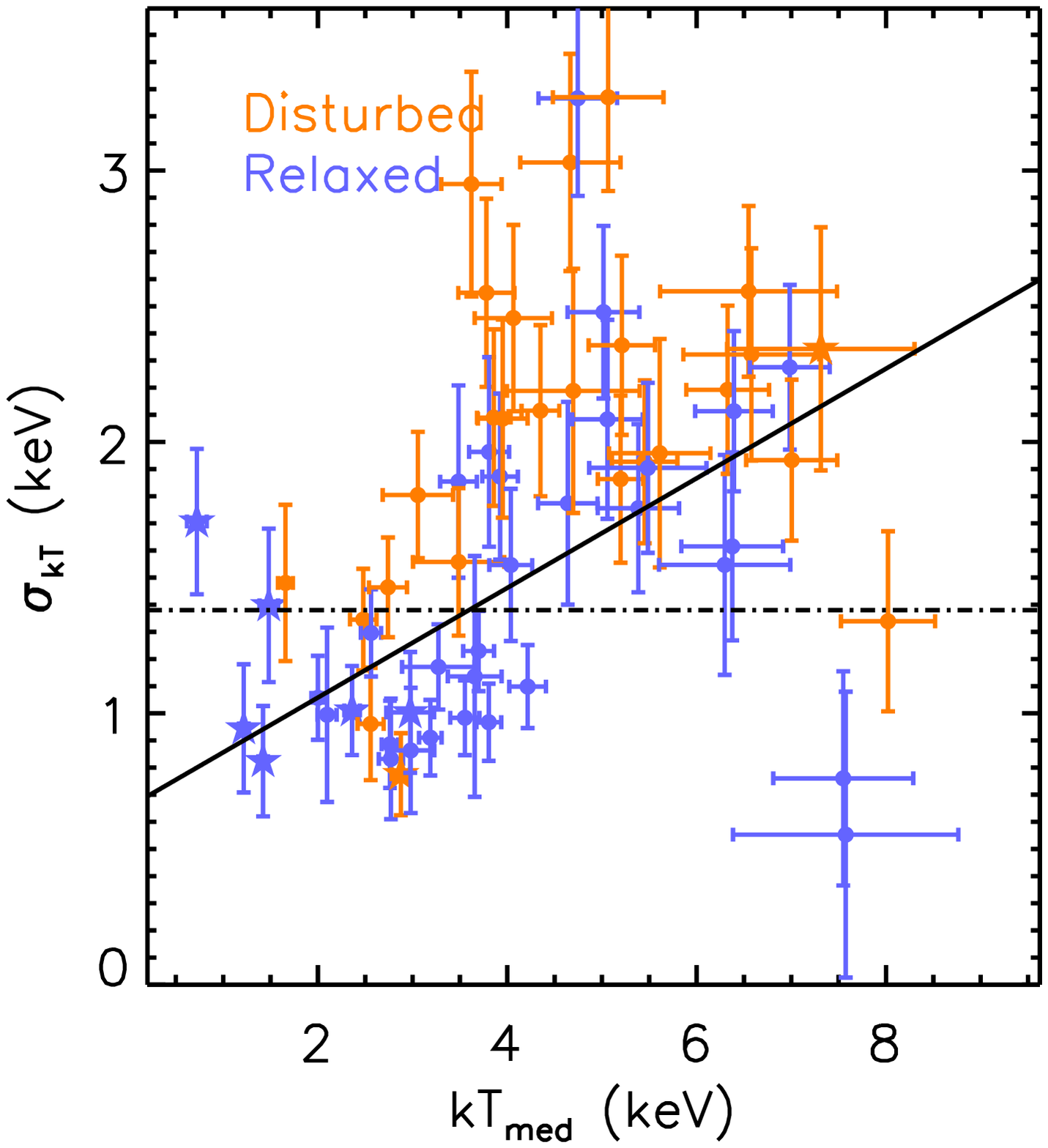}
\includegraphics[width=0.3\textwidth]{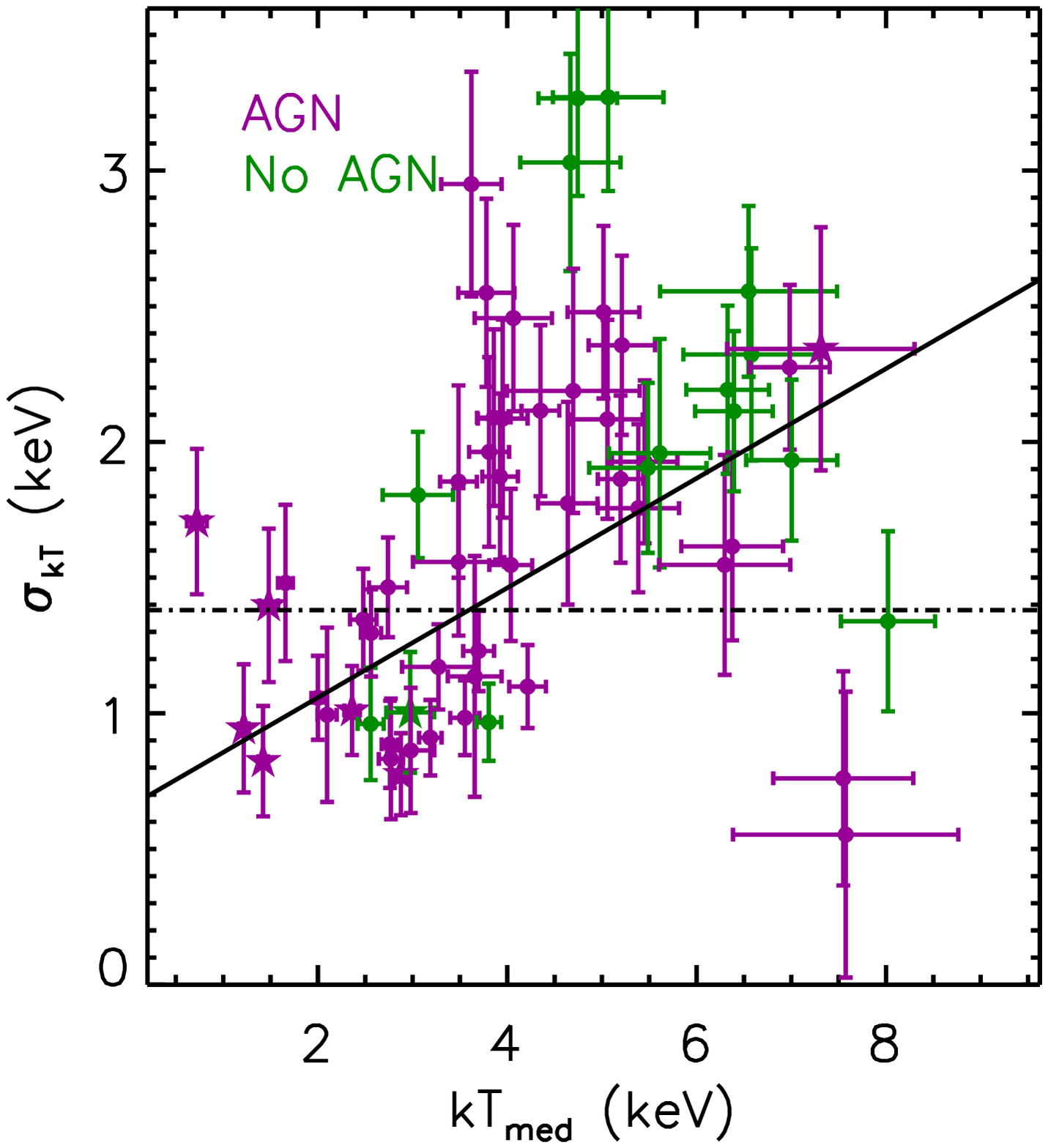}
\includegraphics[width=0.3\textwidth]{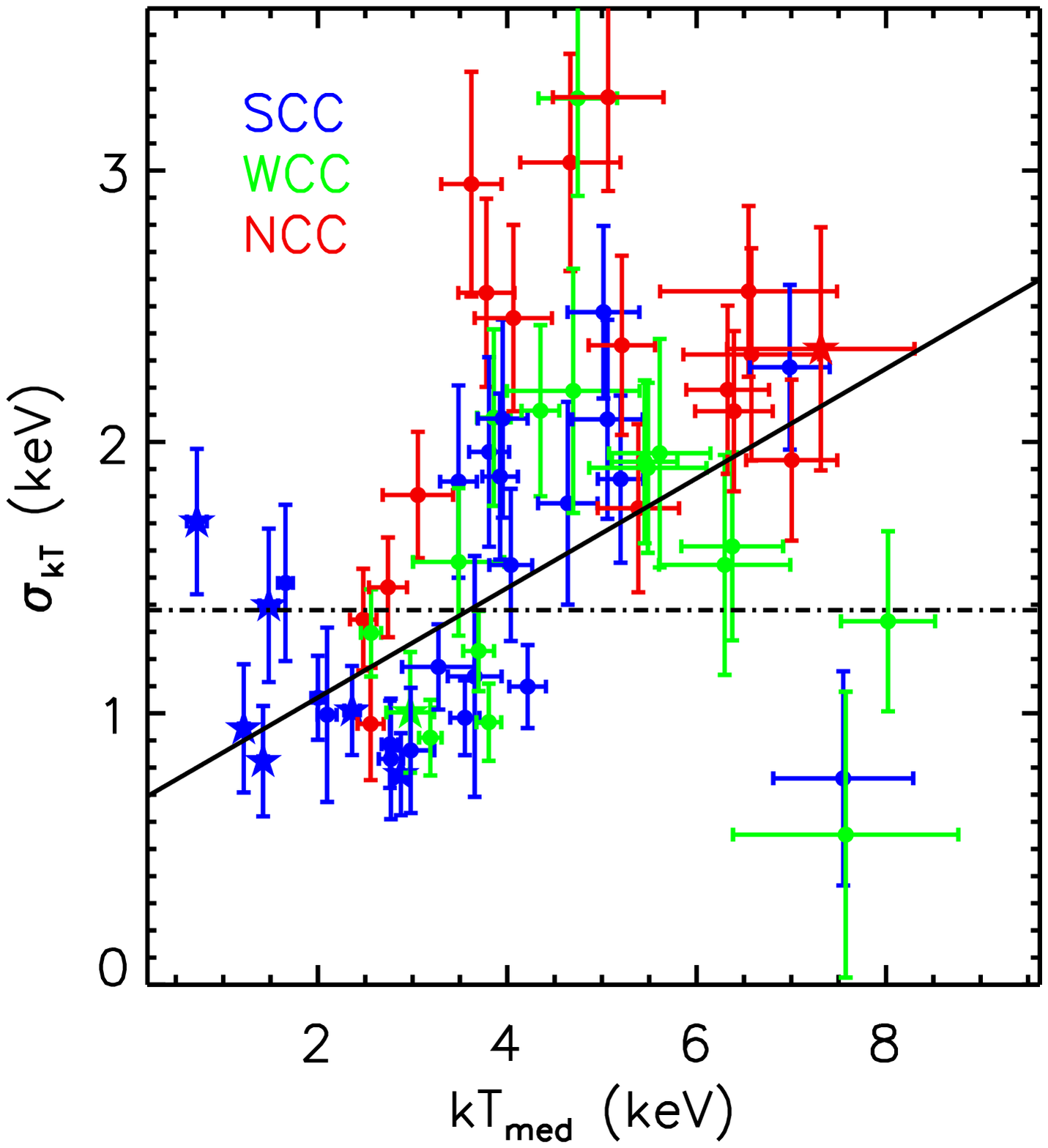}
\caption{\footnotesize $kT_{med} - \sigma_{kT}$ scaling relation, shown with each classification scheme and the associated linear and constant fits, $\sigma_{kT}\sim \kTsigmaalpha kT_{med} + \kTsigmabeta$ (solid) and $\sigma_{kT}\sim\kTsigmaconstant$ (dot-dash).  Stars represent the eight largest clusters for which $r_{2500}$ is beyond the field of view.  Colors are the same as Figure \ref{figure:kThists}.}
\label{figure:sigmakTkT}
\end{center}
\end{figure*}

\begin{figure*}[p]
\begin{center}
\subfigure{\includegraphics[width=0.3\textwidth]{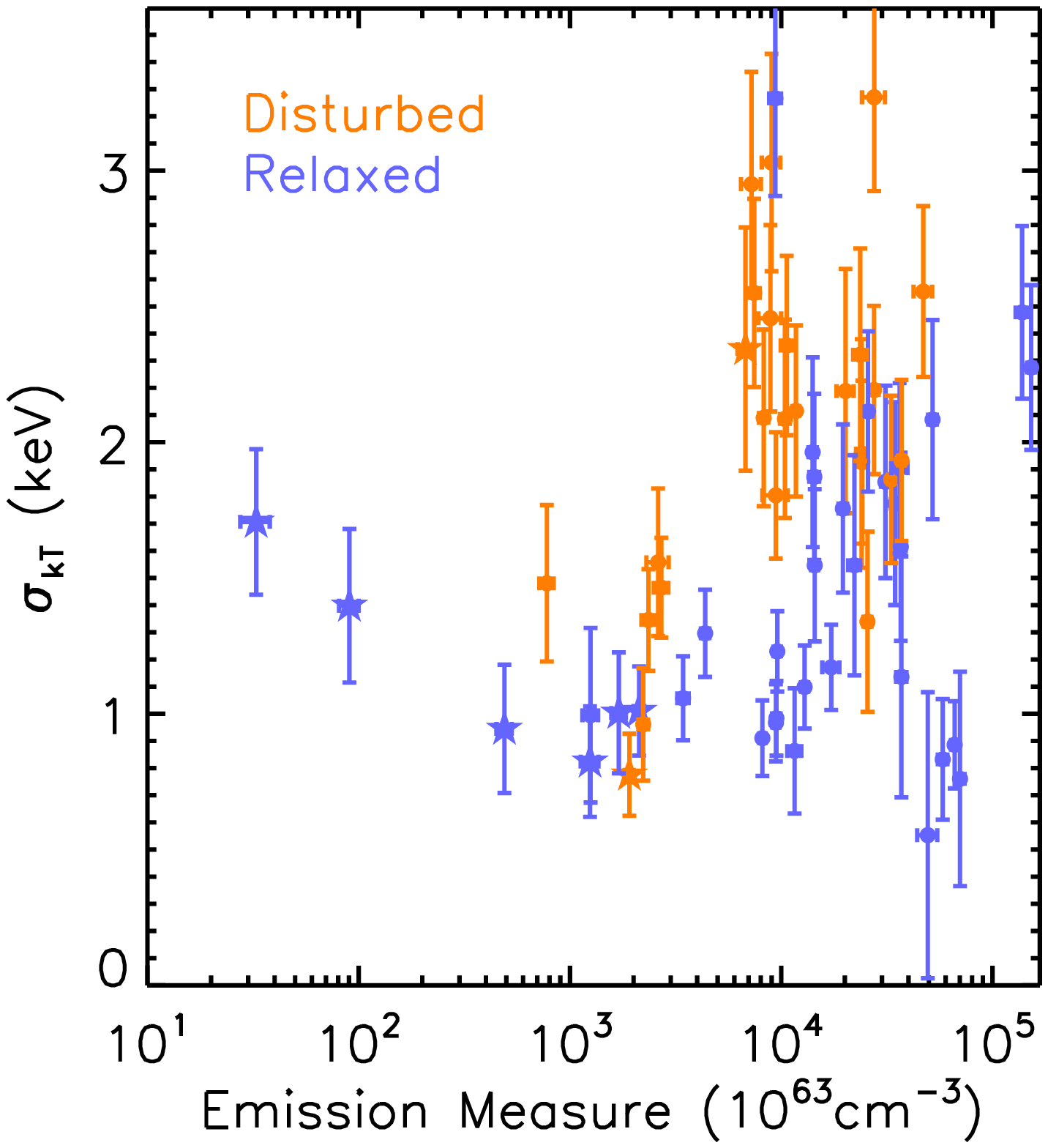}}
\subfigure{\includegraphics[width=0.3\textwidth]{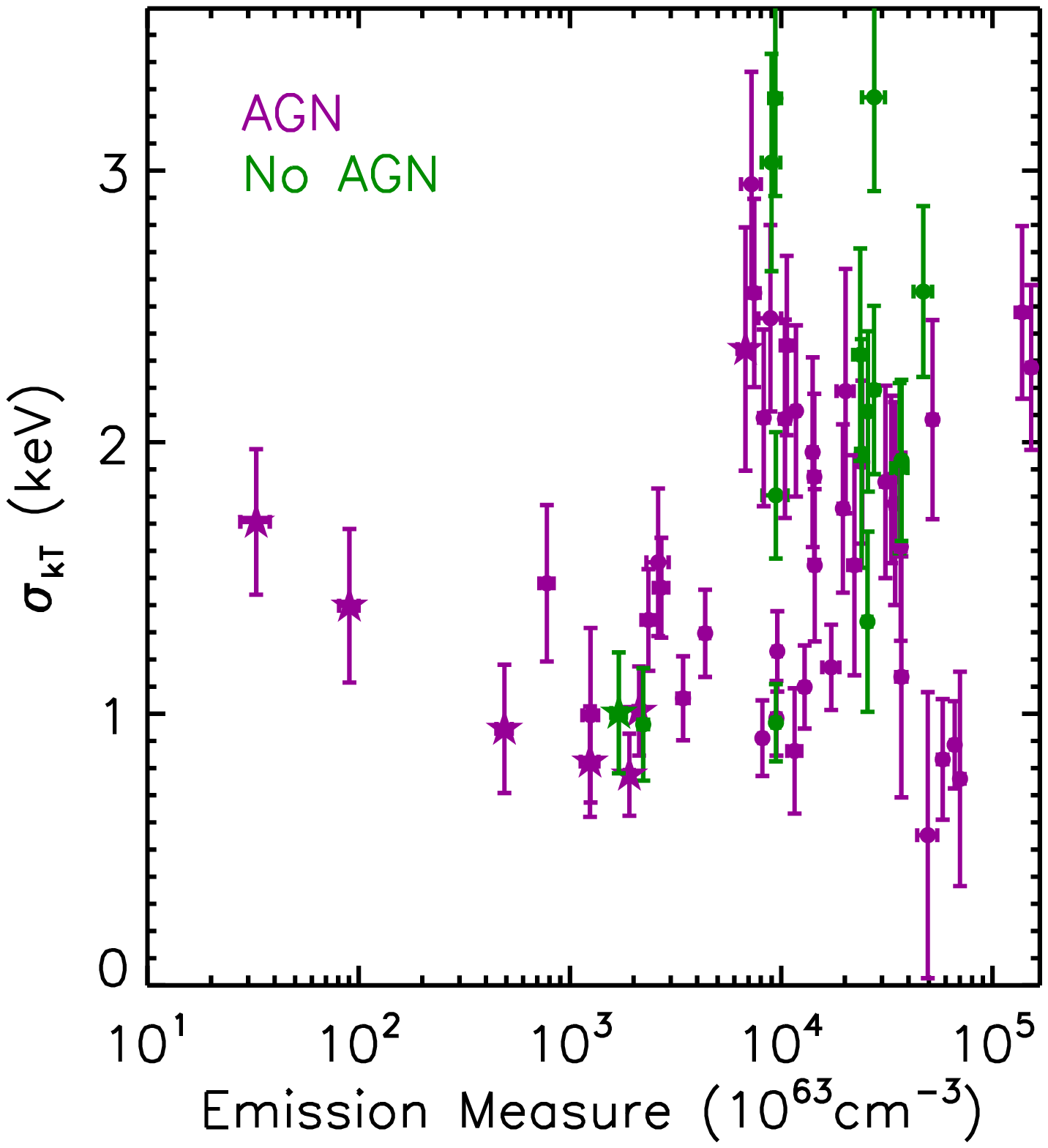}}
\subfigure{\includegraphics[width=0.3\textwidth]{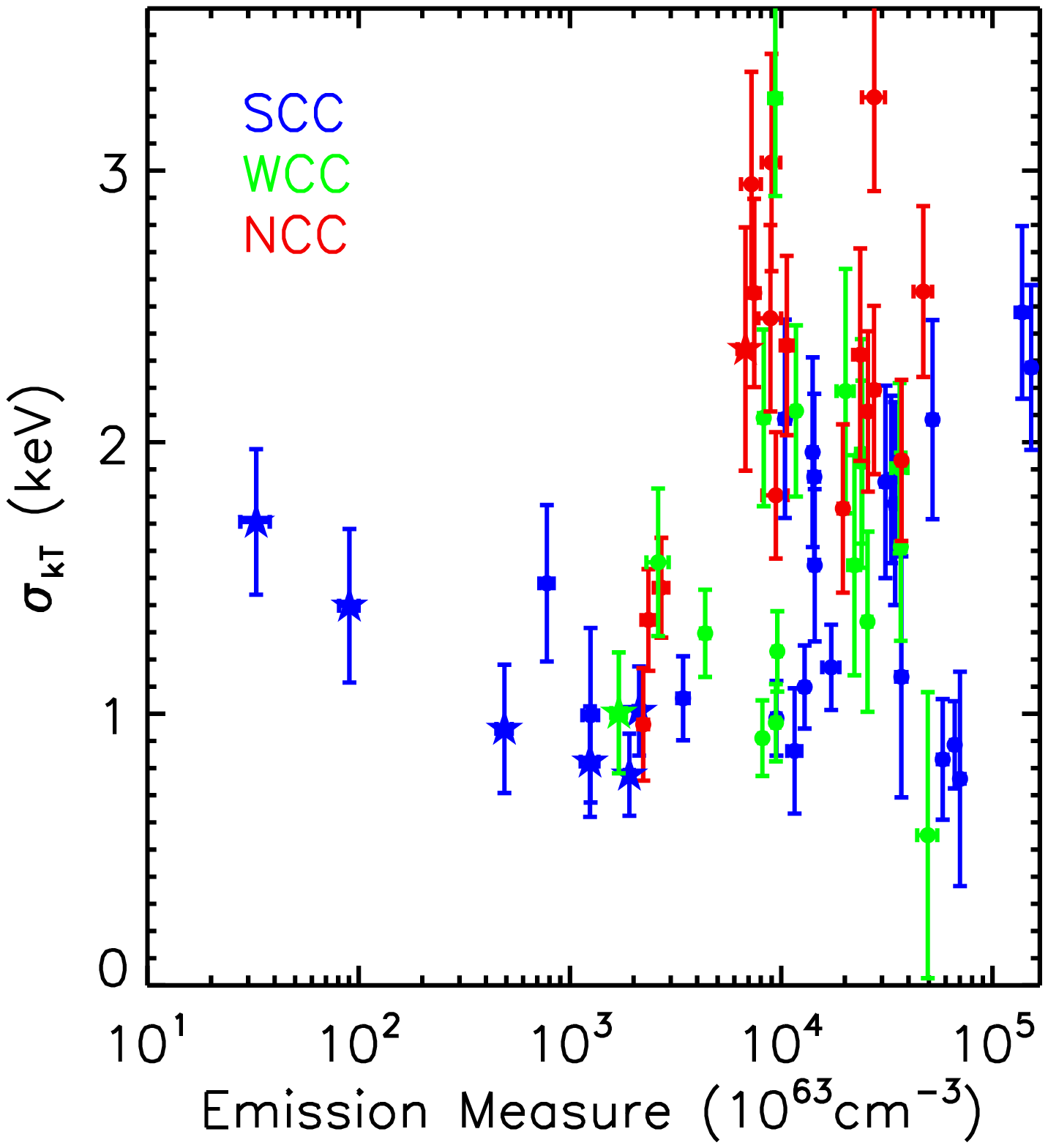}}
\caption{\footnotesize $EM-\sigma_{kT}$ scaling relation, shown with each classification scheme.  Stars represent the eight largest clusters for which $r_{2500}$ is beyond the field of view.  Colors are the same as Figure \ref{figure:kThists}.}
\label{figure:emsigmakT}
\end{center}
\end{figure*}

\begin{figure*}[p]
\begin{center}
\subfigure{\includegraphics[width=0.3\textwidth]{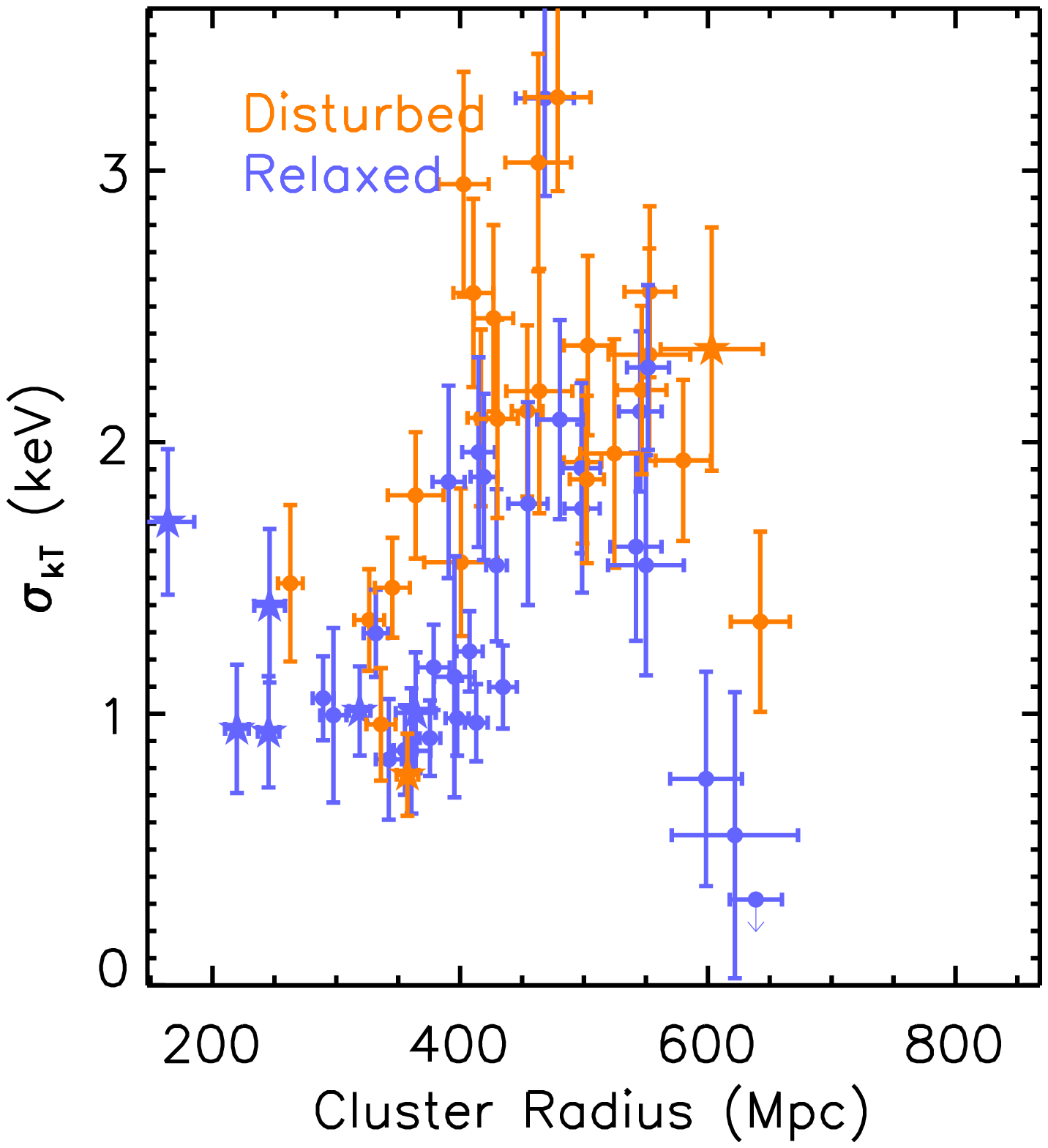}}
\subfigure{\includegraphics[width=0.3\textwidth]{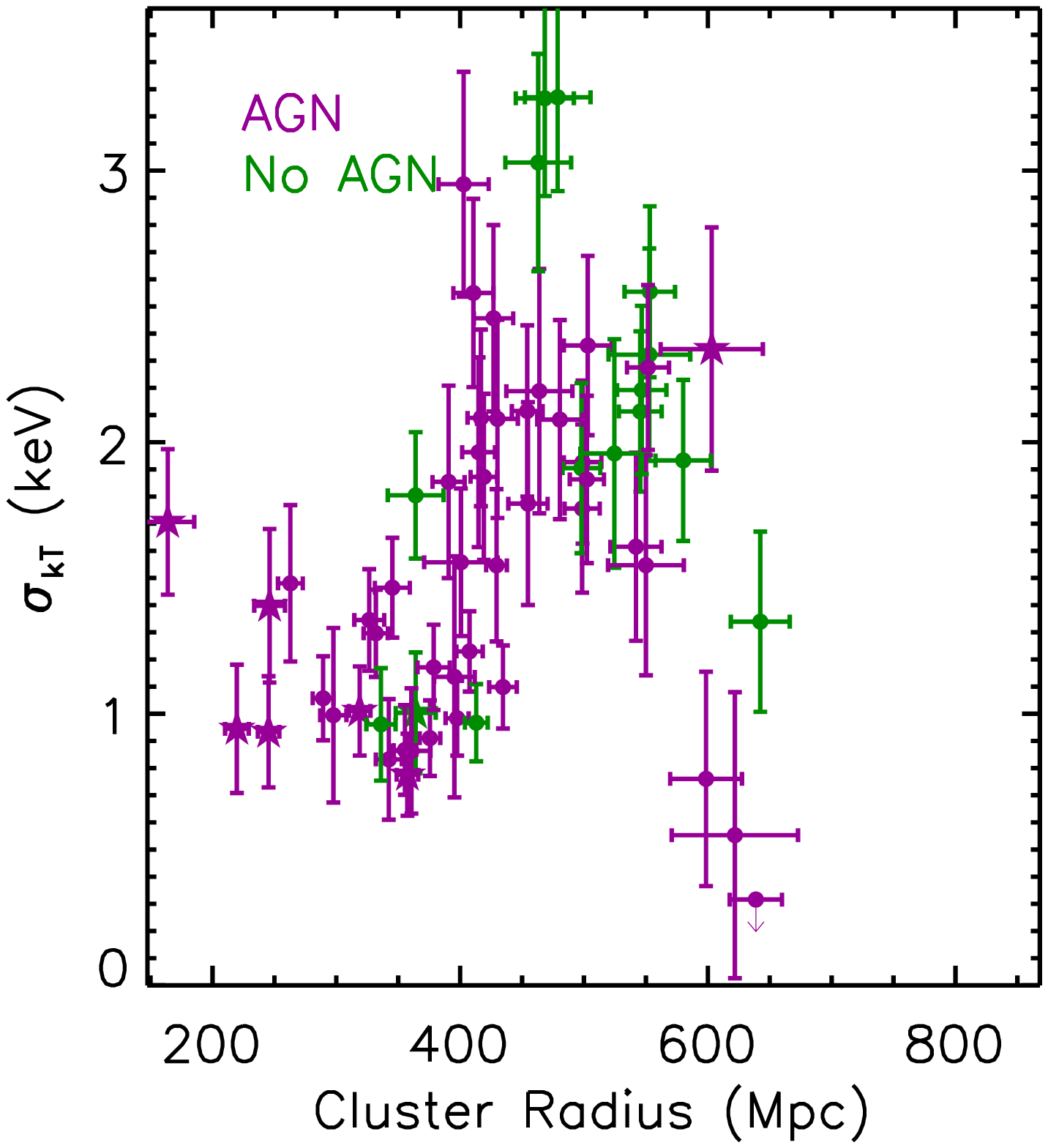}}
\subfigure{\includegraphics[width=0.3\textwidth]{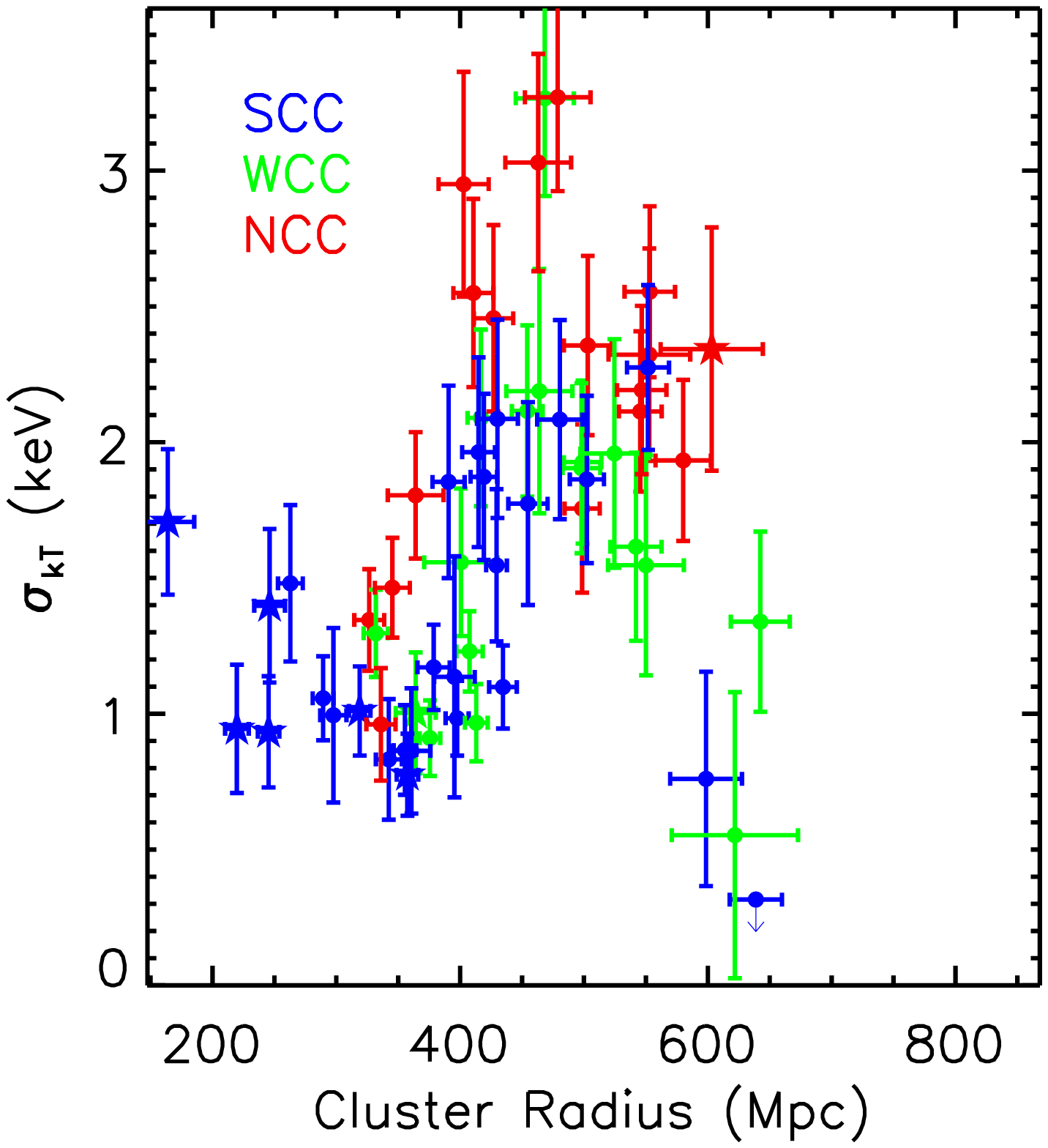}}
\caption{\footnotesize $r_{2500}-\sigma_{kT}$ scaling relation, shown with each classification scheme.  Stars represent the eight largest clusters for which $r_{2500}$ is beyond the field of view.  Colors are the same as Figure \ref{figure:kThists}. }
\label{figure:rsigmakT}
\end{center}
\end{figure*}

Plots of $\sigma_{kT} - kT_{med}$, $\sigma_{kT} - EM$, $\sigma_{kT} - r_{2500}$ and are shown in Figures \ref{figure:sigmakTkT}, \ref{figure:emsigmakT}, and \ref{figure:rsigmakT}, respectively.  To investigate the possibility of a correlation between $\sigma_{kT}$ and $kT_{med}$, the data was fit to two models, a linear relation and a constant.  The best fit constant is $\sigma_{kT}\sim\kTsigmaconstant$ keV with a $\chi^2$ of $\kTsigmaconstantchi$.  The linear fit resulted in a best fit relation of $\sigma_{kT}\sim\kTsigmaalpha kT_{med}+\kTsigmabeta$ with a $\chi^2$ of $\kTsigmalinearchi$.  Though neither the constant nor linear model is a very good fit, the decrease in $\chi^2$ of $\Delta\chi^2=\kTsigmadeltachi$ between the two fits hints at a correlation between $\sigma_{kT}$ and $kT_{med}$.  The linear fit indicates the presence of intrinsic scatter of \intscatter.


\section{Discussion}
\label{section:discussion}

We have found that for relaxed, SCC, WCC, and AGN clusters, $\sigma_{kT}$ is more likely to be small, while for disturbed, NCC, and non-AGN clusters, $\sigma_{kT}$ is typically larger and may be correlated with $kT_{med}$.  Apparently, the establishment of a strong cooling flow, often in conjunction with central AGN activity, leads to a reduction of both $\sigma_{kT}$ and $kT_{med}$.  As the gas cools, the higher temperature gas is removed, turned into lower temperature gas, down to some cutoff which scales as $kT_{med}$ \citep{Peterson2003a}, likely due to AGN heating, which has the effect of lowering $\sigma_{kT}$.  This would also have the effect of lowering $kT_{med}$.  Nearly any disturbance to the ICM, generally from interaction with another cluster, usually disrupts the cooling flow, raising the median temperature.  It also seems plausible that the merging of two clusters, with different temperature distributions, should result in a larger $\sigma_{kT}$.  Even for weaker interactions, where the clusters remain separate, the interaction may give rise to cold fronts, shocks, or at the very least a disruption to the cool cores, all of which could increase $\sigma_{kT}$.  The presence of intrinsic scatter in $\sigma_{kT}$ suggests the exact details of each cluster's merger, cooling, and AGN history probably has a significant effect on $\sigma_{kT}$.  

\section{Summary and Future Work}
\label{section:summary}
We have used the smoothed particle inference method of \citep{Peterson2007} to measure the temperature distributions of the HIFLUGCS sample of galaxy clusters.  The shape of the temperature distributions are found to be dominated by the isothermal spread function, which is approximately log-normal.  The measured temperature distributions tend to be somewhat broader than the ISF and have slight variations from cluster to cluster.  We find the width of the distributions, $\sigma_{kT}$, is inconsistent with isothermality for all \nclusters clusters.  We find a slope for $EM-kT_{med}$ relation that is steeper than predicted from self-similarity, but consistent with other works, $\alpha = \emkTalpha$, which increases to $\alpha = \lowemkTalpha$ for low-temperature clusters.  Separating the clusters according to various criteria, we find that the relaxed, SCC, and AGN groups largely overlap, as do the disturbed, NCC, and non-AGN cluster groups.  The relaxed, SCC, WCC, and AGN clusters tend to have not only lower $kT_{med}$, as expected and found in other works, but also smaller $\sigma_{kT}$ than the disturbed, NCC, and non-AGN clusters.  The $\sigma_{kT}-kT_{med}$ is described better with a linear fit, $\sigma_{kT}\sim\kTsigmaalpha kT_{med}+\kTsigmabeta$, than a constant, indicating that there may be a correlation between $\sigma_{kT}$ and $kT_{med}$.  However, the intrinsic scatter in the $\sigma_{kT}$ measurements, \intscatter, indicates that the exact history of each cluster may have a significant impact on $\sigma_{kT}$ and there is still much to be learned.  

The upper end of the cluster temperature distributions, particularly for massive clusters, is not as well constrained as the lower temperature end.  This is due partly to the fact that XMM-Newton does not have much sensitivity at energies above $\sim 15 keV$, and partly due to a lack of line emission at such high temperatures.  However, it may be possible to somewhat improve temperature constraints at higher temperatures by including hard X-ray data in the analysis, e.g. from Suzaku.  The foremost issue to be addressed with future work is the large amount of scatter in the $\sigma_{kT}$ results, both to confirm or rule out a $\sigma_{kT}-kT_{med}$ correlation and to more accurately determine the amount of intrinsic scatter in $\sigma_{kT}$.  Towards this end, it would be useful to include more clusters in the sample, e.g. the 43 additional clusters in the HIFLUGCS extended sample.  The most useful avenue of research would be to compare our results with those from hydrodynamical simulations of cluster formation.  Measurements of $\sigma_{kT}$ from such simulations would shed light on the amount of intrinsic scatter to be expected.  They may also reveal how complicated cluster physics, such as conduction, AGN feedback, cooling flows, and cluster mergers, might effect $\sigma_{kT}$. 

\acknowledgments
KAF and JRP are supported by NASA grants \#NNX07AH51G, \#NNX09AD15G, \#NNX07AQ30G, and \#NNX08AX45G.  KAF is also grateful for support from the Gary L. Wright Memorial Fellowship.  This research has made use of the NASA/IPAC Extragalactic Database (NED) which is operated by the Jet Propulsion Laboratory, California Institute of Technology, under contract with the National Aeronautics and Space Administration.  The authors would like to thank the referee for many helpful comments and suggestions.

\bibliography{/Users/kafrank/Dropbox/master_refs}

\begin{thebibliography}{100}
\expandafter\ifx\csname natexlab\endcsname\relax\def\natexlab#1{#1}\fi

\bibitem[{Allen \& Fabian(1998)}]{Allen1998a}
Allen, S. \& Fabian, A. 1998, \mnras, 297, L57

\bibitem[{{Anders} \& {Grevesse}(1989)}]{Anders1989}
{Anders}, E. \& {Grevesse}, N. 1989, \gca, 53, 197

\bibitem[{{Andersson} {et~al.}(2007){Andersson}, {Peterson}, \&
  {Madejski}}]{Andersson2007}
{Andersson}, K., {Peterson}, J.~R., \& {Madejski}, G. 2007, \apj, 670, 1010

\bibitem[{{Andersson} {et~al.}(2009){Andersson}, {Peterson}, {Madejski}, \&
  {Goobar}}]{Andersson2009}
{Andersson}, K., {Peterson}, J.~R., {Madejski}, G., \& {Goobar}, A. 2009, \apj,
  696, 1029

\bibitem[{{Andersson} \& {Madejski}(2004)}]{Andersson2004}
{Andersson}, K.~E. \& {Madejski}, G.~M. 2004, \apj, 607, 190

\bibitem[{{Araudo} {et~al.}(2008){Araudo}, {Cora}, \& {Romero}}]{Araudo2008}
{Araudo}, A.~T., {Cora}, S.~A., \& {Romero}, G.~E. 2008, \mnras, 390, 323

\bibitem[{{Arnaud} \& {Evrard}(1999)}]{Arnaud1999}
{Arnaud}, M. \& {Evrard}, A.~E. 1999, \mnras, 305, 631

\bibitem[{{Arnaud} {et~al.}(2005){Arnaud}, {Pointecouteau}, \&
  {Pratt}}]{Arnaud2005a}
{Arnaud}, M., {Pointecouteau}, E., \& {Pratt}, G.~W. 2005, \aap, 441, 893

\bibitem[{{Bacchi} {et~al.}(2003){Bacchi}, {Feretti}, {Giovannini}, \&
  {Govoni}}]{Bacchi2003}
{Bacchi}, M., {Feretti}, L., {Giovannini}, G., \& {Govoni}, F. 2003, \aap, 400,
  465

\bibitem[{{Bagchi} {et~al.}(2006){Bagchi}, {Durret}, {Neto}, \&
  {Paul}}]{Bagchi2006}
{Bagchi}, J., {Durret}, F., {Neto}, G.~B.~L., \& {Paul}, S. 2006, Science, 314,
  791

\bibitem[{{Belsole} {et~al.}(2005){Belsole}, {Sauvageot}, {Pratt}, \&
  {Bourdin}}]{Belsole2005}
{Belsole}, E., {Sauvageot}, J.-L., {Pratt}, G.~W., \& {Bourdin}, H. 2005,
  Advances in Space Research, 36, 630

\bibitem[{{Bildfell} {et~al.}(2008){Bildfell}, {Hoekstra}, {Babul}, \&
  {Mahdavi}}]{Bildfell2008}
{Bildfell}, C., {Hoekstra}, H., {Babul}, A., \& {Mahdavi}, A. 2008, \mnras,
  389, 1637

\bibitem[{{B{\^\i}rzan} {et~al.}(2004){B{\^\i}rzan}, {Rafferty}, {McNamara},
  {Wise}, \& {Nulsen}}]{Birzan2004}
{B{\^\i}rzan}, L., {Rafferty}, D.~A., {McNamara}, B.~R., {Wise}, M.~W., \&
  {Nulsen}, P.~E.~J. 2004, \apj, 607, 800

\bibitem[{{Blanton} {et~al.}(2011){Blanton}, {Randall}, {Clarke}, {Sarazin},
  {McNamara}, {Douglass}, \& {McDonald}}]{Blanton2011}
{Blanton}, E.~L., {Randall}, S.~W., {Clarke}, T.~E., {Sarazin}, C.~L.,
  {McNamara}, B.~R., {Douglass}, E.~M., \& {McDonald}, M. 2011, \apj, 737, 99

\bibitem[{{Bourdin} \& {Mazzotta}(2008)}]{Bourdin2008}
{Bourdin}, H. \& {Mazzotta}, P. 2008, \aap, 479, 307

\bibitem[{{Buote} \& {Tsai}(1996)}]{Buote1996}
{Buote}, D.~A. \& {Tsai}, J.~C. 1996, \apj, 458, 27

\bibitem[{{Burns} {et~al.}(2008){Burns}, {Hallman}, {Gantner}, {Motl}, \&
  {Norman}}]{Burns2008a}
{Burns}, J.~O., {Hallman}, E.~J., {Gantner}, B., {Motl}, P.~M., \& {Norman},
  M.~L. 2008, \apj, 675, 1125

\bibitem[{{Chen} {et~al.}(2007){Chen}, {Reiprich}, {B{\"o}hringer}, {Ikebe}, \&
  {Zhang}}]{Chen2007}
{Chen}, Y., {Reiprich}, T.~H., {B{\"o}hringer}, H., {Ikebe}, Y., \& {Zhang},
  Y.-Y. 2007, \aap, 466, 805

\bibitem[{{Clarke} \& {Ensslin}(2006)}]{Clarke2006a}
{Clarke}, T.~E. \& {Ensslin}, T.~A. 2006, \aj, 131, 2900

\bibitem[{{Crawford} {et~al.}(2005){Crawford}, {Hatch}, {Fabian}, \&
  {Sanders}}]{Crawford2005}
{Crawford}, C.~S., {Hatch}, N.~A., {Fabian}, A.~C., \& {Sanders}, J.~S. 2005,
  \mnras, 363, 216

\bibitem[{{David} {et~al.}(2011){David}, {O'Sullivan}, {Jones}, {Giacintucci},
  {Vrtilek}, {Raychaudhury}, {Nulsen}, {Forman}, {Sun}, \&
  {Donahue}}]{David2011}
{David}, L.~P., {O'Sullivan}, E., {Jones}, C., {Giacintucci}, S., {Vrtilek},
  J., {Raychaudhury}, S., {Nulsen}, P.~E.~J., {Forman}, W., {Sun}, M., \&
  {Donahue}, M. 2011, \apj, 728, 162

\bibitem[{{den Herder} {et~al.}(2001){den Herder}, {Brinkman}, {Kahn},
  {Branduardi-Raymont}, {Thomsen}, {Aarts}, {Audard}, {Bixler}, {den Boggende},
  {Cottam}, {Decker}, {Dubbeldam}, {Erd}, {Goulooze}, {G{\"u}del}, {Guttridge},
  {Hailey}, {Janabi}, {Kaastra}, {de Korte}, {van Leeuwen}, {Mauche},
  {McCalden}, {Mewe}, {Naber}, {Paerels}, {Peterson}, {Rasmussen}, {Rees},
  {Sakelliou}, {Sako}, {Spodek}, {Stern}, {Tamura}, {Tandy}, {de Vries},
  {Welch}, \& {Zehnder}}]{denHerder2001}
{den Herder}, J.~W., {Brinkman}, A.~C., {Kahn}, S.~M., {Branduardi-Raymont},
  G., {Thomsen}, K., {Aarts}, H., {Audard}, M., {Bixler}, J.~V., {den
  Boggende}, A.~J., {Cottam}, J., {Decker}, T., {Dubbeldam}, L., {Erd}, C.,
  {Goulooze}, H., {G{\"u}del}, M., {Guttridge}, P., {Hailey}, C.~J., {Janabi},
  K.~A., {Kaastra}, J.~S., {de Korte}, P.~A.~J., {van Leeuwen}, B.~J.,
  {Mauche}, C., {McCalden}, A.~J., {Mewe}, R., {Naber}, A., {Paerels}, F.~B.,
  {Peterson}, J.~R., {Rasmussen}, A.~P., {Rees}, K., {Sakelliou}, I., {Sako},
  M., {Spodek}, J., {Stern}, M., {Tamura}, T., {Tandy}, J., {de Vries}, C.~P.,
  {Welch}, S., \& {Zehnder}, A. 2001, \aap, 365, L7

\bibitem[{{Dickey} \& {Lockman}(1990)}]{Dickey1990}
{Dickey}, J.~M. \& {Lockman}, F.~J. 1990, \araa, 28, 215

\bibitem[{{Diehl} {et~al.}(2008){Diehl}, {Li}, {Fryer}, \&
  {Rafferty}}]{Diehl2008}
{Diehl}, S., {Li}, H., {Fryer}, C.~L., \& {Rafferty}, D. 2008, \apj, 687, 173

\bibitem[{{Dong} {et~al.}(2010){Dong}, {Rasmussen}, \& {Mulchaey}}]{Dong2010}
{Dong}, R., {Rasmussen}, J., \& {Mulchaey}, J.~S. 2010, \apj, 712, 883

\bibitem[{{Donnelly} {et~al.}(2001){Donnelly}, {Forman}, {Jones}, {Quintana},
  {Ramirez}, {Churazov}, \& {Gilfanov}}]{Donnelly2001}
{Donnelly}, R.~H., {Forman}, W., {Jones}, C., {Quintana}, H., {Ramirez}, A.,
  {Churazov}, E., \& {Gilfanov}, M. 2001, \apj, 562, 254

\bibitem[{{Dupke} {et~al.}(2007){Dupke}, {Mirabal}, {Bregman}, \&
  {Evrard}}]{Dupke2007}
{Dupke}, R.~A., {Mirabal}, N., {Bregman}, J.~N., \& {Evrard}, A.~E. 2007, \apj,
  668, 781

\bibitem[{{Durret} {et~al.}(2000){Durret}, {Adami}, {Gerbal}, \&
  {Pislar}}]{Durret2000}
{Durret}, F., {Adami}, C., {Gerbal}, D., \& {Pislar}, V. 2000, \aap, 356, 815

\bibitem[{{Fabian} {et~al.}(2005){Fabian}, {Sanders}, {Taylor}, \&
  {Allen}}]{Fabian2005}
{Fabian}, A.~C., {Sanders}, J.~S., {Taylor}, G.~B., \& {Allen}, S.~W. 2005,
  \mnras, 360, L20

\bibitem[{{Fabian} {et~al.}(2011){Fabian}, {Sanders}, {Williams}, {Lazarian},
  {Ferland}, \& {Johnstone}}]{Fabian2011}
{Fabian}, A.~C., {Sanders}, J.~S., {Williams}, R.~J.~R., {Lazarian}, A.,
  {Ferland}, G.~J., \& {Johnstone}, R.~M. 2011, \mnras, 417, 172

\bibitem[{{Feretti} {et~al.}(2001){Feretti}, {Fusco-Femiano}, {Giovannini}, \&
  {Govoni}}]{Feretti2001}
{Feretti}, L., {Fusco-Femiano}, R., {Giovannini}, G., \& {Govoni}, F. 2001,
  \aap, 373, 106

\bibitem[{{Finoguenov} {et~al.}(2004){Finoguenov}, {Henriksen}, {Briel}, {de
  Plaa}, \& {Kaastra}}]{Finoguenov2004}
{Finoguenov}, A., {Henriksen}, M.~J., {Briel}, U.~G., {de Plaa}, J., \&
  {Kaastra}, J.~S. 2004, \apj, 611, 811

\bibitem[{{Giacintucci} {et~al.}(2005){Giacintucci}, {Venturi}, {Brunetti},
  {Bardelli}, {Dallacasa}, {Ettori}, {Finoguenov}, {Rao}, \&
  {Zucca}}]{Giacintucci2005}
{Giacintucci}, S., {Venturi}, T., {Brunetti}, G., {Bardelli}, S., {Dallacasa},
  D., {Ettori}, S., {Finoguenov}, A., {Rao}, A.~P., \& {Zucca}, E. 2005, \aap,
  440, 867

\bibitem[{{Giovannini} {et~al.}(1993){Giovannini}, {Feretti}, {Venturi}, {Kim},
  \& {Kronberg}}]{Giovannini1993}
{Giovannini}, G., {Feretti}, L., {Venturi}, T., {Kim}, K.-T., \& {Kronberg},
  P.~P. 1993, \apj, 406, 399

\bibitem[{{Govoni} {et~al.}(2005){Govoni}, {Murgia}, {Feretti}, {Giovannini},
  {Dallacasa}, \& {Taylor}}]{Govoni2005}
{Govoni}, F., {Murgia}, M., {Feretti}, L., {Giovannini}, G., {Dallacasa}, D.,
  \& {Taylor}, G.~B. 2005, \aap, 430, L5

\bibitem[{{Govoni} {et~al.}(2009){Govoni}, {Murgia}, {Markevitch}, {Feretti},
  {Giovannini}, {Taylor}, \& {Carretti}}]{Govoni2009}
{Govoni}, F., {Murgia}, M., {Markevitch}, M., {Feretti}, L., {Giovannini}, G.,
  {Taylor}, G.~B., \& {Carretti}, E. 2009, \aap, 499, 371

\bibitem[{{Hayakawa} {et~al.}(2006){Hayakawa}, {Hoshino}, {Ishida}, {Furusho},
  {Yamasaki}, \& {Ohashi}}]{Hayakawa2006}
{Hayakawa}, A., {Hoshino}, A., {Ishida}, M., {Furusho}, T., {Yamasaki}, N.~Y.,
  \& {Ohashi}, T. 2006, \pasj, 58, 695

\bibitem[{{Henriksen} {et~al.}(2000){Henriksen}, {Donnelly}, \&
  {Davis}}]{Henriksen2000}
{Henriksen}, M., {Donnelly}, R.~H., \& {Davis}, D.~S. 2000, \apj, 529, 692

\bibitem[{{Henry} {et~al.}(2004){Henry}, {Finoguenov}, \& {Briel}}]{Henry2004}
{Henry}, J.~P., {Finoguenov}, A., \& {Briel}, U.~G. 2004, \apj, 615, 181

\bibitem[{{Hudson} {et~al.}(2010){Hudson}, {Mittal}, {Reiprich}, {Nulsen},
  {Andernach}, \& {Sarazin}}]{Hudson2010}
{Hudson}, D.~S., {Mittal}, R., {Reiprich}, T.~H., {Nulsen}, P.~E.~J.,
  {Andernach}, H., \& {Sarazin}, C.~L. 2010, \aap, 513, A37+

\bibitem[{{Hudson} {et~al.}(2006){Hudson}, {Reiprich}, {Clarke}, \&
  {Sarazin}}]{Hudson2006}
{Hudson}, D.~S., {Reiprich}, T.~H., {Clarke}, T.~E., \& {Sarazin}, C.~L. 2006,
  \aap, 453, 433

\bibitem[{{Johnston-Hollitt} {et~al.}(2008){Johnston-Hollitt}, {Sato}, {Gill},
  {Fleenor}, \& {Brick}}]{JohnstonHollitt2008}
{Johnston-Hollitt}, M., {Sato}, M., {Gill}, J.~A., {Fleenor}, M.~C., \&
  {Brick}, A.-M. 2008, \mnras, 390, 289

\bibitem[{{Johnstone} {et~al.}(2002){Johnstone}, {Allen}, {Fabian}, \&
  {Sanders}}]{Johnstone2002}
{Johnstone}, R.~M., {Allen}, S.~W., {Fabian}, A.~C., \& {Sanders}, J.~S. 2002,
  \mnras, 336, 299

\bibitem[{{Kaastra}(1992)}]{Kaastra1992}
{Kaastra}, J.~S. 1992, {An X-Ray Spectral Code for Optically Thin Plasmas},
  Internal SRON-Leiden Report

\bibitem[{{Kanov} {et~al.}(2006){Kanov}, {Sarazin}, \& {Hicks}}]{Kanov2006}
{Kanov}, K.~N., {Sarazin}, C.~L., \& {Hicks}, A.~K. 2006, \apj, 653, 184

\bibitem[{{Kassim} {et~al.}(2001){Kassim}, {Clarke}, {En{\ss}lin}, {Cohen}, \&
  {Neumann}}]{Kassim2001}
{Kassim}, N.~E., {Clarke}, T.~E., {En{\ss}lin}, T.~A., {Cohen}, A.~S., \&
  {Neumann}, D.~M. 2001, \apj, 559, 785

\bibitem[{{Kelly}(2007)}]{Kelly2007}
{Kelly}, B.~C. 2007, \apj, 665, 1489

\bibitem[{{Kempner} {et~al.}(2004){Kempner}, {Blanton}, {Clarke}, {En{\ss}lin},
  {Johnston-Hollitt}, \& {Rudnick}}]{Kempner2004}
{Kempner}, J.~C., {Blanton}, E.~L., {Clarke}, T.~E., {En{\ss}lin}, T.~A.,
  {Johnston-Hollitt}, M., \& {Rudnick}, L. 2004, in The Riddle of Cooling Flows
  in Galaxies and Clusters of galaxies, ed. {T.~Reiprich, J.~Kempner, \&amp;
  N.~Soker} (Published electronically at
  http://www.astro.virginia.edu/coolflow/), 335--+

\bibitem[{{Kempner} {et~al.}(2002){Kempner}, {Sarazin}, \&
  {Ricker}}]{Kempner2002}
{Kempner}, J.~C., {Sarazin}, C.~L., \& {Ricker}, P.~M. 2002, \apj, 579, 236

\bibitem[{{Kuntz} \& {Snowden}(2000)}]{Kuntz2000a}
{Kuntz}, K.~D. \& {Snowden}, S.~L. 2000, \apj, 543, 195

\bibitem[{{Liedahl} {et~al.}(1995){Liedahl}, {Osterheld}, \&
  {Goldstein}}]{Liedahl1995}
{Liedahl}, D.~A., {Osterheld}, A.~L., \& {Goldstein}, W.~H. 1995, \apjl, 438,
  L115

\bibitem[{{Lumb} {et~al.}(2002){Lumb}, {Warwick}, {Page}, \& {De
  Luca}}]{Lumb2002}
{Lumb}, D.~H., {Warwick}, R.~S., {Page}, M., \& {De Luca}, A. 2002, \aap, 389,
  93

\bibitem[{Markevitch(1998)}]{Markevitch1998}
Markevitch, M. 1998, \apj, 504, 27

\bibitem[{{Markevitch} {et~al.}(1999){Markevitch}, {Sarazin}, \&
  {Vikhlinin}}]{Markevitch1999}
{Markevitch}, M., {Sarazin}, C.~L., \& {Vikhlinin}, A. 1999, \apj, 521, 526

\bibitem[{{Markevitch} \& {Vikhlinin}(2001)}]{Markevitch2001}
{Markevitch}, M. \& {Vikhlinin}, A. 2001, \apj, 563, 95

\bibitem[{{Markevitch} \& {Vikhlinin}(2007)}]{Markevitch2007}
---. 2007, \physrep, 443, 1

\bibitem[{{Mazzotta} {et~al.}(2002){Mazzotta}, {Kaastra}, {Paerels},
  {Ferrigno}, {Colafrancesco}, {Mewe}, \& {Forman}}]{Mazzotta2002}
{Mazzotta}, P., {Kaastra}, J.~S., {Paerels}, F.~B., {Ferrigno}, C.,
  {Colafrancesco}, S., {Mewe}, R., \& {Forman}, W.~R. 2002, \apjl, 567, L37

\bibitem[{{McDonald} {et~al.}(2010){McDonald}, {Veilleux}, {Rupke}, \&
  {Mushotzky}}]{McDonald2010}
{McDonald}, M., {Veilleux}, S., {Rupke}, D.~S.~N., \& {Mushotzky}, R. 2010,
  \apj, 721, 1262

\bibitem[{{McNamara} \& {Nulsen}(2007)}]{McNamara2007}
{McNamara}, B.~R. \& {Nulsen}, P.~E.~J. 2007, \araa, 45, 117

\bibitem[{{McNamara} {et~al.}(2000){McNamara}, {Wise}, {Nulsen}, {David},
  {Sarazin}, {Bautz}, {Markevitch}, {Vikhlinin}, {Forman}, {Jones}, \&
  {Harris}}]{McNamara2000}
{McNamara}, B.~R., {Wise}, M., {Nulsen}, P.~E.~J., {David}, L.~P., {Sarazin},
  C.~L., {Bautz}, M., {Markevitch}, M., {Vikhlinin}, A., {Forman}, W.~R.,
  {Jones}, C., \& {Harris}, D.~E. 2000, \apjl, 534, L135

\bibitem[{{McNamara} {et~al.}(2001){McNamara}, {Wise}, {Nulsen}, {David},
  {Carilli}, {Sarazin}, {O'Dea}, {Houck}, {Donahue}, {Baum}, {Voit},
  {O'Connell}, \& {Koekemoer}}]{McNamara2001}
{McNamara}, B.~R., {Wise}, M.~W., {Nulsen}, P.~E.~J., {David}, L.~P.,
  {Carilli}, C.~L., {Sarazin}, C.~L., {O'Dea}, C.~P., {Houck}, J., {Donahue},
  M., {Baum}, S., {Voit}, M., {O'Connell}, R.~W., \& {Koekemoer}, A. 2001,
  \apjl, 562, L149

\bibitem[{{Mewe} {et~al.}(1985){Mewe}, {Gronenschild}, \& {van den
  Oord}}]{Mewe1985}
{Mewe}, R., {Gronenschild}, E.~H.~B.~M., \& {van den Oord}, G.~H.~J. 1985,
  \aaps, 62, 197

\bibitem[{{Mewe} {et~al.}(1986){Mewe}, {Lemen}, \& {van den Oord}}]{Mewe1986}
{Mewe}, R., {Lemen}, J.~R., \& {van den Oord}, G.~H.~J. 1986, \aaps, 65, 511

\bibitem[{{Mittal} {et~al.}(2009){Mittal}, {Hudson}, {Reiprich}, \&
  {Clarke}}]{Mittal2009}
{Mittal}, R., {Hudson}, D.~S., {Reiprich}, T.~H., \& {Clarke}, T. 2009, \aap,
  501, 835

\bibitem[{{Morrison} \& {McCammon}(1983)}]{Morrison1983}
{Morrison}, R. \& {McCammon}, D. 1983, \apj, 270, 119

\bibitem[{{Murgia} {et~al.}(2010){Murgia}, {Govoni}, {Feretti}, \&
  {Giovannini}}]{Murgia2010}
{Murgia}, M., {Govoni}, F., {Feretti}, L., \& {Giovannini}, G. 2010, \aap, 509,
  A86+

\bibitem[{{Nevalainen} {et~al.}(2001){Nevalainen}, {Kaastra}, {Parmar},
  {Markevitch}, {Oosterbroek}, {Colafrancesco}, \& {Mazzotta}}]{Nevalainen2001}
{Nevalainen}, J., {Kaastra}, J., {Parmar}, A.~N., {Markevitch}, M.,
  {Oosterbroek}, T., {Colafrancesco}, S., \& {Mazzotta}, P. 2001, \aap, 369,
  459

\bibitem[{{Ohto} {et~al.}(2003){Ohto}, {Kawano}, \& {Fukazawa}}]{Ohto2003}
{Ohto}, A., {Kawano}, N., \& {Fukazawa}, Y. 2003, \pasj, 55, 819

\bibitem[{{Owers} {et~al.}(2009){Owers}, {Nulsen}, {Couch}, \&
  {Markevitch}}]{Owers2009}
{Owers}, M.~S., {Nulsen}, P.~E.~J., {Couch}, W.~J., \& {Markevitch}, M. 2009,
  \apj, 704, 1349

\bibitem[{{Paolillo} {et~al.}(2003){Paolillo}, {Fabbiano}, {Peres}, \&
  {Kim}}]{Paolillo2003}
{Paolillo}, M., {Fabbiano}, G., {Peres}, G., \& {Kim}, D.-W. 2003, \apj, 586,
  850

\bibitem[{{Peterson} {et~al.}(2003){Peterson}, {Kahn}, {Paerels}, {Kaastra},
  {Tamura}, {Bleeker}, {Ferrigno}, \& {Jernigan}}]{Peterson2003a}
{Peterson}, J.~R., {Kahn}, S.~M., {Paerels}, F.~B.~S., {Kaastra}, J.~S.,
  {Tamura}, T., {Bleeker}, J.~A.~M., {Ferrigno}, C., \& {Jernigan}, J.~G. 2003,
  \apj, 590, 207

\bibitem[{{Peterson} {et~al.}(2007){Peterson}, {Marshall}, \&
  {Andersson}}]{Peterson2007}
{Peterson}, J.~R., {Marshall}, P.~J., \& {Andersson}, K. 2007, \apj, 655, 109

\bibitem[{{Rafferty} {et~al.}(2006){Rafferty}, {McNamara}, {Nulsen}, \&
  {Wise}}]{Rafferty2006}
{Rafferty}, D.~A., {McNamara}, B.~R., {Nulsen}, P.~E.~J., \& {Wise}, M.~W.
  2006, \apj, 652, 216

\bibitem[{{Reiprich} \& {B{\"o}hringer}(2002)}]{Reiprich2002a}
{Reiprich}, T.~H. \& {B{\"o}hringer}, H. 2002, \apj, 567, 716

\bibitem[{{Reiprich} {et~al.}(2004){Reiprich}, {Sarazin}, {Kempner}, \&
  {Tittley}}]{Reiprich2004}
{Reiprich}, T.~H., {Sarazin}, C.~L., {Kempner}, J.~C., \& {Tittley}, E. 2004,
  \apj, 608, 179

\bibitem[{{Rossetti} {et~al.}(2007){Rossetti}, {Ghizzardi}, {Molendi}, \&
  {Finoguenov}}]{Rossetti2007}
{Rossetti}, M., {Ghizzardi}, S., {Molendi}, S., \& {Finoguenov}, A. 2007, \aap,
  463, 839

\bibitem[{{Rossetti} \& {Molendi}(2010)}]{Rossetti2010}
{Rossetti}, M. \& {Molendi}, S. 2010, \aap, 510, A83+

\bibitem[{{Rottgering} {et~al.}(1994){Rottgering}, {Snellen}, {Miley}, {de
  Jong}, {Hanisch}, \& {Perley}}]{Roettgering1994}
{Rottgering}, H., {Snellen}, I., {Miley}, G., {de Jong}, J.~P., {Hanisch},
  R.~J., \& {Perley}, R. 1994, \apj, 436, 654

\bibitem[{{R{\"o}ttgering} {et~al.}(1997){R{\"o}ttgering}, {Wieringa},
  {Hunstead}, \& {Ekers}}]{Roettgering1997}
{R{\"o}ttgering}, H.~J.~A., {Wieringa}, M.~H., {Hunstead}, R.~W., \& {Ekers},
  R.~D. 1997, \mnras, 290, 577

\bibitem[{{Sakelliou} \& {Ponman}(2004)}]{Sakelliou2004}
{Sakelliou}, I. \& {Ponman}, T.~J. 2004, \mnras, 351, 1439

\bibitem[{{Sakelliou} \& {Ponman}(2006)}]{Sakelliou2006}
---. 2006, \mnras, 367, 1409

\bibitem[{{Sanders} {et~al.}(2008){Sanders}, {Fabian}, {Allen}, {Morris},
  {Graham}, \& {Johnstone}}]{Sanders2008}
{Sanders}, J.~S., {Fabian}, A.~C., {Allen}, S.~W., {Morris}, R.~G., {Graham},
  J., \& {Johnstone}, R.~M. 2008, \mnras, 385, 1186

\bibitem[{{Sanders} {et~al.}(2010){Sanders}, {Fabian}, {Frank}, {Peterson}, \&
  {Russell}}]{Sanders2010}
{Sanders}, J.~S., {Fabian}, A.~C., {Frank}, K.~A., {Peterson}, J.~R., \&
  {Russell}, H.~R. 2010, \mnras, 402, 127

\bibitem[{{Sanders} {et~al.}(2009{\natexlab{a}}){Sanders}, {Fabian}, \&
  {Taylor}}]{Sanders2009}
{Sanders}, J.~S., {Fabian}, A.~C., \& {Taylor}, G.~B. 2009{\natexlab{a}},
  \mnras, 396, 1449

\bibitem[{{Sanders} {et~al.}(2009{\natexlab{b}}){Sanders}, {Fabian}, \&
  {Taylor}}]{Sanders2009a}
---. 2009{\natexlab{b}}, \mnras, 393, 71

\bibitem[{{Schuecker} {et~al.}(2001){Schuecker}, {B{\"o}hringer}, {Reiprich},
  \& {Feretti}}]{Schuecker2001}
{Schuecker}, P., {B{\"o}hringer}, H., {Reiprich}, T.~H., \& {Feretti}, L. 2001,
  \aap, 378, 408

\bibitem[{{Shurkin} {et~al.}(2008){Shurkin}, {Dunn}, {Gentile}, {Taylor}, \&
  {Allen}}]{Shurkin2008}
{Shurkin}, K., {Dunn}, R.~J.~H., {Gentile}, G., {Taylor}, G.~B., \& {Allen},
  S.~W. 2008, \mnras, 383, 923

\bibitem[{{Snowden} {et~al.}(2008){Snowden}, {Mushotzky}, {Kuntz}, \&
  {Davis}}]{Snowden2008}
{Snowden}, S.~L., {Mushotzky}, R.~F., {Kuntz}, K.~D., \& {Davis}, D.~S. 2008,
  \aap, 478, 615

\bibitem[{{Sun} {et~al.}(2003){Sun}, {Forman}, {Vikhlinin}, {Hornstrup},
  {Jones}, \& {Murray}}]{Sun2003}
{Sun}, M., {Forman}, W., {Vikhlinin}, A., {Hornstrup}, A., {Jones}, C., \&
  {Murray}, S.~S. 2003, \apj, 598, 250

\bibitem[{{Sun} \& {Murray}(2002)}]{Sun2002}
{Sun}, M. \& {Murray}, S.~S. 2002, \apj, 576, 708

\bibitem[{{Sun} {et~al.}(2002){Sun}, {Murray}, {Markevitch}, \&
  {Vikhlinin}}]{Sun2002a}
{Sun}, M., {Murray}, S.~S., {Markevitch}, M., \& {Vikhlinin}, A. 2002, \apj,
  565, 867

\bibitem[{{Takahashi} \& {Yamashita}(2003)}]{Takahashi2003}
{Takahashi}, S. \& {Yamashita}, K. 2003, \pasj, 55, 1105

\bibitem[{{Takizawa} {et~al.}(2003){Takizawa}, {Sarazin}, {Blanton}, \&
  {Taylor}}]{Takizawa2003}
{Takizawa}, M., {Sarazin}, C.~L., {Blanton}, E.~L., \& {Taylor}, G.~B. 2003,
  \apj, 595, 142

\bibitem[{{Tamura} {et~al.}(2001){Tamura}, {Kaastra}, {Peterson}, {Paerels},
  {Mittaz}, {Trudolyubov}, {Stewart}, {Fabian}, {Mushotzky}, {Lumb}, \&
  {Ikebe}}]{Tamura2001}
{Tamura}, T., {Kaastra}, J.~S., {Peterson}, J.~R., {Paerels}, F.~B.~S.,
  {Mittaz}, J.~P.~D., {Trudolyubov}, S.~P., {Stewart}, G., {Fabian}, A.~C.,
  {Mushotzky}, R.~F., {Lumb}, D.~H., \& {Ikebe}, Y. 2001, \aap, 365, L87

\bibitem[{{Ventimiglia} {et~al.}(2008){Ventimiglia}, {Voit}, {Donahue}, \&
  {Ameglio}}]{Ventimiglia2008}
{Ventimiglia}, D.~A., {Voit}, G.~M., {Donahue}, M., \& {Ameglio}, S. 2008,
  \apj, 685, 118

\bibitem[{{Vikhlinin} {et~al.}(2009){Vikhlinin}, {Burenin}, {Ebeling},
  {Forman}, {Hornstrup}, {Jones}, {Kravtsov}, {Murray}, {Nagai}, {Quintana}, \&
  {Voevodkin}}]{Vikhlinin2009a}
{Vikhlinin}, A., {Burenin}, R.~A., {Ebeling}, H., {Forman}, W.~R., {Hornstrup},
  A., {Jones}, C., {Kravtsov}, A.~V., {Murray}, S.~S., {Nagai}, D., {Quintana},
  H., \& {Voevodkin}, A. 2009, \apj, 692, 1033

\bibitem[{{Vikhlinin} {et~al.}(2005){Vikhlinin}, {Markevitch}, {Murray},
  {Jones}, {Forman}, \& {Van Speybroeck}}]{Vikhlinin2005}
{Vikhlinin}, A., {Markevitch}, M., {Murray}, S.~S., {Jones}, C., {Forman}, W.,
  \& {Van Speybroeck}, L. 2005, \apj, 628, 655

\bibitem[{{Zhang} {et~al.}(2011){Zhang}, {Andernach}, {Caretta}, {Reiprich},
  {Boehringer}, {Puchwein}, {Sijacki}, \& {Girardi}}]{Zhang2011}
{Zhang}, Y., {Andernach}, H., {Caretta}, C.~A., {Reiprich}, T.~H.,
  {Boehringer}, H., {Puchwein}, E., {Sijacki}, D., \& {Girardi}, M. 2011, \aap,
  526, A105+

\bibitem[{{Zhang} {et~al.}(2009){Zhang}, {Reiprich}, {Finoguenov}, {Hudson}, \&
  {Sarazin}}]{Zhang2009}
{Zhang}, Y., {Reiprich}, T.~H., {Finoguenov}, A., {Hudson}, D.~S., \&
  {Sarazin}, C.~L. 2009, \apj, 699, 1178

\bibitem[{{Zhang} {et~al.}(2007){Zhang}, {Finoguenov}, {B{\"o}hringer},
  {Kneib}, {Smith}, {Czoske}, \& {Soucail}}]{Zhang2007}
{Zhang}, Y.-Y., {Finoguenov}, A., {B{\"o}hringer}, H., {Kneib}, J.-P., {Smith},
  G.~P., {Czoske}, O., \& {Soucail}, G. 2007, \aap, 467, 437

\end{thebibliography}
\clearpage

\begin{deluxetable*}{ccccccccc}[p]
\tabletypesize{\footnotesize}
\tablecaption{Observation Parameters \label{table:observations}}
\tablewidth{0pt}
\tablehead{\colhead{Cluster} & \colhead{Obs. ID} & \colhead{Redshift} & \colhead{$n_H$} & \colhead{PN} & \colhead{M1} & \colhead{M2} & \colhead{R1} & \colhead{R2} \\
(1) & (2) & (3) & (4) & (5) & (6) & (7) & (8) & (9) }
\startdata
NGC4636 & 0111190701 & 0.00313 & 1.75 & 50.9 & 58.6 & 58.6 & 55.3 & 55.3 \\
NGC5044 & 0554680101 & 0.00928 & 4.91 & 71.8 & 103.8 & 103.8 & 108.7 & 108.9 \\
NGC1550 & 0152150101 & 0.01239 & 18.00 & 19.0 & 21.4 & 21.9 & 20.5 & 23.4 \\ 
FORNAX & 0400620101 & 0.00460 & 1.45 & 68.7 & 114.9 & 114.9 & 118.6 & 118.6 \\
NGC507 & 0080540101 & 0.01646 & 5.25 & 26.4 & 34.2 & 34.2 &\ldots&\ldots\\
A3581 & 0504780301 & 0.02300 & 4.30 & 32.7 & 53.3 & 53.3 & 79.5 & 79.5 \\
MKW4 & 0093060101 & 0.02000 & 1.86 & 9.4 & 14.1 & 14.1 & 14.5 & 14.5 \\
A262 & 0504780101 & 0.01630 & 5.52 & 8.3 & 35.6 & 35.7 & 113.5 & 113.5 \\
A400 & 0404010101 & 0.02440 & 9.38 & 20.5 & 33.6 & 34.4 & 32.7 & 32.7 \\
A1367 & 0061740101 & 0.02200 & 2.55 & 17.6 & 30.9 & 31.5 &\ldots&\ldots\\
Zw54 & 0505230401 & 0.02900 & 16.68 & 29.6 & 45.3 & 45.6 & 48.8 & 48.8 \\
MKW8 & 0300210701 & 0.02700 & 2.60 & 16.5 & 23.1 & 23.1 & 23.5 & 23.5 \\
AS1101 & 0147800101 & 0.05800 & 1.85 & 80.0 & 102.1 & 102.8 & 100.5 & 100.5 \\
A3526 & 0406200101 & 0.01140 & 8.25 & 84.3 & 114.4 & 114.4 & 116.3 & 116.3 \\
2A0335 & 0147800201 & 0.03490 & 24.00 & 70.6 & 79.9 & 80.6 & 106.1 & 106.1 \\ 
A2052 & 0109920101 & 0.03549 & 2.90 & 23.0 & 30.2 & 30.2 & 25.8 & 25.8 \\
A1060 & 0206230101 & 0.01260 & 4.92 & 28.9 & 42.1 & 42.1 & 42.8 & 42.8 \\
A1736 & 0505210201 & 0.04580 & 5.36 & 2.5 & 10.5 & 10.5 & 12.0 & 15.0 \\
A4038 & 0204460101 & 0.03000 & 1.55 & 25.0 & 29.4 & 29.4 & 29.7 & 29.7 \\
MKW3S & 0109930101 & 0.04500 & 3.15 & 31.0 & 32.0 & 32.0 & 31.4 & 31.4 \\
A2634 & 0002960101 & 0.03139 & 5.17 & 6.0 & 9.9 & 9.9 &\ldots&\ldots\\
EXO0422 & 0300210401 & 0.03970 & 6.40 & 31.0 & 40.6 & 40.6 & 39.2 & 39.3 \\
HYDA & 0504260101 & 0.05390 & 4.86 & 51.9 & 82.4 & 82.4 & 95.6 & 95.6 \\
A2063 & 0550360101 & 0.03494 & 2.92 & 12.7 & 23.8 & 23.8 & 26.5 & 26.7 \\
A2597 & 0147330101 & 0.08520 & 2.50 & 48.3 & 58.1 & 58.2 & 77.1 & 77.1 \\
A2589 & 0204180101 & 0.04140 & 4.39 & 21.8 & 32.8 & 32.9 & 32.0 & 32.0 \\
A2147 & 0505210601 & 0.03500 & 3.29 & 5.5 & 10.7 & 10.7 & 11.6 & 11.6 \\
A496 & 0506260301 & 0.03290 & 5.68 & 39.5 & 59.0 & 59.1 & 61.4 & 61.4 \\
A3376 & 0151900101 & 0.04560 & 5.01 & 21.0 & 21.0 & 21.0 & 27.2 & 29.4 \\
A576 & 0504320101 & 0.03890 & 5.69 & 17.4 & 33.8 & 33.8 & 16.8 & 32.1 \\
A4059 & 0109950101 & 0.04750 & 1.10 & 12.6 & 12.6 & 12.6 & 15.3 & 15.3 \\
A2199 & 0008030201 & 0.03015 & 0.84 & 12.1 & 13.9 & 13.9 & 13.8 & 14.2 \\
A1644 & 0010420201 & 0.04730 & 5.33 & 11.1 & 13.9 & 14.6 &\ldots&\ldots\\
A0133 & 0144310101 & 0.05660 & 1.60 & 16.0 & 22.1 & 22.1 & 22.1 & 24.1 \\
A3395 & 0400010301 & 0.05060 & 8.49 & 22.3 & 29.2 & 29.2 & 29.8 & 0.0 \\
A3562 & 0105261301 & 0.04900 & 3.91 & 34.6 & 38.8 & 39.2 &\ldots&\ldots\\
A3112 & 0603050101 & 0.07525 & 2.53 & 63.2 & 87.0 & 87.1 & 99.9 & 100.2 \\
A3558 & 0107260101 & 0.04800 & 3.63 & 35.2 & 43.6 & 43.6 & 37.5 & 40.0 \\
A0119 & 0505211001 & 0.04420 & 3.10 & 6.8 & 8.8 & 11.0 & 12.2 & 12.2 \\
A1795 & 0097820101 & 0.06247 & 1.20 & 30.4 & 35.9 & 36.0 & 36.5 & 36.5 \\
A2657 & 0402190301 & 0.04020 & 5.27 & 2.0 & 22.8 & 23.7 & 24.3 & 24.3 \\
A0085 & 0065140101 & 0.05506 & 3.58 & 9.0 & 12.4 & 12.4 &\ldots&\ldots\\
A3158 & 0300210201 & 0.05970 & 1.06 & 10.4 & 21.3 & 21.3 & 19.9 & 19.9 \\
A3391 & 0505210401 & 0.05140 & 5.42 & 16.0 & 26.0 & 26.0 &\ldots&\ldots\\
A1650 & 0093200101 & 0.08384 & 1.54 & 30.6 & 38.1 & 38.1 & 34.6 & 35.6 \\
A2065 & 0202080201 & 0.07260 & 2.84 & 14.6 & 20.6 & 20.6 & 20.7 & 20.7 \\
A399 & 0112260101 & 0.07181 & 10.58 & 5.6 & 13.1 & 13.3 & 9.0 & 11.0 \\
A3667 & 0206850101 & 0.05560 & 4.59 & 47.8 & 56.9 & 56.9 &\ldots&\ldots\\
A3571 & 0086950201 & 0.03910 & 3.93 & 12.2 & 24.1 & 24.1 & 24.8 & 24.8 \\
A1651 & 0203020101 & 0.08495 & 1.71 & 7.0 & 10.7 & 11.0 & 12.9 & 12.9 \\
A2256 & 0141380201 & 0.05810 & 4.02 & 10.8 & 12.5 & 12.5 & 11.4 & 11.8 \\
ZwCl1215 & 0300211401 & 0.07500 & 1.64 & 16.9 & 26.7 & 26.7 & 26.4 & 26.4 \\
A401 & 0112260301 & 0.07366 & 10.19 & 8.1 & 13.0 & 13.0 & 12.9 & 12.9 \\
A2255 & 0112260801 & 0.08060 & 2.51 & 4.5 & 12.2 & 12.8 & 11.5 & 11.5 \\
A754 & 0136740101 & 0.05420 & 4.59 & 11.5 & 13.8 & 14.2 & 15.6 & 15.6 \\
A2204 & 0306490201 & 0.15216 & 5.94 & 13.6 & 15.3 & 15.3 & 17.3 & 17.3 \\
A2029 & 0551780401 & 0.07728 & 3.07 & 17.9 & 24.5 & 24.5 & 44.0 & 44.0 \\
A2142 & 0111870301 & 0.09090 & 4.05 & 4.5 & 17.9 & 17.9 &\ldots&\ldots\\
COMA & 0124711401 & 0.02310 & 0.89 & 14.2 & 17.4 & 17.4 & 19.1 & 19.1 \\
A3266 & 0105260901 & 0.05890 & 1.48 & 15.0 & 23.6 & 23.8 &\ldots&\ldots\\
A478 & 0109880101 & 0.08810 & 29.00 & 59.0 & 59.0 & 65.4 & 95.0 & 103.6 \\ 
A2163 & 0112230601 & 0.20300 & 12.27 & 5.1 & 10.3 & 10.5 & 7.8 & 7.9
\enddata
\tablecomments{\footnotesize{(1) Cluster Name; (2) XMM-Newton Observation ID; (3) Optical redshift, from the NASA/IPACExtragalactic Database (NED); (4) Galactic hydrogen column density, in $10^{20}$cm$^{-2}$, from \citet{Dickey1990}, with the exception of A478, NGC1550, and 2A0334, for which the value comes from \citet{Hudson2010}; (5)-(9) Net detector exposure times, in ks, after flare screening.}}
\end{deluxetable*}

\begin{deluxetable*}{llc}[p]
\tabletypesize{\footnotesize}
\tablecolumns{3}
\tablecaption{Model Parameters \label{table:modelparams}}
\tablewidth{0pt}
\tablehead{
Parameter & Value & Global}
\startdata
\multicolumn{3}{c}{ICM}\\
\hline
$n_{H}$\tablenotemark{a} & $(0.8n_{H,gal} , 1.2n_{H,gal})$ & yes \\
$log~kT$ (keV), $kT_{avg}<4$ keV & $(-1.0, 1.0)$ & no \\
$log~kT$ (keV), $kT_{avg}\ge4$ keV & $(-1.0, 1.3)$ & no \\
$Z$ $(Z_\odot)$\tablenotemark{b} & $(0.0, 2.0)$ & yes \\
$z$\tablenotemark{c} & $z_{NED}$ & yes \\
$ln~\sigma$ (arcsec) & $(0.5 ,5.5)$ & no \\
$x$ (arcmin)\tablenotemark{d} & $(-10.0,10.0)$  & no \\
$x$ (arcmin) EPIC, large clusters\tablenotemark{d},\tablenotemark{e} & $(-15.0,15.0)$  & no \\
$y$ (arcmin) RGS\tablenotemark{d}  & $(-2.3, 2.3)$  & no \\
$y$ (arcmin) EPIC\tablenotemark{d} & $(-10.0, 10.0)$ & no \\
$y$ (arcmin) EPIC, large clusters\tablenotemark{d},\tablenotemark{e} & $(-15.0, 15.0)$ & no \\
\cutinhead{Galactic X-ray Background}
$kT$ (keV) & $0.16$ & yes \\
 $Z$ $(Z_\odot)$\tablenotemark{b} & $1.0$ & yes \\
\cutinhead{Cosmic X-ray Background}
$\Gamma$ & $1.47$ & yes \\
$n_{H}$\tablenotemark{a} & $n_{H,gal}$ & yes \\
\cutinhead{EPIC Instrumental Background\tablenotemark{f}}
Particle $\Gamma$ & $-0.205$ & yes \\
$E_{i}$ & $158$ eV & yes \\
\cutinhead{RGS Instrumental Background\tablenotemark{g}}
Particle $\Gamma$ & $-0.35$ & yes \\
$E_{i}$ & $500$ eV & yes
\enddata
\tablecomments{\footnotesize Summary of model parameters and their allowed ranges (all priors are uniform).  Values in parentheses represent the minimum and maximum values of the free parameters.  All other parameters are fixed at the given values.  Global parameters are the same for all smoothed particles in a given iteration. }
\tablenotetext{a}{\footnotesize{The absorbing column densities, $n_{H,gal}$ for each cluster, listed in Table \ref{table:observations}.}}
\tablenotetext{b}{\footnotesize{Metallicities are with respect to solar metallicities of \citet{Anders1989}. }}
\tablenotetext{c}{\footnotesize{The redshift is fixed for each cluster at the optical NED value.}}
\tablenotetext{d}{\footnotesize{The $x$ and $y$ coordinates are with respect to the cluster center.}}
\tablenotetext{e}{\footnotesize{For clusters which are large compared to the EPIC FOV (e.g. Coma), the $x$ and $y$ prior ranges were extended to 15' from the center to coincide with the spatial cuts to the data (see \S \ref{section:spi}).}}
\tablenotetext{f}{\footnotesize{All EPIC instrumental background parameters are from \citet{Andersson2007}.}}
\tablenotetext{g}{\footnotesize{RGS instrumental background parameters are from \citet{Peterson2003a}.}}   

\end{deluxetable*}

\LongTables
\begin{deluxetable}{ccccccccc}[p]
\tabletypesize{\footnotesize}
\tablecolumns{8}
\tablecaption{Cluster Properties \label{table:clusterparameters}}
\tablewidth{0pt}
\tablehead{\colhead{Cluster} & \colhead{$kT_{med}$ (keV)} & \colhead{$\sigma_{kT}$ (keV)} & \colhead{EM ($10^{65}$cm$^{-3}$)} & \colhead{$r_{2500}$ (kpc)} &\colhead{AGN} & \colhead{Relaxed or Disturbed} & \colhead{Cool-Core Status} }
\startdata
NGC4636 & 0.72$\pm$0.06 & 1.71$\pm$0.27 & 0.33$\pm$0.05 & 163.9$\pm$21.4 & Y & Relaxed\tablenotemark{1}$^,$\tablenotemark{2} & SCC \\
NGC5044 & 1.22$\pm$0.02 & 0.94$\pm$0.24 & 4.90$\pm$0.41 & 219.7$\pm$9.5 & Y & Relaxed\tablenotemark{3}$^,$\tablenotemark{2} & SCC \\
NGC1550 & 1.42$\pm$0.01 & 0.82$\pm$0.20 & 12.44$\pm$1.20 & 239.6$\pm$8.8 & Y & Relaxed\tablenotemark{4} & SCC \\
FORNAX & 1.48$\pm$0.03 & 1.40$\pm$0.28 & 0.90$\pm$0.10 & 246.1$\pm$12.4 & Y & Relaxed\tablenotemark{5} & SCC \\
NGC507 & 1.66$\pm$0.02 & 1.48$\pm$0.29 & 7.77$\pm$0.59 & 262.9$\pm$10.1 & Y & Disturbed\tablenotemark{6}$^,$\tablenotemark{2} & SCC \\
A3581 & 2.00$\pm$0.01 & 1.06$\pm$0.15 & 34.31$\pm$2.10 & 289.3$\pm$8.4 & Y & Relaxed\tablenotemark{7}$^,$\tablenotemark{2} & SCC \\
MKW4 & 2.10$\pm$0.02 & 0.99$\pm$0.32 & 12.51$\pm$1.06 & 297.6$\pm$10.6 & Y & Relaxed\tablenotemark{8} & SCC \\
A262 & 2.36$\pm$0.01 & 1.01$\pm$0.16 & 21.24$\pm$1.21 & 318.9$\pm$8.7 & Y & Relaxed\tablenotemark{9}$^,$\tablenotemark{2} & SCC \\
A400 & 2.48$\pm$0.02 & 1.35$\pm$0.19 & 23.52$\pm$1.69 & 326.6$\pm$12.1 & Y & Disturbed\tablenotemark{10} & NCC \\
A1367 & 2.56$\pm$0.02 & 0.96$\pm$0.21 & 22.22$\pm$1.16 & 336.1$\pm$11.8 & N & Disturbed\tablenotemark{11} & NCC \\
Zw54 & 2.56$\pm$0.02 & 1.30$\pm$0.16 & 43.64$\pm$1.98 & 331.9$\pm$9.8 & Y & Relaxed\tablenotemark{12} & WCC \\
MKW8 & 2.74$\pm$0.03 & 1.46$\pm$0.18 & 27.07$\pm$2.15 & 345.3$\pm$14.1 & Y & Disturbed\tablenotemark{2} & NCC \\
2A0335 & 2.76$\pm$0.01 & 0.89$\pm$0.16 & 662.01$\pm$25.88 & 345.5$\pm$8.5 & Y & Relaxed\tablenotemark{13} & SCC \\
AS1101 & 2.77$\pm$0.02 & 0.83$\pm$0.22 & 581.18$\pm$29.58 & 342.4$\pm$10.5 & Y & Relaxed\tablenotemark{14}$^,$\tablenotemark{2} & SCC \\
A3526 & 2.88$\pm$0.01 & 0.78$\pm$0.15 & 19.10$\pm$0.93 & 357.4$\pm$8.7 & Y & Disturbed\tablenotemark{15}$^,$\tablenotemark{2} & SCC \\
A1060 & 2.98$\pm$0.04 & 1.00$\pm$0.22 & 17.04$\pm$1.35 & 364.1$\pm$16.2 & N & Relaxed\tablenotemark{16} & WCC \\
A2052 & 2.98$\pm$0.03 & 0.86$\pm$0.23 & 115.64$\pm$8.85 & 360.7$\pm$15.3 & Y & Relaxed\tablenotemark{17}$^,$\tablenotemark{2} & SCC \\
A1736 & 3.06$\pm$0.05 & 1.80$\pm$0.23 & 94.30$\pm$12.81 & 364.1$\pm$22.5 & N & Disturbed\tablenotemark{2} & NCC \\
A4038 & 3.19$\pm$0.02 & 0.91$\pm$0.14 & 81.44$\pm$2.71 & 375.6$\pm$8.3 & Y & Relaxed\tablenotemark{18} & WCC \\
MKW3S & 3.27$\pm$0.05 & 1.17$\pm$0.16 & 172.40$\pm$15.81 & 378.6$\pm$13.0 & Y & Relaxed\tablenotemark{19}$^,$\tablenotemark{2} & SCC \\
HYDA & 3.49$\pm$0.02 & 1.85$\pm$0.35 & 311.31$\pm$14.54 & 390.7$\pm$12.9 & Y & Relaxed\tablenotemark{20}$^,$\tablenotemark{2} & SCC \\
A2634 & 3.49$\pm$0.06 & 1.56$\pm$0.27 & 26.15$\pm$3.16 & 400.6$\pm$29.7 & Y & Disturbed\tablenotemark{12} & WCC \\
EXO0422 & 3.55$\pm$0.02 & 0.98$\pm$0.14 & 94.88$\pm$4.20 & 397.6$\pm$9.3 & Y & Relaxed\tablenotemark{21} & SCC \\
A2147 & 3.62$\pm$0.04 & 2.95$\pm$0.41 & 72.29$\pm$7.80 & 402.8$\pm$20.2 & Y & Disturbed\tablenotemark{12} & NCC \\
A2597 & 3.66$\pm$0.03 & 1.14$\pm$0.44 & 370.47$\pm$20.52 & 395.4$\pm$16.4 & Y & Relaxed\tablenotemark{22}$^,$\tablenotemark{2} & SCC \\
A2063 & 3.70$\pm$0.02 & 1.23$\pm$0.15 & 95.81$\pm$3.83 & 407.6$\pm$10.6 & Y & Relaxed\tablenotemark{23}$^,$\tablenotemark{2} & WCC \\
A3376 & 3.78$\pm$0.03 & 2.55$\pm$0.35 & 74.49$\pm$4.49 & 410.7$\pm$16.2 & Y & Disturbed\tablenotemark{24}$^,$\tablenotemark{25} & NCC \\
A2589 & 3.80$\pm$0.01 & 0.97$\pm$0.14 & 94.39$\pm$3.53 & 413.0$\pm$9.2 & N & Relaxed\tablenotemark{26} & WCC \\
A496 & 3.81$\pm$0.02 & 1.96$\pm$0.35 & 140.83$\pm$6.31 & 414.9$\pm$13.2 & Y & Relaxed\tablenotemark{27} & SCC \\
A576 & 3.86$\pm$0.02 & 2.09$\pm$0.33 & 82.62$\pm$4.21 & 416.8$\pm$10.8 & Y & Disturbed\tablenotemark{28} & WCC \\
A4059 & 3.92$\pm$0.02 & 1.87$\pm$0.31 & 143.36$\pm$7.53 & 419.1$\pm$10.7 & Y & Relaxed\tablenotemark{9}$^,$\tablenotemark{2} & SCC \\
A1644 & 3.95$\pm$0.03 & 2.09$\pm$0.37 & 104.18$\pm$5.53 & 430.1$\pm$16.4 & Y & Disturbed\tablenotemark{29} & SCC \\
A2199 & 4.04$\pm$0.02 & 1.55$\pm$0.28 & 144.02$\pm$5.47 & 429.4$\pm$8.4 & Y & Relaxed\tablenotemark{2}$^,$\tablenotemark{30} & SCC \\
A3395 & 4.06$\pm$0.04 & 2.46$\pm$0.34 & 88.90$\pm$11.03 & 426.9$\pm$15.9 & Y & Disturbed\tablenotemark{31} & NCC \\
A0133 & 4.22$\pm$0.02 & 1.10$\pm$0.15 & 129.01$\pm$4.91 & 434.6$\pm$11.2 & Y & Relaxed\tablenotemark{2} & SCC \\
A3562 & 4.35$\pm$0.02 & 2.12$\pm$0.32 & 117.42$\pm$4.20 & 454.2$\pm$12.4 & Y & Disturbed\tablenotemark{32}$^,$\tablenotemark{33} & WCC \\
A3112 & 4.64$\pm$0.03 & 1.77$\pm$0.37 & 345.89$\pm$15.73 & 454.8$\pm$15.8 & Y & Relaxed\tablenotemark{34} & SCC \\
A0119 & 4.67$\pm$0.05 & 3.03$\pm$0.40 & 89.93$\pm$9.04 & 463.1$\pm$26.6 & N & Disturbed\tablenotemark{35} & NCC \\
A3558 & 4.70$\pm$0.06 & 2.19$\pm$0.45 & 201.56$\pm$18.60 & 463.9$\pm$26.7 & Y & Disturbed\tablenotemark{36} & WCC \\
A2657 & 4.74$\pm$0.04 & 3.27$\pm$0.36 & 93.74$\pm$6.00 & 468.4$\pm$23.5 & N & Relaxed\tablenotemark{2} & WCC \\
A478 & 5.02$\pm$0.03 & 2.48$\pm$0.32 & 1380.62$\pm$93.77 & 472.4$\pm$17.4 & Y & Relaxed\tablenotemark{4} & SCC \\
A1795 & 5.06$\pm$0.03 & 2.08$\pm$0.37 & 520.42$\pm$24.13 & 480.5$\pm$18.3 & Y & Relaxed\tablenotemark{14}$^,$\tablenotemark{37} & SCC \\
A399 & 5.07$\pm$0.05 & 3.27$\pm$0.35 & 275.46$\pm$34.27 & 478.8$\pm$26.5 & N & Disturbed\tablenotemark{38}$^,$\tablenotemark{39} & NCC \\
A0085 & 5.20$\pm$0.02 & 1.86$\pm$0.31 & 331.70$\pm$10.77 & 502.3$\pm$13.8 & Y & Disturbed\tablenotemark{40} & SCC \\
A3391 & 5.21$\pm$0.03 & 2.36$\pm$0.33 & 106.25$\pm$6.80 & 503.1$\pm$19.3 & Y & Disturbed\tablenotemark{21} & NCC \\
A3158 & 5.38$\pm$0.03 & 1.76$\pm$0.31 & 195.77$\pm$9.17 & 498.5$\pm$14.2 & Y & Relaxed\tablenotemark{41} & NCC \\
A2065 & 5.45$\pm$0.03 & 1.93$\pm$0.30 & 241.96$\pm$10.28 & 498.9$\pm$15.1 & Y & Disturbed\tablenotemark{42} & WCC \\
A1650 & 5.49$\pm$0.05 & 1.90$\pm$0.31 & 363.58$\pm$31.82 & 497.9$\pm$15.3 & N & Relaxed\tablenotemark{43} & WCC \\
A3667 & 5.61$\pm$0.04 & 1.96$\pm$0.42 & 240.34$\pm$14.10 & 524.6$\pm$27.5 & N & Disturbed\tablenotemark{44}$^,$\tablenotemark{45} & WCC \\
A3571 & 6.30$\pm$0.05 & 1.55$\pm$0.41 & 221.98$\pm$14.34 & 550.0$\pm$30.6 & Y & Relaxed\tablenotemark{46} & WCC \\
A2256 & 6.33$\pm$0.03 & 2.19$\pm$0.31 & 274.92$\pm$13.74 & 546.7$\pm$19.9 & N & Disturbed\tablenotemark{47}$^,$\tablenotemark{48}$^,$\tablenotemark{49} & NCC \\
A1651 & 6.38$\pm$0.04 & 1.62$\pm$0.35 & 368.09$\pm$19.46 & 542.0$\pm$20.6 & Y & Relaxed\tablenotemark{50} & WCC \\
ZwCl1215 & 6.39$\pm$0.03 & 2.11$\pm$0.30 & 257.40$\pm$12.37 & 545.5$\pm$17.3 & N & Relaxed\tablenotemark{2} & NCC \\
A401 & 6.55$\pm$0.06 & 2.55$\pm$0.31 & 470.33$\pm$47.68 & 553.3$\pm$20.5 & N & Disturbed\tablenotemark{51}$^,$\tablenotemark{39} & NCC \\
A2255 & 6.58$\pm$0.05 & 2.32$\pm$0.39 & 236.52$\pm$17.47 & 552.8$\pm$33.0 & N & Disturbed\tablenotemark{52}$^,$\tablenotemark{53} & NCC \\
A2204 & 6.98$\pm$0.03 & 2.28$\pm$0.30 & 1524.08$\pm$64.87 & 551.7$\pm$16.9 & Y & Relaxed\tablenotemark{26} & SCC \\
A754 & 7.00$\pm$0.03 & 1.93$\pm$0.30 & 371.03$\pm$13.61 & 580.2$\pm$22.3 & N & Disturbed\tablenotemark{54}$^,$\tablenotemark{55} & NCC \\
COMA & 7.31$\pm$0.06 & 2.34$\pm$0.45 & 67.74$\pm$5.60 & 603.2$\pm$41.2 & Y & Disturbed\tablenotemark{56}$^,$\tablenotemark{2} & NCC \\
A2029 & 7.55$\pm$0.04 & 0.76$\pm$0.39 & 701.87$\pm$34.93 & 598.6$\pm$29.0 & Y & Relaxed\tablenotemark{57}$^,$\tablenotemark{58} & SCC \\
A2142 & 7.57$\pm$0.06 & 0.55$\pm$0.53 & 494.12$\pm$54.14 & 621.9$\pm$51.0 & Y & Relaxed\tablenotemark{59} & WCC \\
A3266 & 8.02$\pm$0.03 & 1.34$\pm$0.33 & 256.00$\pm$11.02 & 642.4$\pm$23.8 & N & Disturbed\tablenotemark{60} & WCC \\
A2163 & 13.77$\pm$0.04 & 2.79$\pm$0.45 & 1378.54$\pm$108.04 & 789.3$\pm$30.1 & N & Disturbed\tablenotemark{61}$^,$\tablenotemark{62} & NCC \\
\enddata
\tablecomments{\footnotesize{Measured cluster parameters and classifications for our cluster sample.  Quoted uncertainties are $1\sigma$ errors. AGN classification is from \citet{Mittal2009}. Cool-core status is from \citet{Hudson2010}. }}
\tablerefs{\tablenotemark{1}{\footnotesize{\citet{Ohto2003}}} \tablenotemark{2}{\footnotesize{\citet{Zhang2011}}} \tablenotemark{3}{\footnotesize{\citet{David2011}}} \tablenotemark{4}{\footnotesize{\citet{Sun2003}}} \tablenotemark{5}{\footnotesize{\citet{Shurkin2008}}} \tablenotemark{6}{\footnotesize{\citet{Paolillo2003}}} \tablenotemark{7}{\footnotesize{\citet{Dong2010}}} \tablenotemark{8}{\footnotesize{\citet{Vikhlinin2005}}} \tablenotemark{9}{\footnotesize{\citet{Birzan2004}}} \tablenotemark{10}{\footnotesize{\citet{Hudson2006}}} \tablenotemark{11}{\footnotesize{\citet{Sun2002}}} \tablenotemark{12}{\footnotesize{\citet{Hudson2010}}} \tablenotemark{13}{\footnotesize{\citet{Sanders2009}}} \tablenotemark{14}{\footnotesize{\citet{Diehl2008}}} \tablenotemark{15}{\footnotesize{\citet{Fabian2005}}} \tablenotemark{16}{\footnotesize{\citet{Hayakawa2006}}} \tablenotemark{17}{\footnotesize{\citet{Blanton2011}}} \tablenotemark{18}{\footnotesize{\citet{Vikhlinin2009a}}} \tablenotemark{19}{\footnotesize{\citet{Mazzotta2002}}} \tablenotemark{20}{\footnotesize{\citet{McNamara2000}}} \tablenotemark{21}{\footnotesize{\citet{Zhang2009}}} \tablenotemark{22}{\footnotesize{\citet{McNamara2001}}} \tablenotemark{23}{\footnotesize{\citet{Kanov2006}}} \tablenotemark{24}{\footnotesize{\citet{Bagchi2006}}} \tablenotemark{25}{\footnotesize{\citet{Araudo2008}}} \tablenotemark{26}{\footnotesize{\citet{Buote1996}}} \tablenotemark{27}{\footnotesize{\citet{Durret2000}}} \tablenotemark{28}{\footnotesize{\citet{Dupke2007}}} \tablenotemark{29}{\footnotesize{\citet{Reiprich2004}}} \tablenotemark{30}{\footnotesize{\citet{Johnstone2002}}} \tablenotemark{31}{\footnotesize{\citet{Donnelly2001}}} \tablenotemark{32}{\footnotesize{\citet{Giacintucci2005}}} \tablenotemark{33}{\footnotesize{\citet{Finoguenov2004}}} \tablenotemark{34}{\footnotesize{\citet{Takizawa2003}}} \tablenotemark{35}{\footnotesize{\citet{Rossetti2010}}} \tablenotemark{36}{\footnotesize{\citet{Rossetti2007}}} \tablenotemark{37}{\footnotesize{\citet{Tamura2001}}} \tablenotemark{38}{\footnotesize{\citet{Murgia2010}}} \tablenotemark{39}{\footnotesize{\citet{Sakelliou2004}}} \tablenotemark{40}{\footnotesize{\citet{Kempner2002}}} \tablenotemark{41}{\footnotesize{\citet{JohnstonHollitt2008}}} \tablenotemark{42}{\footnotesize{\citet{Belsole2005}}} \tablenotemark{43}{\footnotesize{\citet{Takahashi2003}}} \tablenotemark{44}{\footnotesize{\citet{Roettgering1997}}} \tablenotemark{45}{\footnotesize{\citet{Markevitch1999}}} \tablenotemark{46}{\footnotesize{\citet{Nevalainen2001}}} \tablenotemark{47}{\footnotesize{\citet{Roettgering1994}}} \tablenotemark{48}{\footnotesize{\citet{Clarke2006a}}} \tablenotemark{49}{\footnotesize{\citet{Sun2002a}}} \tablenotemark{50}{\footnotesize{\citet{Schuecker2001}}} \tablenotemark{51}{\footnotesize{\citet{Bacchi2003}}} \tablenotemark{52}{\footnotesize{\citet{Govoni2005}}} \tablenotemark{53}{\footnotesize{\citet{Sakelliou2006}}} \tablenotemark{54}{\footnotesize{\citet{Kassim2001}}} \tablenotemark{55}{\footnotesize{\citet{Henry2004}}} \tablenotemark{56}{\footnotesize{\citet{Giovannini1993}}} \tablenotemark{57}{\footnotesize{\citet{Govoni2009}}} \tablenotemark{58}{\footnotesize{\citet{Bourdin2008}}} \tablenotemark{59}{\footnotesize{\citet{Owers2009}}} \tablenotemark{60}{\footnotesize{\citet{Henriksen2000}}} \tablenotemark{61}{\footnotesize{\citet{Feretti2001}}} \tablenotemark{62}{\footnotesize{\citet{Markevitch2001}}}}
\end{deluxetable}

\begin{figure*}[p]
\begin{center}
\includegraphics[width=\textwidth]{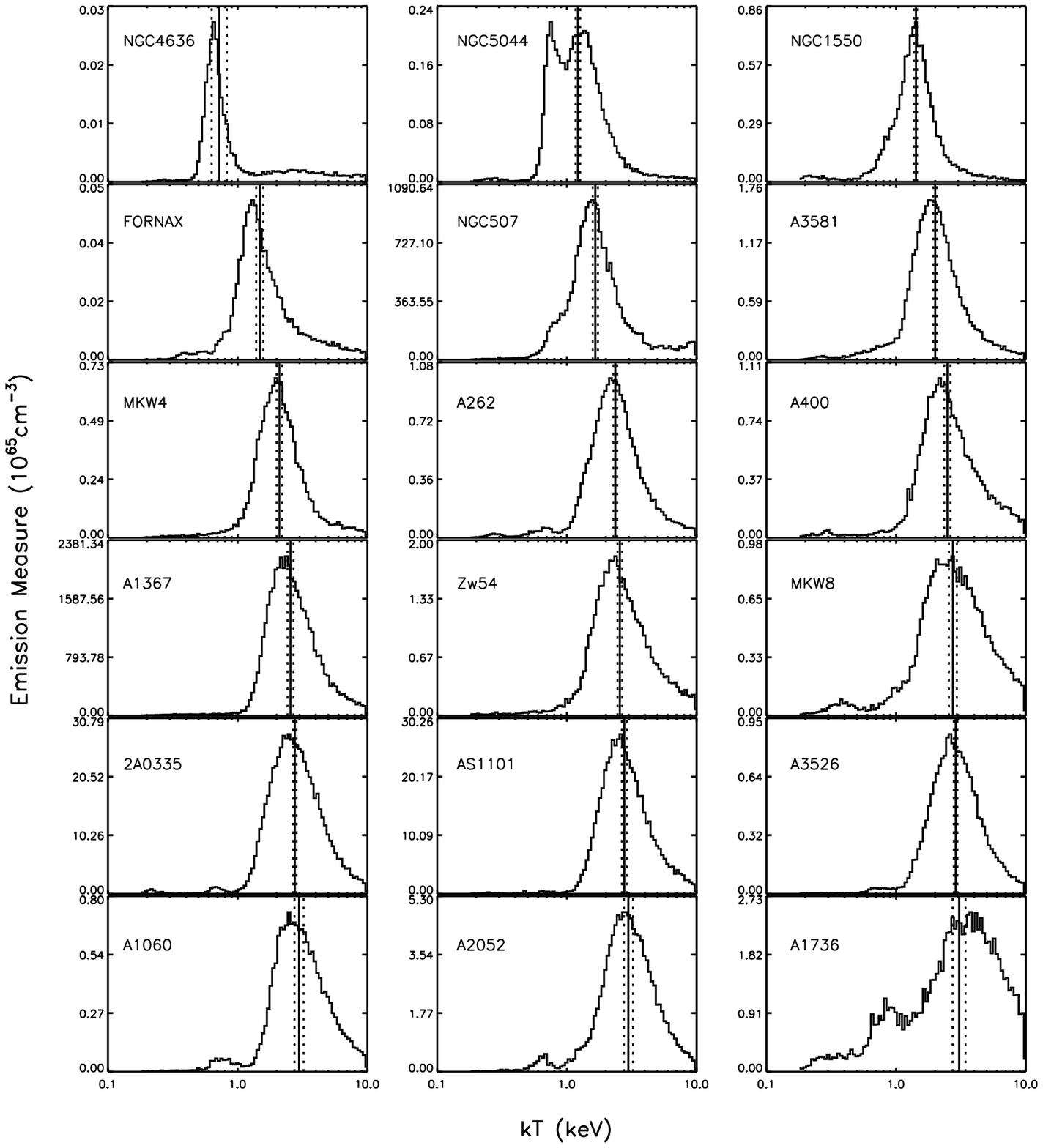}
\caption{\footnotesize Temperature distributions of clusters with $0-10$ keV prior for the smoothed particle temperatures.  The vertical line marks $kT_{med}$ and the dashed lines represent the $1\sigma$ confidence interval of $kT_{med}$.  Clusters are arranged in increasing order of $kT_{med}$.}
\label{figure:tempdists0}
\end{center}
\end{figure*}

\begin{figure*}[p]
\begin{center}
\includegraphics[width=\textwidth]{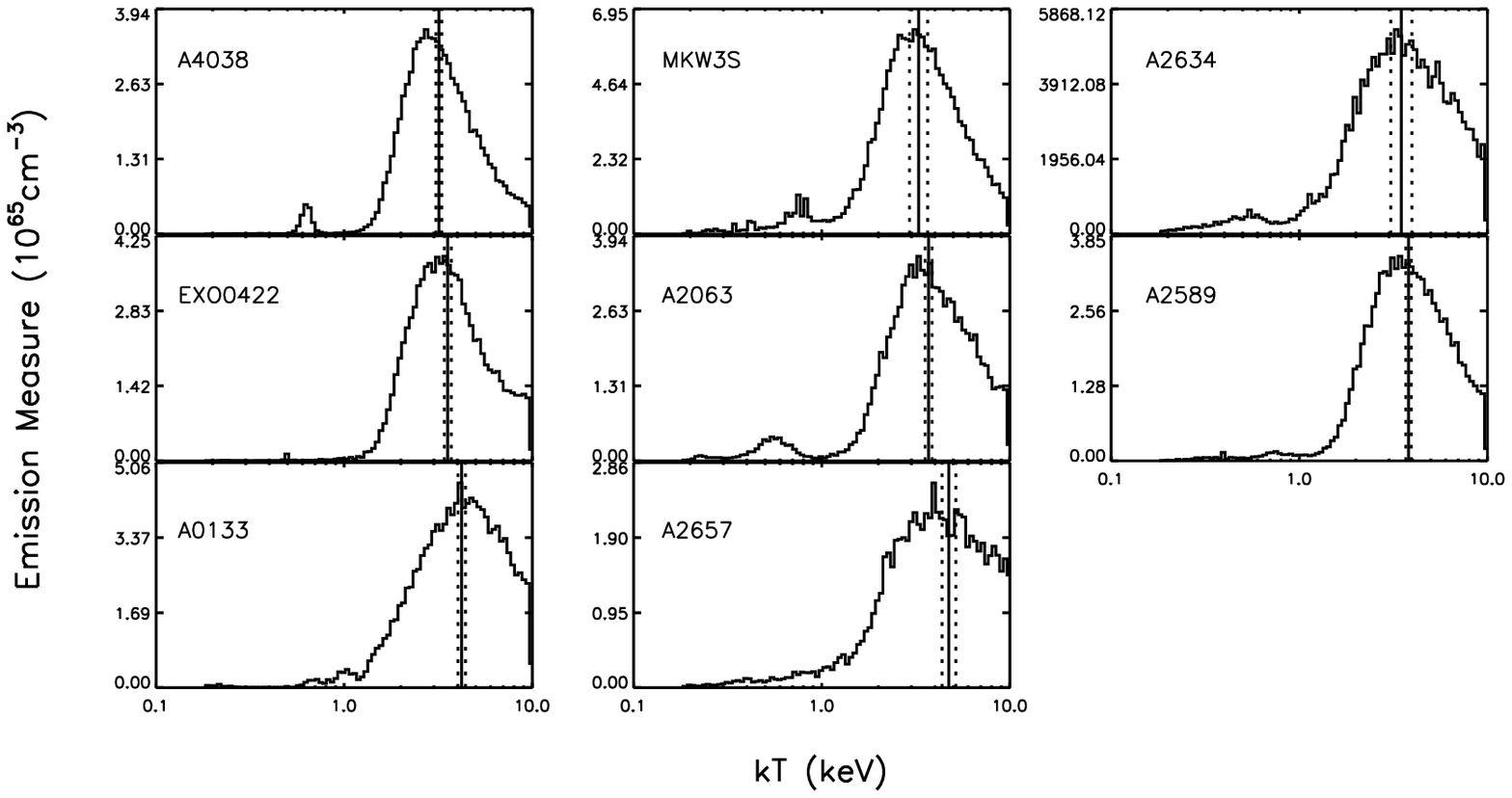}
\caption{\footnotesize Continuation of  Figure \ref{figure:tempdists0}.}
\label{figure:tempdists1}
\end{center}
\end{figure*}

\begin{figure*}[p]
\begin{center}
\includegraphics[width=\textwidth]{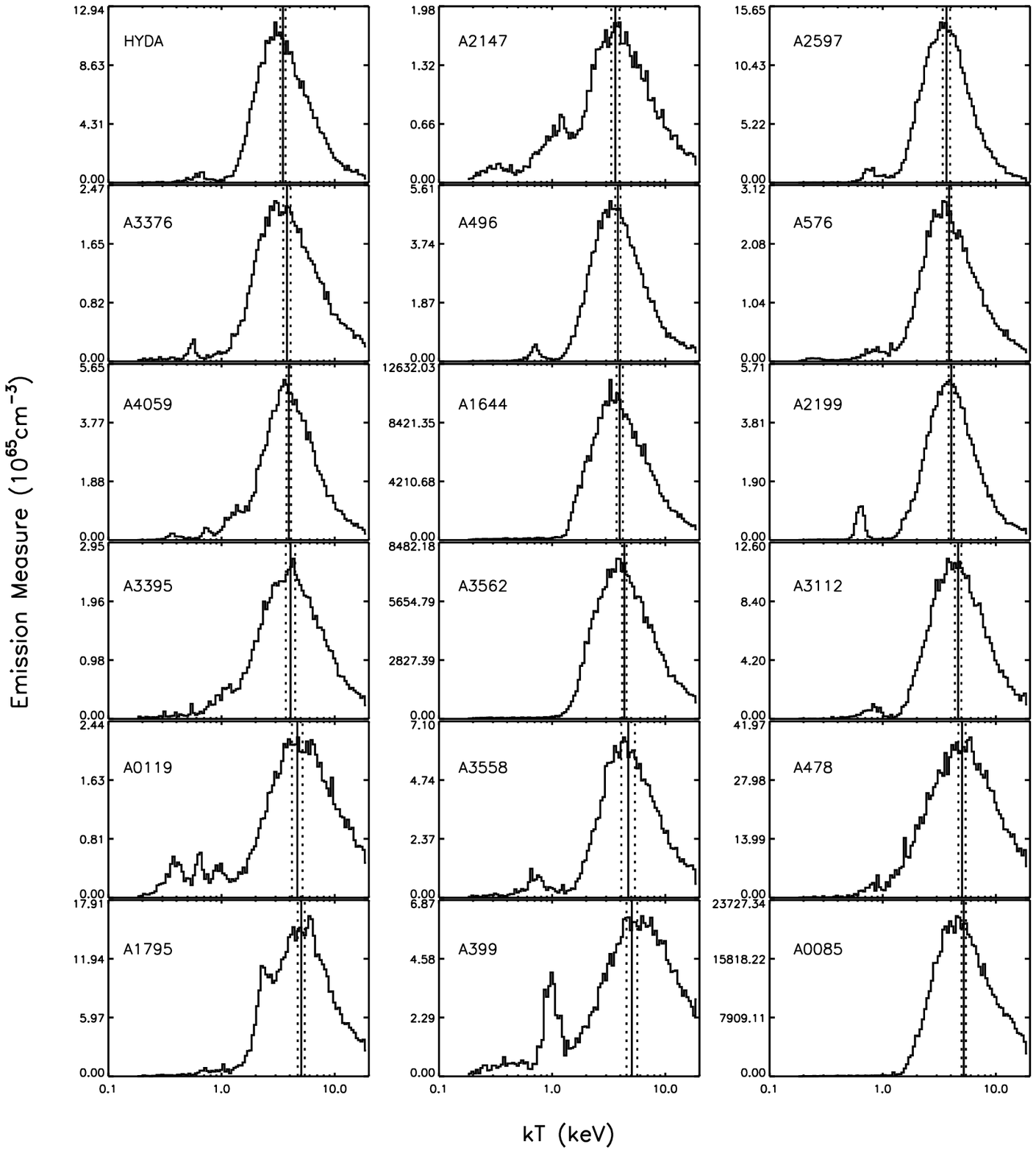}
\caption{\footnotesize Temperature distributions of clusters with $0-20$ keV prior for the smoothed particle temperatures.  The vertical line marks $kT_{med}$ and the dashed lines represent the $1\sigma$ confidence interval of $kT_{med}$.  Clusters are arranged in increasing order of $kT_{med}$. }
\label{figure:tempdists2}
\end{center}
\end{figure*}

\begin{figure*}[p]
\begin{center}
\includegraphics[width=\textwidth]{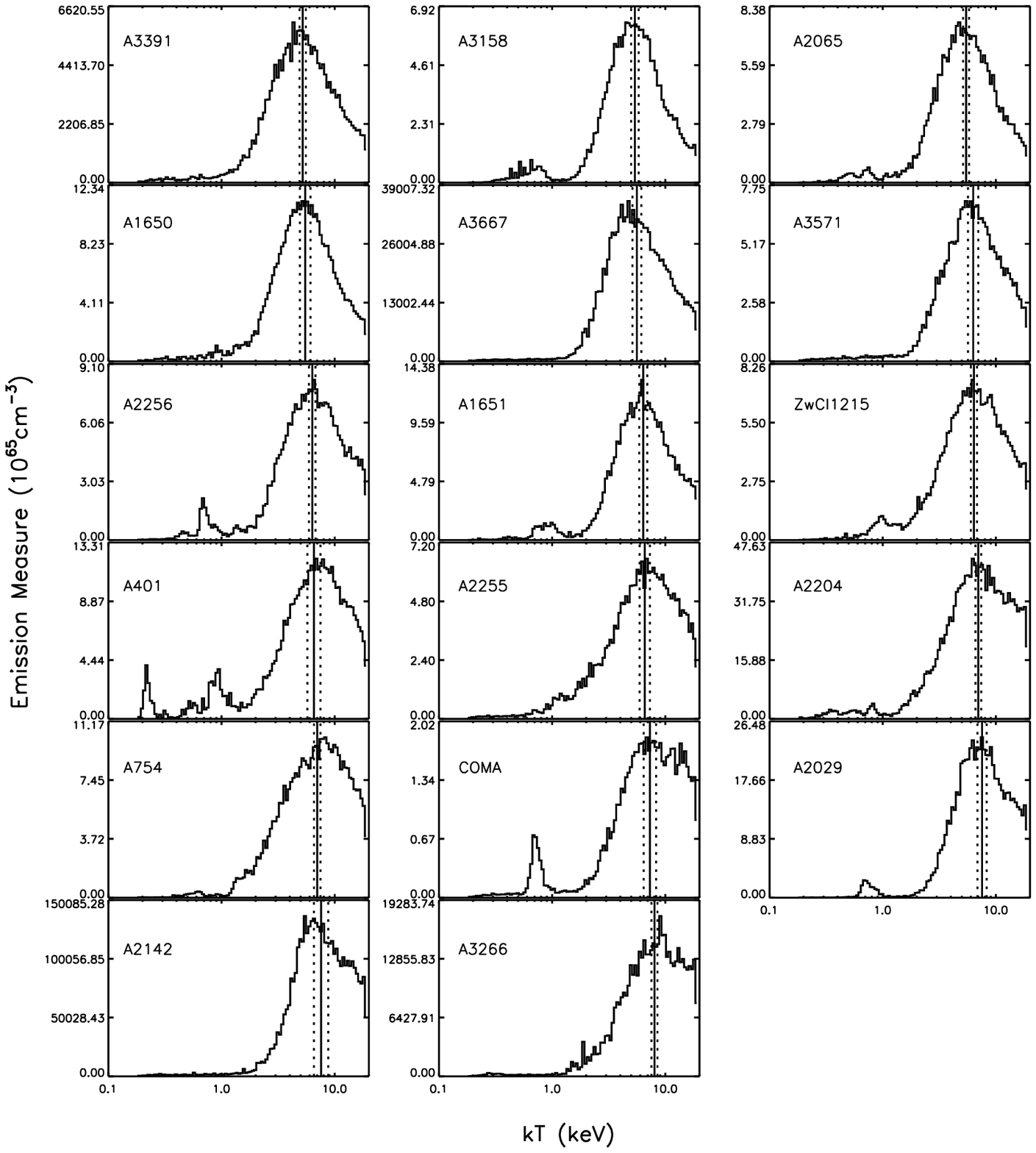}
\caption{\footnotesize Continuation of Figure \ref{figure:tempdists2}.}
\label{figure:tempdists3}
\end{center}
\end{figure*}

\begin{figure*}[p]
\begin{center}
\includegraphics[width=\textwidth]{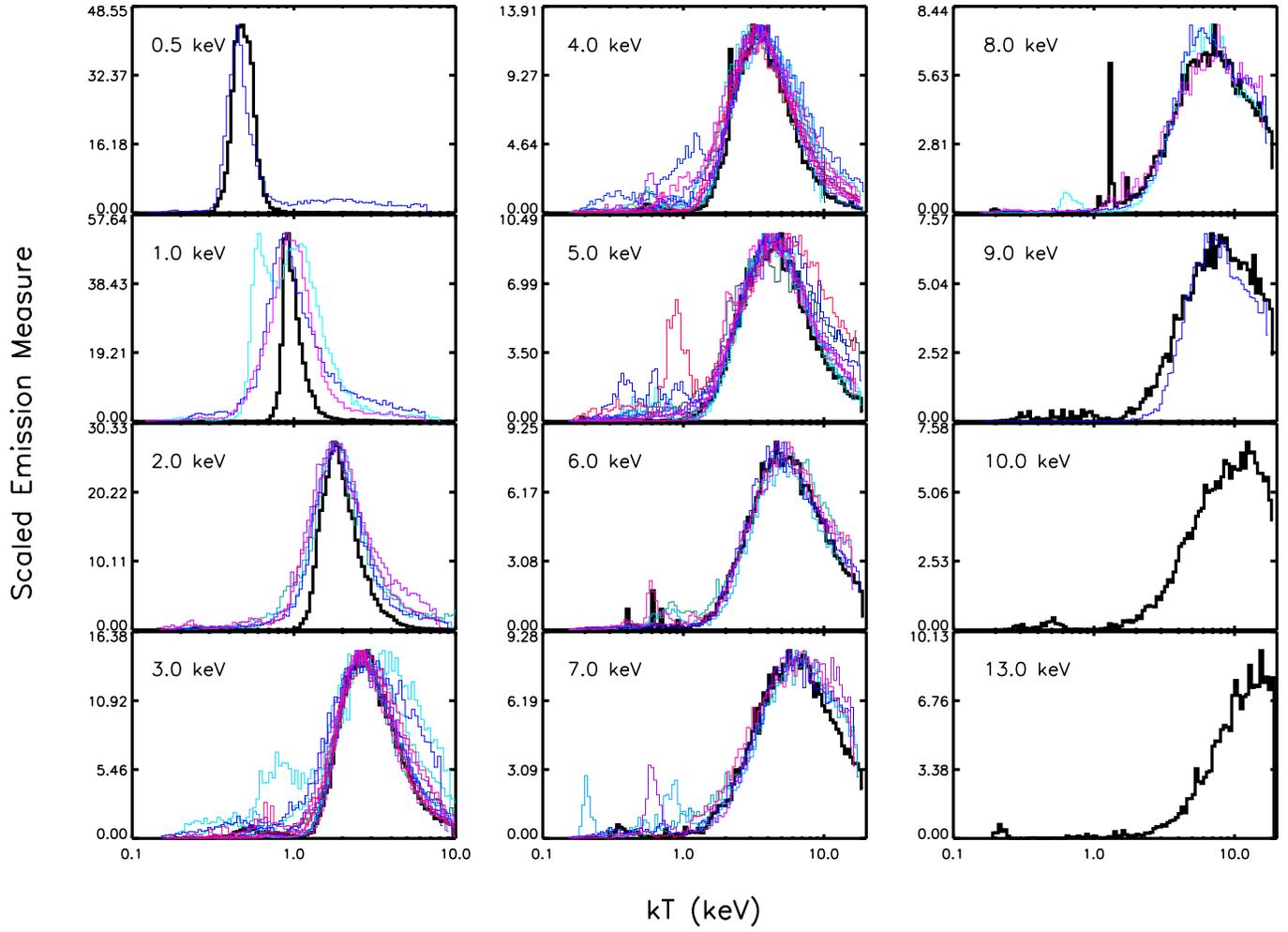}
\caption{\footnotesize Temperature distributions of the simulated isothermal clusters (black), with temperature distributions from the HIFLUGCS sample (the same as temperature distributions shown in Figures \ref{figure:tempdists0} - \ref{figure:tempdists3}) also shown for comparison (color).  Each simulated cluster is shown with all clusters in our sample having $kT_{med}$ within $0.5$ keV of the simulated cluster.  The HIFLUGCS temperature distributions are vertically scaled to match the simulations and horizontally shifted such that the median temperatures are lined up.}
\label{figure:tempdistscomp}
\end{center}
\end{figure*}

\end{document}